\documentclass[aps,prd,a4paper,showpacs,twocolumn]{revtex4-1}
\usepackage{amsmath,amssymb,bm,natbib}
\usepackage{gensymb}\usepackage{nicefrac} 
\usepackage{epsfig}
\usepackage{graphicx}
\usepackage{slashed}
\usepackage{xfrac}
\usepackage{epsfig} \usepackage{graphicx} \usepackage{color}
\usepackage{mathrsfs} \usepackage{amssymb} \usepackage{amsmath} \usepackage{url}

\usepackage{xspace} 
\usepackage[bookmarks, breaklinks, colorlinks,urlcolor=cyan, citecolor=red, 
linkcolor=blue]{hyperref}

\def\dsp{\displaystyle}
\def\be {\begin{equation}}
\def\ee {\end{equation}}
\def\bea {\begin{eqnarray}}
\def\eea {\end{eqnarray}}
\def\bc {\begin{center}}
\def\ec {\end{center}}
\def\nn {\nonumber}
\def\sss{\scriptscriptstyle}
\def\gev{\ensuremath{\mathrm{Ge\kern -0.1em V}}}
\def \thl {{\theta_\ell}}
\def \thK {{\theta_{K}}}
\def \alamL{{{\cal A}_\lambda^L}}
\def \alamR{{{\cal A}_\lambda^R}}
\def \azeL{{{\cal A}_0^L}}
\def \azeR{{{\cal A}_0^R}}
\def \apaL{{{\cal A}_\parallel^L}}
\def \apaR{{{\cal A}_\parallel^R}}
\def \apeL{{{\cal A}_\perp^L}}
\def \apeR{{{\cal A}_\perp^R}}
\def \Re{\text{Re}}
\def \Im{\text{Im}}
\def \at{{{\cal A}_t}}
\def \kstar{{K^{\!*}}}

\def \eff{{\text{eff }}}
\def \braket#1#2#3{\langle #1|#2| #3\rangle}
\def\AFB{A_{\text{FB}}}
\def\AFBo{A^\o_{\text{FB}}}
\def\o{\text{o}}
\def\hel#1{{\sss{#1}}}

\def\eff{\text{eff}}
\def\AFB{A_{\text{FB}}}
\def\AFBo{A^{\o}_{\text{FB}}}
\def\Gf{\Gamma_{\!\! f}}
\def\qmax{q^2_{\text{max}}}
\def\qmin{q^2_{\text{min}}}

\def\invfb   {\ensuremath{\mbox{\,fb}^{-1}}\xspace}

\def\ket|#1>{\left|#1 \right>}

\def\bra<#1|{\left< #1 \right|}

\def\bracket<#1|#2>{\setbox0=\vbox{\hbox{$#1$$#2$}}\left<#1\kern1pt
	\vrule  height\ht0\kern2pt #2\right>} 

\def\abs#1{\left| #1 \right|}

\def\dirmat<#1|#2|#3>{\setbox0=\vbox{\hbox{$#1$$#2$$#3$}}\left<#1\kern1pt
	\vrule height\ht0\kern1pt#2\kern1pt \vrule height\ht0\kern1pt
	#3\right>}

\allowdisplaybreaks

\begin{document}

\begin{flushright}
{\vspace*{-15mm}
\small IMSc/2014/8/7, DO-TH-14/19, QFET-2014-15
} 
\end{flushright}
\phantom{}
\vspace*{-17mm}


\title{Testing New Physics Effects in $\mathbf{B\to
    \kstar\ell^+\ell^-}$}

\author{Rusa Mandal}\email{rusam@imsc.res.in} 
\affiliation{The Institute of Mathematical
  Sciences, Taramani, Chennai 600113, India} 
\author{Diganta Das}\email{diganta.das@tu-dortmund.de} \affiliation{Institut 
f{\"u}r Physik, Technische 
Universit{\"a}t Dortmund, D-44221 Dortmund, Germany} 
\author{Rahul Sinha}\email{sinha@imsc.res.in}\affiliation{The Institute of 
Mathematical Sciences, 
Taramani, Chennai 600113, India}

\date{\today}

\begin{abstract}
It is generally believed that the decay mode $B\to\kstar\ell^+\ell^-$ is one of
the best modes to search for physics beyond the standard model. The angular
distribution enables the independent measurement of several observables as a
function of the dilepton invariant mass. The plethora of observables so obtained
enable unique tests of the standard model contributions. We start  by writing
the most general parametric form of the standard model amplitude for $B\to
\kstar\ell^+\ell^-$ taking into account comprehensively all contributions within
SM. These include all short-distance and long-distance effects, factorizable and
non-factorizable contributions, complete electromagnetic corrections to hadronic
operators up to all orders, resonance contributions and the finite lepton and
quark masses. The parametric form of the amplitude in the standard model results
a new relation involving all the $CP$ conserving observables. The derivation of
this relation only needs the parametric form of the amplitude and not a detailed
calculation of it. Hence, we make no approximations, however, innocuous. The
violation of this relation will provide a smoking gun signal of new physics. We
use the $1\invfb$ LHC$b$ data to explicitly show how our relation can be used to
test standard model and search for new physics that might contribute to this
decay.
\end{abstract}


\maketitle

\section{Introduction} 
\label{sec:Intro} 

It is a historical fact that several discoveries in particle physics were
preceded by indirect evidence through quantum loop contributions. It is for this
reason that significant attention is devoted to studying loop processes. The
muon magnetic moment is one of the best examples of such a process where
precision calculations have been done in order to search for new physics by
comparing the theoretical expectation with experimental observation. It is a
testimony to such searches for New Physics (NP) beyond the Standard Model (SM)
that both theoretical estimates and experimental observation have reached a
precision where the hadronic effects even for the lepton magnetic moment
dominate the discrepancy between theory and observation. Indirect searches for
new physics often involve precision measurement of a single quantity that is
compared to a theoretical estimate that also needs to be very accurately
calculated. Unfortunately, hadronic estimates involve calculation of long
distance QCD effects which cannot easily be done accurately, limiting the scope
of such searches. There exist, however, certain decay modes which involve the
measurement of several observables that can be related to each other with
minimal assumptions and completely calculable QCD contributions within the SM .
The break down of such relation(s) between observables would unambiguously
signal the presence of NP. Such tests are by nature not limited by incalculable
hadronic effects and hence provide an unambiguous signal of NP.  A well known
example~\cite{Kruger:1999xa,Das:2012kz} of such a process is the semileptonic
penguin decay $B\to \kstar \ell^+\ell^-$, where $\ell$ is either the electron or
the muon. In this paper we will show how this decay, which occurs in multiple
partial waves, can be used to obtain reliable tests of NP.

Flavor changing neutral current transitions are well known to be sensitive to NP
contributions. However, hadronic flavor changing neutral current receive short
and long distance QCD contributions that are not easy to estimate reliably. It
is evident from the data collected by the Belle, Babar and CMS collaborations at
the B-factories, CLEO, CDF, Tevatron and LHC$b$ that NP does not show up as a
large and unambiguous effect. This has bought into focus the need for approaches
that are theoretically cleaner i.e., where the hadronic uncertainties are much
smaller than the effects of NP that are being probed. Hence, to effectively
search for NP it is crucial to separate the effect of new physics from hadronic
uncertainties that can contribute to the decay. The decay mode $B\to \kstar
\ell^+\ell^-$ is regarded~\cite{Das:2012kz} as significant in this attempt. The
full angular analysis of the final state gives rise to a multitude of
observables~\cite{Sinha:1996sv, Kruger:1999xa} that are related as they arise
from the same decay mode. In addition, each of these observables can be measured
as function of the dilepton  invariant mass. In Ref.~\cite{Das:2012kz} an
interesting relation between the various observables that can be measured in
this mode was derived. The derivation was based on a few assumptions that are
reasonable. These included ignoring the mass of the lepton $\ell$ and the
$s$-quark that appears in the short distance Hamiltonian describing the decay.
The decay amplitude was assumed to be real, thereby ignoring the extremely tiny
$CP$ violation, the small imaginary contribution to the amplitude that arises
from the Wilson coefficient $C_9$ which is complex in general and the dilepton
resonances which were presumed to be removed from the experimental analysis.
These assumptions reduced the number of non-zero observables to only six. In
this paper, we carefully {\em redo the analysis without making any kind of
approximation, however, innocuous}. Our approach once again is to derive the
most general parametric form of the decay amplitude, which results in a relation
between the several related observables.

In this paper we generalize the derivation in Refs.~\cite{Das:2012kz} to
incorporate a complex decay amplitude, eliminating the need to ignore imaginary
contributions arising from $C_9$  and ensuring that the new relation is valid
even when resonance contributions are not excluded from the (experimental)
analysis. This implies that the new relation derived in this paper involves all
the nine $CP$ conserving observables that can be measured using this mode.  The
derivation of the new relation does not depend on theoretical values of the
Wilson coefficients and does not require making any assumptions on the
form-factors; in particular we do not limit the form-factor to any power of
$\Lambda_{\text{QCD}}/m_b$ expansion in heavy quark effect theory
(HQET)~\cite{Isgur-Wise}. In fact, the derivation of the new relation itself
does not require HQET. The new derivation parametrically incorporates all
short-distance and long-distance effects including resonance contributions, as
well as, factorizable and non-factorizable contributions. We also include
complete electromagnetic corrections to hadronic operators up to all orders.
Finally, we retain the lepton mass and the $s$-quark mass. We envisage that the
derivation to be exact in all respects and the new relation obtained here to be
one of the cleanest tests of the SM in $B$ decays.

The LHC$b$ collaboration has measured~\cite{Aaij:2013iag} all the possible $CP$
conserving observables through an angular analysis. These independent
measurements consist of the differential decay rate with respect to the dilepton
invariant mass, two independent helicity fractions and six angular asymmetries.
Three of the asymmetries are zero unless there exist imaginary contributions to
the decay amplitudes. If these asymmetries are measured to be zero in the
future, the relation between the observables would be free from any hadronic
parameter as derived in Ref.~\cite{Das:2012kz}. While these asymmetries are
currently measured to be small and consistent with zero, there could, however,
exist contributions from wide resonances which might still be permitted within
statistical errors. Including these asymmetries in the analysis to account for
complex amplitudes results in a modification of the relation purely between
observables. The modifying terms now involve a single hadronic parameter  in
addition to being proportional to the three asymmetries. Hence, SM can be tested
or equivalently NP contributions can be probed reliably with the knowledge of
just one hadronic parameter. It is interesting that all effort to estimate long
and short distance QCD contributions now need to be focused only on accurately
estimating this single parameter. Since the asymmetries involved in modifying
terms (which arise from complex amplitudes) are already constrained to be small,
the results are not very sensitive to the single hadronic parameter. We find
that the inclusion of imaginary contributions to the amplitude must always
reduce the parameter space. This would enhance any discrepancy that may be
observed even when the imaginary part of the amplitudes are ignored.  We use the
$1\invfb$ LHC$b$ data to show how our relation can be used to test Standard
model and find new physics that might contribute to this decay.

In this paper we review the theoretical framework required to describe $B\to
\kstar \ell^+\ell^-$ and derive the most general parametric form of the
amplitude describing the decay in Sec.~\ref{sec:framework}. The amplitude
written is notionally exact in all respects. In Sec.~\ref{sec:angular} we
construct all the observables in terms of the amplitude derived in
Sec.~\ref{sec:framework}. Here we retain the lepton mass as well as the strange
quark mass that appears in the short-distance Hamiltonian describing this decay.
A new relation between observables is derived in Sec.~\ref{sec:massless_case}
under the assumption of massless lepton, but retaining all other effects and
contribution. In Sec.~\ref{sec:massive_case} we generalize the new relation
derived in Sec.~\ref{sec:massless_case} to include the mass of the lepton that
had been ignored earlier. We re-derive two simple limits of the relation between
observables that hold at zero crossings of other asymmetries such as the
forward-backward asymmetry. The values of all the observables at kinematic
endpoints of the dilepton invariant mass are easily understood in
Sec.~\ref{sec:observables}. A numerical analysis is presented in
Sec.~\ref{sec:NPAnalysis} that tests the validity of the relation derived
assuming SM. We discuss the constraints already imposed by the $1\invfb$  LHC$b$
data~\cite{Aaij:2013iag}, but refrain from drawing even the obvious conclusions
given that results for $3\invfb$ data will soon be presented.  In
Sec.~\ref{sec:conclusions} we summarize the significant results obtained in our
paper.

\section{Theoretical Framework}
\label{sec:framework}

In this section we will discuss the most general from of the amplitude that can
describe the exclusive decay mode $B\to \kstar\ell^+\ell^-$ in the SM. The
description of the decay $B\to \kstar\ell^+\ell^-$ requires as the first step
the separation of short-distance effects which involve perturbative QCD and weak
interaction from the long distance QCD contributions in an effective
Hamiltonian.  As is well explained in literature, exclusive decay modes are a 
challenge to describe theoretically. This difficulty arises not only in the
need to know hadronic form-factors accurately but also from the existence of
``non-factorizable'' contributions that do not correspond to form-factors. These
contributions originate from electromagnetic corrections to the matrix element
of purely hadronic operators in the effective Hamiltonian. It has been
demonstrated~\cite{Beneke:2001at} that these non- factorizable corrections can
be computed allowing exclusive decay such as $B \to \kstar \gamma$ and $B\to
\kstar\ell^+\ell^-$ to be treated in a systematically much as their inclusive
decay counterparts. It is based on this theoretical understanding that we will
write the most general from of the amplitude for $B\to \kstar\ell^+\ell^-$ in
the SM. Our approach will be to examine the various factorizable and 
non-factorizable  contributions to the process and write the
most general parametric form of the amplitude without making any attempt to
evaluate it. 

The decays $B\to \kstar \ell^+\ell^-$ occurs at the quark level via a
$b\to s \ell^+\ell^-$ flavor changing neutral current transition.  The
short distance effective Hamiltonian for the inclusive process $b\to s
\ell^+\ell^-$ is given in the SM
by~\cite{Bobeth:1999mk,Altmannshofer:2008dz,Buchalla:1995vs},
\begin{widetext}
\begin{align}
  \label{eq:Hamiltonian}
  \mathcal{H}_\eff=&-4\frac{G_F}{\sqrt{2}} \Big[ V_{tb}V_{ts}^*
  \Big(C_1 {\cal O}_1^c+C_2 {\cal O}_2^c+\sum_{i=3}^{10} C_i {\cal
    O}_i \Big)
  + V_{ub}V_{us}^* \Big(C_1 ({\cal O}_1^c-{\cal O}_1^u)+C_2 ({\cal
    O}_2^c-{\cal O}_2^u)\Big) \Big].
\end{align}
\end{widetext}
The local operators ${\cal O}_i$ are as given in
Ref.~\cite{Altmannshofer:2008dz}, however, for completeness we present
the relevant operators that are dominant:
\begin{eqnarray}
  {\cal O}_7&=& \frac{e}{g^2} \big[\bar{s}\sigma_{\mu\nu}(m_b P_R+m_s
  P_L) b\big]F^{\mu\nu} ,\nn\\ 
  {\cal O}_9&=& \frac{e^2}{g^2}(\bar{s}\gamma_\mu  P_L
  b)\,\bar{\ell}\gamma^\mu \ell , \nn\\ 
  {\cal O}_{10}&=& \frac{e^2}{g^2}(\bar{s}\gamma_\mu P_L b)\,
  \bar{\ell}\gamma^\mu\gamma_5\ell\nn ,
\end{eqnarray}
where $g\,(e)$ is the strong (electromagnetic)coupling constant,
$P_{L,R}=(1\mp\gamma_5)/2$ are the left and right chiral projection
operators and $m_b\,(m_s)$ are the running $b\,(s)$ quark mass in the
$\overline{\text{MS}}$ scheme. The Wilson coefficients $C_i$ encode
all the short distance effects and are calculated in perturbation
theory at a matching scale $\mu=M_W$ up to desired order in the strong
coupling constant $\alpha_s$ before being evolved down to the scale
$\mu=m_b\approx 4.8\gev$. All NP contributions to $B\to \kstar\ell^+\ell^-$
contribute exclusively to $C_i$; this includes new Wilson coefficients
corresponding to new operators that arise from NP. 

Significant effort (see Ref.~\cite{Hurth:2010tk,Buras:2011we} for reviews) has
gone into evaluating the Wilson coefficients up to NNLO order.  As has been
stressed earlier~\cite{Buras:2011we} it is important to remember that ``the
construction of the effective Hamiltonian by means of operator product expansion
and renormalization group methods can be done fully in the perturbative
framework. The fact that the decaying hadron are bound states of quarks is
irrelevant for this construction.'' This implies that the $C_i$ are decay mode
independent. The dependence on the mode enters only through the matrix element
of local bilinear quark operators ${\cal O}_i$, i.e. $\langle f |{\cal
O}_i|B\rangle$, which encodes the long distance contributions. Since the decay
amplitude cannot depend on the scale $\mu$, $\langle f |{\cal O}_i|B\rangle$
must depend on the scale $\mu$ as well. The cancellation of $\mu$ dependence
generally involves several terms in the operator product expansion.  Since the
calculation of the hadronic matrix element involves long distance contributions,
non-perturbative methods are required. Much progress has been made in these
calculations using HQET as a tool. However, the dominant theoretical error in
the amplitude arises due to the lack of reliable calculations of the hadron
matrix element.

The simple picture of the decay presented above is unfortunately not accurate
enough; there exist several corrections making a reliable estimate of the decay
amplitude a challenge. The difficulty goes beyond accurately estimating the
form-factors involved in the hadron matrix element. There
exist~\cite{Beneke:2001at} additional non-factorizable and long-distance
contributions which arise from electromagnetic corrections to the matrix
elements of purely hadronic operators in the Hamiltonian that cannot be absorbed
into hadronic form-factors. These contributions are generated by current-current
operators ${\cal O}_{1,2}$ and penguin operators ${\cal O}_{3-6,8}$, combined
with electromagnetic interaction of quarks to produce $\ell^+\ell^-$. The
complication in dealing with these corrections is that the average distances
between the photon emission and the weak interaction points are not necessarily
short resulting in essentially non-local contributions to the decay amplitude
which cannot be reduced to local form-factors. A further challenge is that each
such contribution has to identified and estimated one by one. The intermediate
charm quark (and in principle the up quark) loops can couple to lepton pairs via
a virtual photon and even though these effects are sub-dominant numerically in
certain kinematical regions, they cannot be completely neglected. The other
quarks contribute negligibly (except for resonant contribution which we will
discuss later) to ${\cal O}_{1,2}$ and penguin operators ${\cal O}_{3-6,8}$ for
$B\to \kstar \ell^+\ell^-$ as they are either CKM suppressed or have small
accompanying Wilson coefficient. A remarkable effort
\cite{Buras:1994dj,exclusiveB} has gone into understanding the details of the
hadronic contributions in $B$ decays and in particular to $B\to \kstar
\ell^+\ell^-$. It is fortunate that the remarkable progress made so far, enables
us to write a completely accurate parametric form of the amplitude for this mode
in the SM.

LHC$b$ has observed a broad peaking structure~\cite{Aaij:2013pta,Lyon:2014hpa}
in the dimuon spectrum of $B\to K\ell^+\ell^-$. It would be of interest to see
if this observation of broad resonances has implication on
$B\to\kstar\ell^+\ell^-$ mode, since long distance effects would have to be
included systematically. The decay mode $B\to\kstar\ell^+\ell^-$ carries more
information~\cite{Sinha:1996sv,Kruger:1999xa} on the dynamics as compared to the
counterpart pseudoscalar mode $B\to K\ell^+\ell^-$, since the $\kstar$
polarization can also be measured. In order to study the dependence of the
amplitude on the helicity of the $\kstar$ we further consider the decay $
\kstar\to K\pi$ or the decay process $B\to\kstar\ell^+\ell^-\to
(K\pi)_{\sss\kstar}\ell^+\ell^-$. This further step itself does not complicate
matters. The decay amplitude in terms of hadronic matrix elements must therefore
include direct contributions proportional to $C_7$, $C_9$ and $C_{10}$
multiplied by $B\to \kstar$ form-factors and contributions from non-local
hadronic matrix elements ${\cal H}_i$ such that~\cite{Khodjamirian:2010vf,
Grinstein:2004},
\begin{widetext}
\begin{align}
\label{eq:FullAmp}
A(B(p)\to \kstar(k) \ell^+\ell^-)=&\frac{G_F\alpha}{\sqrt{2}\pi}V_{tb}V_{ts}^*
\bigg[\bigg\{\widehat{C}_9\braket{\kstar}{\bar{s}\gamma^{\mu}P_Lb}{\bar B}
-\frac{2\widehat{C}_7}{q^2}\braket{\kstar}{\bar{s} i\sigma^{\mu\nu}q_{\nu}
(m_b  P_R+m_s P_L)b} {\bar  B}\nn \\ 
&\qquad\qquad -\frac{16\pi^2}{q^2}\sum_{i=\{1-6,8\}}\widehat{C}_i 
{\cal H}_i^\mu \bigg\}\, \bar{\ell}\gamma_\mu \ell + 
\widehat{C}_{10}\braket{\kstar}
  {\bar{s}\gamma^{\mu}P_Lb}{\bar B} \,
  \bar{\ell}\gamma_\mu\gamma_5\ell \bigg],
\end{align}
\end{widetext}
where, $p=q+k$ with  $q$ being the dilepton invariant momentum  and the 
non-local hadron matrix element ${\cal H}_i^\mu$ is given by 
\begin{equation*}
\label{eq:non-local-el}
{\cal H}_i^\mu = \braket{\kstar(k)}{i\int d^4x\, e^{iq\cdot
    x}T\{j_{em}^\mu(x),{\cal O}_i(0)\}}{\bar B(p)}. 
\end{equation*}
In Eq.~\eqref{eq:FullAmp}, we have introduced {\em new notional theoretical
parameters $\widehat{C}_7$, $\widehat{C}_9$ and $\widehat{C}_{10}$ to indicate
the true values of Wilson coefficients, which are by definition not dependent on
the order of the perturbative calculation to which they are evaluated}. Our
definition is explicit and should not be confused with those defined earlier in
literature. The amplitude expressed in Eq.~\eqref{eq:FullAmp} is {\em notionally
complete and free from any approximations}. In this paper we do not attempt to
estimate the hardronic matrix element involved in Eq.~\eqref{eq:FullAmp},
instead we use Lorentz invariance to write out the most general form of the
hadron matrix elements $\braket{\kstar}{\bar{s}\gamma^{\mu}P_Lb}{\bar B(p)}$ and
$\braket{\kstar}{\bar{s} i \sigma^{\mu\nu}q_{\nu}P_{R,L}b} {\bar  B(p)}$ which
may be defined as
\begin{widetext}
\begin{eqnarray}
\label{eq:formfactor1}
&\braket{\kstar\!(\epsilon^*,k))}{\bar{s}\gamma^{\mu}P_L 
b}{B(p)}=\epsilon^*_\nu 
\Big(\mathcal{X}_0\,q^\mu q^\nu
+\mathcal{X}_1\,(g^{\mu\nu}\!-\!\dsp\frac{q^\mu 
q^\nu}{q^2})+\mathcal{X}_2\,(k^\mu\!-\! 
\dsp\frac{k.q}{q^2}q^\mu)q^\nu+i \mathcal{X}_3\,\epsilon^{\mu\nu\rho\sigma}\, 
k_{\rho}q_{\sigma}\Big),\\[2ex]
\label{eq:formfactor2}
&\braket{\kstar(\epsilon^*,k))}{i\bar{s}\sigma^{\mu\nu}q_{\nu}P_{R,L} b}{B(p)}= 
\epsilon^*_\nu\Big(\pm
\mathcal{Y}_1\,(g^{\mu\nu}\!-\!\dsp\frac{q^\mu q^\nu}{q^2}) \pm \mathcal{Y}_2\, 
(k^\mu\!-\!\dsp\frac{k.q}{q^2}q^\mu) q^\nu+i
\mathcal{Y}_3\, \epsilon^{\mu\nu\rho\sigma}\, k_{\rho}q_{\sigma}\Big).
\end{eqnarray}
We have written Eq.~\eqref{eq:formfactor1} such that the vector part of the
current in $\braket{\kstar\!(\epsilon^*,k))}{\bar{s}\gamma^{\mu}P_L b}{B(p)}$ is
conserved and only the $\mathcal{X}_0$ term in the divergence of the axial part
survives. Eq.~\eqref{eq:formfactor2} is also written so as to ensure that
$\braket{\kstar}{i\bar{s}\sigma^{\mu\nu}q_{\nu}P_{R,L} b}{B} q_\mu=0$. The
relations between $\mathcal{X}_{0,1,2,3}$ and $\mathcal{Y}_{1,2,3}$ and the
form-factors conventionally defined for on-shell $K^*$ are discussed in
Appendix~\ref{sec:form-factors}. It should be noted that form-factors
$\mathcal{X}_{0,1,2,3}$ and $\mathcal{Y}_{1,2,3}$  are functions of $q^2$ and
$k^2$, but we suppress the explicit dependence for simplicity of notation. The
subsequent decay of the $\kstar$, i.e., $\kstar(k) \to K(k_1) \pi(k_2)$ can be
easily taken into account~\cite{Kruger:1999xa,Altmannshofer:2008dz} resulting in
the hadronic matrix element
$\braket{[K(k_1)\pi(k_2)]_{\sss\kstar}}{\bar{s}\gamma^{\mu}P_L b}{B(p)}$ being
written as
\begin{eqnarray}
\label{eq:formfactor11}
&\braket{[K(k_1)\pi(k_2)]_{\sss\kstar}}{\bar{s}\gamma^{\mu}P_L 
b}{B(p)}=D_{\!\sss\kstar}\!(k^2) W_\nu 
\Big(\mathcal{X}_0\,q^\mu q^\nu
+\mathcal{X}_1\!(g^{\mu\nu}\!-\!\dsp\frac{q^\mu 
q^\nu}{q^2})+\mathcal{X}_2\,(k^\mu\!- \!
\dsp\frac{k.q}{q^2}q^\mu)q^\nu+i \mathcal{X}_3\,\epsilon^{\mu\nu\rho\sigma}\, 
k_{\rho}q_{\sigma}\Big) ,\\[2ex] 
\label{eq:formfactor21}
&\braket{[K(k_1)\pi(k_2)]_{\sss\kstar}}{i\bar{s}\sigma^{\mu\nu}q_{\nu}P_{R,L} 
b}{B(p)}= 
D_{\!\sss\kstar}\!(k^2) W_\nu \Big(\pm
\mathcal{Y}_1\,(g^{\mu\nu}\!-\!\dsp\frac{q^\mu q^\nu}{q^2}) \pm \mathcal{Y}_2\, 
(k^\mu \!-\!\dsp\frac{k.q}{q^2}q^\mu) q^\nu+i
\mathcal{Y}_3\, \epsilon^{\mu\nu\rho\sigma}\, k_{\rho}q_{\sigma}\Big),
\end{eqnarray}
\end{widetext}
where, the subscript $\kstar$ in $[K(k_1)\pi(k_2)]_{\sss\kstar}$ indicates that 
the final sate is produced by the decay of a $\kstar$, $D_{\!\sss\kstar}\!(k^2) 
$ is the $\kstar$ propagator, so that
\begin{equation}
\label{eq:prop}
|D_{\!\sss\kstar}\!(k^2)|^2= \frac{g^2_{\sss\kstar \! \sss K \sss\pi}}
  {(k^2-m^2_\kstar)^2+(m_\kstar \Gamma_\kstar)^2}, 
\end{equation}
 with $g_{\sss\kstar \! \sss K \sss\pi}$ being the $\kstar \! K\pi$ coupling 
 and the other parameters introduced are
\begin{equation*}
W_\nu=K_\nu-\xi k_\nu,~K=\!k_1-k_2,~k=\!k_1+k_2,~\xi=\frac{k_1^2-k_2^2}{k^2}.
\end{equation*}
The most general expression for the hadronic matrix element $\mathcal{H}_i^\mu$ can also 
be
written using Lorentz invariance. Since this hadronic matrix element arises from
non-local contributions at the quark level, it involves introducing ``new'' form
factors $\mathcal{Z}_1^i$, $\mathcal{Z}_2^i$ and $\mathcal{Z}_3^i$ 
corresponding 
to 
non-factorizable
contribution from each $\mathcal{H}_i^\mu$ in analogy with those  introduced in
Eq.~\eqref{eq:formfactor1} as follows:
\begin{widetext}
\begin{align}
\label{eq:Hmu}
{\cal H}_i^\mu &= \braket{\kstar\!(\epsilon^*,k)}{i\int d^4x\, e^{iq\cdot
    x}T\{j_{em}^\mu(x),{\cal O}_i(0)\}}{\bar B(p)} \nn \\
    &=\epsilon^*_\nu \Big(\mathcal{Z}_1^i\,(g^{\mu\nu}
    -\dsp\frac{q^\mu q^\nu}{q^2})+\mathcal{Z}_2^i\,(k^\mu - 
    \dsp\frac{k.q}{q^2}q^\mu)q^\nu +i 
    \mathcal{Z}_3^i\,\epsilon^{\mu\nu\rho\sigma}\, 
    k_{\rho}q_{\sigma}\Big).
\end{align}
\end{widetext}
Our definition follows Ref.~\cite{Beneke:2001at} of ``non-factorizable'' and
includes those corrections that are not contained in the definition of
form-factors introduced in Eqs.~\eqref{eq:formfactor1} and
\eqref{eq:formfactor2}. Here the most general form of ${\cal H}_i^\mu$ is
written to ensure the conservation of EM current i.e, $~q_\mu {\cal H}_i^\mu=0$.

The non-local effects represented by $\mathcal{H}_i^\mu$ can be taken into account by
absorbing the contributions into redefined $\widehat{C}_9$ and modifying the
contribution from the electromagnetic dipole operator ${\cal O}_7$. The
electromagnetic corrections to operators ${\cal O}_{1-6,8}$ can also contribute
to $B\to\kstar\gamma$ at $q^2=0$. Since, only the Wilson coefficient
$\widehat{C}_7$ contributes to $B\to\kstar\gamma$, the charm-loops at $q^2=0$
must contribute to $\widehat{C}_7$ in order for the Wilson coefficient to be
process independent. It is easily seen that the effect of this is to modify the
$\widehat{C}_7\braket{K\pi}{\bar{s}i \sigma^{\mu\nu}q_{\nu}(m_b P_R+m_sP_L)b}
{\bar  B}$ terms such that the form-factors and Wilson coefficients mix in an
essentially inseparable fashion. This holds true even for the leading
logarithmic contributions~\cite{Beneke:2001at,Grinstein:1988me}. Both
factorizable and non-factorizable contributions arising from electromagnetic
corrections to hadronic operators up to all orders can in principle be included
in this approach. The remaining contributions can easily be absorbed into a
redefined ``effective'' Wilson coefficient $\widehat{C}_9$ defined such that
\begin{equation}
\widehat{C}_9\to \widetilde{C}_9^{(j)}=\widehat{C}_9+\Delta 
C_9^{\text{(fac)}}(q^2)+\Delta C_9^{{(j)}{\text{,(non-fac)}}}(q^2)
\end{equation}
where, $j=1,2,3$ and $\Delta C_9^{\text{(fac)}}(q^2)$, $\Delta 
C_9^{\text{(non-fac)}}(q^2)$
correspond to factorizable and soft gluon non-factorizable contributions. Note
that the non-factorizable contributions necessitates the introduction of new
form-factors $\mathcal{Z}_j$ and the explicit dependence on 
$\mathcal{Z}_j/\mathcal{X}_j$ is absorbed   in
defining 
\begin{align}
\Delta C_9^{\text{(fac)}}+\Delta C_9^{{(j)}{\text{,(non-fac)}}}= - 
\frac{16 \pi^2}{q^2}\!\!\!\sum_{i=\{1-6,8\}} \!\!\widehat{C}_i 
\,\frac{\mathcal{Z}^i_j}{\mathcal{X}_j},
\end{align}
resulting in the $j$ dependence of the term as indicated. We also mention that
there is no non-factorizable correction term in Eq.~\eqref{eq:Hmu} analogous to 
$\mathcal{X}_0$ (in
Eq.~\eqref{eq:formfactor1}) due EM current conservation as discussed above.

The corresponding corrections to $\widehat{C}_7$ are taken into by the
replacement,
\begin{equation}
\!\frac{2(m_b\!+\!m_s)}{q^2}\widehat{C}_7\, \mathcal{Y}_j \to
\mathcal{\widetilde{Y}}_j=\frac{2(m_b+m_s)}{q^2}\widehat{C}_7\, 
\mathcal{Y}_j+\cdots,
\end{equation}
where the dots indicate other factorizable and  non-factorizable contributions
and the factor $2(m_b+m_s)/q^2$ has been absorbed in  the form-factors
$\mathcal{\widetilde{Y}}_j$.  Note that the $\mathcal{\widetilde{Y}}_j$'s are in
general complex because of the non-factorizable contributions to the Wilson
coefficient $\widehat{C}_7$, but on-shell quarks and resonances do not contribute
to them. It should be noted that $\widetilde{C}^{{(j)}}_9$ includes
contributions from both factorizable and non-factorizable effects, whereas
$\widehat{C}_{10}$ is unaffected by strong interaction effects coming from
electromagnetic corrections to hadronic operators.   The use of a `widetilde'
versus  `widehat'  throughout the paper is also meant as a notation to indicate
this fact. It should be noted that $\widehat{C}_{10}$ is real in the SM,
whereas, $\widetilde{C}_9^{(j)}$ and $\mathcal{\widetilde{Y}}_j$ are in general
complex within the SM. The amplitude in Eq.~\eqref{eq:FullAmp} can therefore be
written as
\begin{widetext}
\begin{align}
\label{eq:FullAmp-1} A\big(B(p)\to&
[K(k_1)\pi(k_2)]_{\sss\kstar}\ell^+\ell^-\big)=
\frac{G_F\alpha}{\sqrt{2}\pi}V_{tb}V_{ts}^*D_{\!\sss\kstar}\!(k^2) 
\nn \\ &\bigg[\bigg\{\Big(C_L W.q \mathcal{X}_0 \,q^\mu +C_L^1
\mathcal{X}_1\,(K^\mu-\frac{W.q}{q^2}q^\mu-\xi k^\mu)
+C_L^2 W.q \mathcal{X}_2 ( k^\mu-\frac{k.q}{q^2}q^\mu) +i C_L^3 \mathcal{X}_3\,
\epsilon^{\mu\nu\rho\sigma}\,K_\nu k_{\rho}q_{\sigma}\Big) \nn \\
&-\Big(\zeta\,\mathcal{\widetilde{Y}}_1\,
(K^\mu-\frac{W.q}{q^2}q^\mu-\xi k^\mu) +  \zeta W.q\, \mathcal{\widetilde{Y}}_2 
( k^\mu
-\frac{k.q}{q^2}\,q^\mu) +i\, \mathcal{\widetilde{Y}}_3\,
\epsilon^{\mu\nu\rho\sigma}\,K_\nu k_{\rho}q_{\sigma}\Big)\bigg\}\,
\bar{\ell}\gamma_\mu\,P_L \ell + L\to R \bigg],
\end{align}
\end{widetext}
where,
$C_{L,R}=\widehat{C}_9\mp \widehat{C}_{10}$,
$C_{L,R}^{(j)}=\widetilde{C}_9^{(j)}\mp \widehat{C}_{10}$
 and
$\zeta=(m_b-m_s)/(m_b+m_s)$.
It may be noted that no  assumption are made in  obtaining
Eq.~\eqref{eq:FullAmp-1} from Eq.~\eqref{eq:FullAmp}. The form-factors defined
are not limited by power corrections in Heavy Quark Effective Theory
(HQET)~\cite{BF00}. We emphasize that Eq.~\eqref{eq:FullAmp-1} continues to be
notionally exact. In our approach we will relate observables, hence, we do not
need to evaluate the Wilson coefficients and form-factors. Only in doing so
approximations need to be made.  In Appendix~\ref{sec:form-factors} comparative
relation between the amplitude in Eq.~\eqref{eq:FullAmp-1} and the leading order
expression excluding non-factorization contribution used widely in literature
are presented. These approximations are unnecessary for the discussions in this
paper and are presented only as clarification of our notation and as ready
reference for readers wanting to examine Eq.~\eqref{eq:FullAmp-1} in limiting
conditions.

\section{Angular Distribution and observables.}
\label{sec:angular}
\begin{widetext}
The decay $\bar{B}(p)\to \kstar(k)\ell^+(q_1)\ell^-(q_2)$,with $\kstar(k)\to
K(k_1)\pi(k_2)$ on the mass shell, is completely describe by four independent
kinematic variables. These kinematic variables are the lepton-pair invariant
mass squared $q^2=(q_1+q_2)^2$, and the three angles $\phi$, $\theta_\ell$ and
$\theta_K$. The angle $\phi$ is the angle between the decay planes formed by
$\ell^+\ell^-$ and $K\pi$. The angles $\theta_\ell$ and $\theta_K$ are defined
as follows: assuming that the $\kstar$ has a momentum along the positive $z$
direction in $B$ rest frame, $\theta_K$ is the angle between the $K$ and the
$+z$ axis and $\theta_\ell$ is the angle of the $\ell^-$ with the $+z$ axis. 
The differential decay distribution of $B\to \kstar\ell^+\ell^-$ is written as %

\begin{align}
 \label{eq:helicity}
  \frac{d^4\Gamma(B\to \kstar\ell^+\ell^-)}{dq^2\, d\cos\thl\,
	d\cos\thK\, d\phi}
& =
  I(q^2,\thl,\thK,\phi)= \frac{9}{32\pi}\Big[ I_1^s \sin^2\thK + I_1^c
  \cos^2\thK + (I_2^s \sin^2\thK + I_2^c \cos^2\thK) \cos 2\thl
  \nonumber \\ &+ I_3 \sin^2\thK \sin^2\thl \cos 2\phi + I_4 \sin
  2\thK \sin 2\thl \cos\phi + I_5 \sin 2\thK \sin\thl \cos\phi + I_6^s
  \sin^2\thK \cos\thl\nonumber \\ &+ I_7 \sin 2\thK \sin\thl \sin\phi
  + I_8 \sin 2\thK \sin 2\thl \sin\phi + I_9 \sin^2\thK \sin^2\thl
  \sin 2\phi\Big]\,.
\end{align}
\end{widetext}
The angular coefficients $I$'s, which can be measured from the study of 
the angular distribution, are $q^2$ dependent. But for convenience 
we will suppress the explicit $q^2$ dependence.

The $I$'s are conveniently expressed in terms of ``seven'' amplitudes. These
compromise of the six transversity amplitudes that survive in the massless
lepton limit and an amplitude ${\cal A}_t$ that contributes only if the mass $m$
of the lepton is finite. The six transversity amplitudes ${\cal
A}_{\perp,\parallel,0}^{L,R}$, where $\perp$, $\|$ and $0$ represent the
polarizations of the on shell $\kstar$ and $L$, $ R$ denote the chirality of the
lepton current.  The explicit expression for $I$'s in terms of the transversity
amplitudes ${\cal A}_{\perp,\parallel,0}^{L,R}$ and ${\cal A}_t$ are
\begin{subequations}
\begin{align}
\label{eq:I1s}
  I_1^s & = \frac{(2+\beta^2)}{4} \Big[|\apeL|^2 + |\apaL|^2 +
    (L\to R) \Big]  \nn \\
    &+\frac{4m^2}{q^2}\Re(\apeL^{}\apeR^{*}+\apaL^{}\apaR^{*}), \\ 
\label{eq:I1c}
  I_1^c & = |\azeL|^2 \!+\!|\azeR|^2\!+\!\frac{4m^2}{q^2}\big[\abs{{\cal A}_t}^2
  \!+\!2\Re(\azeL^{}\azeR^{*})\big],\\
  I_2^s & = \frac{\beta^2}{4}\Big[ |\apeL|^2+ |\apaL|^2 + (L\to
    R)\Big],\\ 
  I_2^c & = -\beta^2 \Big[|\azeL|^2 + (L\to R)\Big],\\
  I_3 & = \frac{\beta^2}{2}\Big[ |\apeL|^2 - |\apaL|^2  + (L\to
    R)\Big],\\  
  I_4 & = \frac{\beta^2}{\sqrt{2}}\Big[\Re (\azeL^{}\apaL^*) +
    (L\to R)\Big],\\
  I_5 & = \sqrt{2}\beta\Big[\Re(\azeL^{}\apeL^*) - (L\to R)
\Big], \\
  I_6^s  & = 2\beta\Big[\Re (\apaL^{}\apeL^*) - (L\to R) \Big],\\
  I_7 & = \sqrt{2}\beta\Big[\Im (\azeL^{}\apaL^*) - (L\to R)
  \Big],\\
  I_8 & = \frac{1}{\sqrt{2}}\beta^2\Big[\Im(\azeL^{}\apeL^*) +
    (L\to R)\Big],\\
  I_9 & =\beta^2\Big[\Im (\apaL^{*}\apeL) + (L\to R)\Big],
\end{align}
\end{subequations}
where $$\beta=\sqrt{1-\frac{4\,m^2}{q^2}}.$$ We 
have dropped the explicit $q^2$ dependence of the transversity amplitudes ${\cal
A}_{\perp,\parallel,0}^{L,R}$ and ${\cal A}_t$ for notational simplicity.

The seven amplitudes can be written in terms of the form-factors 
$\mathcal{X}_{0,1,2,3}$ and $ \mathcal{Y}_{1,2,3}$ as follows:
\begin{widetext}
\begin{subequations}
\begin{align}
\label{eq:trans-amp1} 
&{\cal A}_{\perp}^{L,R} \!= N\sqrt{2} \lambda^{\nicefrac{1}{2}}\!(m_B^2, 
m_\kstar^2,q^2)
\Big[ (\widetilde{C}_9^{(3)} \mp \widehat{C}_{10}) \mathcal{X}_3- 
\mathcal{\widetilde{Y}}_3\Big],\\
\label{eq:trans-amp2}
&{\cal A}_{\parallel}^{L,R}  = 2\sqrt{2} N \Big[ 
(\widetilde{C}_9^{(1)} \mp \widehat{C}_{10})   
\mathcal{X}_1- \zeta\,\mathcal{\widetilde{Y}}_1 \Big],\\
  \label{eq:trans-amp3}
&{\cal A}_{0}^{L,R}  =  \frac{N}{2 m_\kstar \sqrt{q^2}} 
  \Big[ (\widetilde{C}_9^{(2)}\kappa \mp \widehat{C}_{10})
  \big\{ 4 k.q \mathcal{X}_1 +\lambda(m_B^2, m_\kstar^2,q^2) 
  \mathcal{X}_2\big\}  
  -\zeta\big\{ 4 k.q\mathcal{\widetilde{Y}}_1 +\lambda(m_B^2, 
 m_\kstar^2,q^2) 
 \mathcal{\widetilde{Y}}_2\big\}\Big], \\
\label{eq:trans-ampt}
&{\cal A}_t = -\frac{N}{m_{K}^*}\sqrt{q^2}\lambda^{1/2}(m_B^2, m_\kstar^2,q^2)
  \widehat{C}_{10}~\mathcal{X}_0,
\end{align}
\end{subequations}
\end{widetext}
where, 
\begin{equation*}
\kappa=1+\frac{\widetilde{C}_9^{(1)} - 
\widehat{C}_{9}^{(2)}}{\widehat{C}_{9}^{(2)}} \frac{4 k.q \mathcal{X}_1}{4 k.q 
\mathcal{X}_1 
+\lambda(m_B^2, m_\kstar^2,q^2)\mathcal{X}_2},
\end{equation*}
$\lambda(a, 
b,c)\equiv  a^2 + b^2 + c^2 -2 (a b + b c + a c)$
and $N$ is the normalization constant. In the narrow width approximation 
for the $\kstar$, $|D_{\!\sss\kstar}\!(k^2)|^2$  simplifies to
\begin{equation}
|D_{\!\sss\kstar}\!(k^2)|^2=\frac{48 
\pi^2 m^4_\kstar} 
{\lambda^{3/2}(m^2_\kstar,m^2_K,m^2_\pi)}\delta(k^2-m^2_\kstar).
\end{equation}
This results in simplifying $N$ to,
\begin{equation*}
N= V_{tb}^{\vphantom{*}}V_{ts}^* \left[\frac{G_F^2 \alpha^2}{3\cdot 2^{10}
\pi^5 m_B^3} q^2 \sqrt{\lambda(m_B^2, m_\kstar^2,q^2)}\beta\right]^{1/2}.
\end{equation*} 
We note that in principle the effect of finite $\kstar$ resonance width can
easily be taken into account, however, we make no attempt to do so as the value
of the normalization constant is not going to be used anywhere in our
calculation.

The six transversity amplitudes described by Eqs.~\eqref{eq:trans-amp1} --
\eqref{eq:trans-amp3} which survive in the massless lepton case, can be
rewritten in a short-form notation by introducing new form-factors
$\mathcal{F}_\lambda$ and $\widetilde{\mathcal{G}}_\lambda$ as follows,
\begin{equation}
\mathcal{A}_\lambda^{L,R}=C_{L,R}^{\sss\lambda}\,\mathcal{F}_\lambda 
-\widetilde{\mathcal{G}}_\lambda
\label{eq:amp-def1}
=\big(\widetilde{C}_9^{\sss\lambda}\mp 
\widehat{C}_{10})\mathcal{F}_\lambda -\widetilde{\mathcal{G}}_\lambda.
\end{equation}
The expressions of $\mathcal{F}_\lambda$ and $\widetilde{\mathcal{G}}_\lambda$
can be obtained by comparing Eq.~{(\ref{eq:amp-def1}) with
Eqs.~(\ref{eq:trans-amp1}) -- (\ref{eq:trans-amp3}) and are given in
Appendix-(\ref{sec:form-factors}). $\mathcal{F}_\lambda$ and
$\widetilde{\mathcal{G}}_\lambda$ are $q^2$ dependent form-factors, suitably
defined to include both factorizable and non-factorizable corrections to all 
orders
\cite{Das:2012kz}. The form-factor dependence of
$\widetilde{C}_9^{(j)}$ indicated by `$j$' in Eqs.~(\ref{eq:trans-amp1}) --
(\ref{eq:trans-amp3}) is now translated to an effective helicity `$\lambda$'
dependence of Wilson coefficient $\widetilde{C}_9^{\sss\lambda}$ as
\begin{eqnarray}
\widetilde{C}_9^{\sss\perp} \equiv \widetilde{C}_9^{(3)},~~
 \widetilde{C}_9^{\sss\|} \equiv \widetilde{C}_9^{(1)},~~
\widetilde{C}_9^{\sss 0} \equiv \widetilde{C}_9^{(2)}\kappa  .
\end{eqnarray} 
It is easily seen that $\mathcal{F}_\lambda$  and
$\widetilde{\mathcal{G}}_\lambda$ are proportional to $\mathcal{X}_j$ and
$\mathcal{\widetilde{Y}}_j$ respectively. Thus $\mathcal{F}_\lambda$'s are 
completely
real and $\widetilde{\mathcal{G}}_\lambda$'s are complex in SM. All imaginary
contributions to the amplitude arise from the complex
$\widetilde{C}_9^{\sss\lambda}$ and $\widetilde{\mathcal{G}}_\lambda$. An
interesting observation is that $\mathcal{A}_\lambda^{L,R}$ remains unchanged if
the non-factorizable contributions between $\widetilde{\mathcal{G}}_\lambda$ and
$\widetilde{C}_9^\hel{\lambda}$ are rearranged. This observation differs from
the conclusion obtained in Ref.~\cite{Das:2012kz} because
$\widetilde{C}_9^\hel{\lambda}$ are now helicity dependent and implies that
$\widetilde{\mathcal{G}}_\lambda$ and $\widetilde{C}_9^\hel{\lambda}$ cannot be
individually extracted. 

Using very general arguments it is easy to see that the form of the amplitude
described in Eq.~\eqref{eq:amp-def1} is the most general possible and the full
decay amplitude can be completely described by them for the
massless case. The amplitude must be described by the helicity of the $\kstar$
and can be divided into two parts one that depends on the chirality of the
lepton and another that does not. It is easily noted that the term described by
${\cal F}_\lambda$ is chirality dependent whereas the contribution corresponding
to the effective photon vertex  $\widetilde{{\cal G}}_\lambda$ is not. The form
factors $\mathcal{F}_\lambda$ and $\widetilde{\mathcal{G}}_\lambda$ depend only
on the helicity and the chirality dependence is absorbed completely into the
Wilson coefficients.  The coefficient of chirality dependent terms proportional 
to ${\cal F}_\lambda$ can themselves  either depend on helicity or be 
independent of it. Hence, the amplitudes in Eq.~\eqref{eq:amp-def1} are  
parameterized in terms of three terms. Throughout the rest of the paper we will 
use only the form of the amplitudes in Eq.~\eqref{eq:amp-def1}, which is the 
most general possible in the SM.

It is obvious from Eq.~(\ref{eq:helicity}) that a complete study of the angular
distribution involves eleven orthogonal terms  allowing us to measure `eleven'
observables. In the limit of massless lepton there exist two relations between
the coefficient $I$'s, i.e. $I_1^c=-I_2^c$ and $I_1^s=3 I_2^s$. This
reduces the number of independent observables to `nine'. 
We will divide our discussion into two parts. In Sec.~\ref{sec:massless_case} we
will restrict our discussion by assuming that the lepton is massless
and in Sec.~\ref{sec:massive_case} we will generalize the discussion to the
massive lepton case. In a previous paper~\cite{Das:2012kz} the mode $B\to \kstar
\ell^+\ell^-$ was studied in the limit of massless lepton and under the
assumption of vanishing $CP$ violation and absence of resonance contributions in
the $q^2$ domains considered. Under these approximations $I_{7,8,9}=0$ and the
number of useful observables reduce to only `six'. In this paper we carefully
examine each of these assumptions and in particular take into account resonance
contributions and the effect of massless lepton. As emphasized in
Sec.~\ref{sec:framework} we have taken into account charm loop effects. The 
charm
loop effect and other resonance contributions can make the amplitude complex. In
the discussions that pursue we will assume that the amplitude is complex and
ensure that all SM contributions, both factorizable and non-factorizable, are
taken into  account completely when writing the most general parameterized
amplitude.

Within SM, $CP$-violation is expected to be extremely tiny and essentially
unobservable~\cite{Sinha:1996sv,Kruger:1999xa} at the current level of
experimental accuracy. In Ref.~\cite{Kruger:1999xa} the  $CP$ violating
asymmetry was evaluated to be $\sim 3\times 10^{-4}$. This would imply that one
need at the very least $10^7$ reconstructed events in this decay channel to
observe the asymmetry at $1\sigma$. Given this we have justifiably ignored $CP$
violation in this channel and any observation of $CP$ violation at the current
level of experimental sensitivity would constitute an unambiguous signal of NP.
In view of this, we ignore $CP$ violation hence forth. It may be noted that $CP$
violation can be easily included in our approach. However, we ignore it because
it is not central to our discussion and we do not wish to complicate our
notation accounting for unobservable effects within the SM. Under the assumption
of vanishing $CP$ violation the conjugate mode $\bar{B}\to\bar{K}^*\ell^+\ell^-$
has an identical decay distribution except that $I_{5,6,8,9}$ switch signs to
become $-I_{5,6,8,9}$ in the differential decay distribution~\cite{Sinha:1996sv,
Kruger:1999xa}.

Integration over $\cos\thK$, $\cos\thl$ and $\phi$ results in the differential
decay rate with respect to the invariant lepton mass:
\begin{equation}
  \label{eq:DGamma}
  \frac{d\Gamma}{dq^2}=\sum_{\lambda=0,\|,\perp}(|{\cal
	A}_\lambda^L|^2+|{\cal  A}_\lambda^R|^2).
\end{equation}
We define the relevant observables to be the three helicity fractions defined as
\begin{subequations}
\begin{align}
  \label{eq:FL}
  F_L&= \dsp \frac{|\azeL|^2 +|\azeR|^2}{\dsp \Gamma_{\!f}}~,\\
  \label{eq:Fparallel}
  F_\|&= \dsp\frac{|\apaL|^2 +|\apaR|^2}{\dsp\Gamma_{\!f}}~,\\
  \label{eq:Fperp}
  F_\perp&= \frac{|\apeL|^2 +|\apeR|^2}{\dsp\Gamma_{\!f}}~,
\end{align}
\end{subequations}
where $ \Gamma_{\!\!f}\equiv\sum_\lambda(|{\cal A}_\lambda^L|^2+|{\cal
A}_\lambda^R|^2)$ and $F_L+F_\|+F_\perp=1$. The other observables are the six
asymmetries defined below. The well known forward--backward asymmetry
$\AFB$ is defined conventionally as,
\begin{equation}
  \label{eq:AFB}
  \AFB=\dsp\frac{\Big[\dsp\int_0^1-\dsp \int_{-1}^0\Big]\dsp
    d\cos\thl \frac{d^2 
      (\Gamma-\bar{\Gamma})}{d q^2d\cos\thl}}{\dsp\int_{-1}^1 \dsp d\cos\thl
    \frac{d^2 (\Gamma+\bar{\Gamma})}{d q^2d\cos\thl}}~,
\end{equation}
and isolates the contribution from the $I_6$ term in Eq.~\eqref{eq:helicity}.
\begin{widetext}
Contributions from $I_4$ and $I_5$  in Eq.~\eqref{eq:helicity} are extracted by
the two angular asymmetries,

\begin{align}
  \label{eq:A4}
  A_{4}=&\frac{\Big[\dsp\int_{-\pi/2}^{\pi/2}-\dsp\int_{\pi/2}^{3\pi/2} 
  \Big]d\phi
    \Big[\dsp\int_0^1 -\dsp\int_{-1}^0 \Big] d\cos\thK 
    \Big[\int_0^1 -\int_{-1}^0 \Big] d\cos\thl \dsp
    \frac{d^4(\Gamma+\bar{\Gamma})}{dq^2d\cos\thl d\cos\thK d\phi}} 
  {\dsp\int_0^{2\pi}d\phi\int_{-1}^1d\cos\thK \int_{-1}^1d\cos\thl\,
    \frac{d^4(\Gamma+\bar{\Gamma})}{dq^2d\cos\thl d\cos\thK d\phi}}~,\\
  \label{eq:A5}
  A_{5}=&\frac{\Big[\dsp\int_{-\pi/2}^{\pi/2} -\dsp\int_{\pi/2}^{3\pi/2} 
  \Big]d\phi
    \Big[\int_0^1 -\int_{-1}^0 \Big] d\cos\thK 
    \dsp\int_{-1}^1 d\cos\thl~
    \dsp\frac{d^4(\Gamma-\bar{\Gamma})}{dq^2d\cos\thl d\cos\thK d\phi} }
  {\dsp\int_0^{2\pi}d\phi\int_{-1}^1d\cos\thK \int_{-1}^1d\cos\thl\,
      \frac{d^4(\Gamma+\bar{\Gamma})}{dq^2d\cos\thl d\cos\thK d\phi}}~.
\end{align}
The three new observables not considered in Ref.~\cite{Das:2012kz} are $A_7$, 
$A_8$ and $A_9$. These are non-zero if the amplitude is complex. They may be 
described in analogy as,

\begin{align}
  \label{eq:A7} 
  A_{7}=&\frac{\Big[\dsp\int_{0}^{\pi}-\int_{\pi}^{2\pi} \Big]d\phi 
  \Big[\int_0^1 -\int_{-1}^0 \Big]d\cos\thK 
  \dsp\int_{-1}^1 d\cos\thl~
  \dsp\frac{d^4(\Gamma+\bar{\Gamma})}{dq^2d\cos\thl d\cos\thK d\phi} }
  {\dsp\int_0^{2\pi}d\phi\int_{-1}^1d\cos\thK \int_{-1}^1d\cos\thl\,
      \frac{d^4(\Gamma+\bar{\Gamma})}{dq^2d\cos\thl d\cos\thK d\phi}}~,\\
  \label{eq:A8}
  A_{8}=&\frac{\Big[\dsp\int_{0}^{\pi} -\dsp\int_{\pi}^{2\pi}\Big] d\phi 
  \Big[\dsp\int_0^1 -\dsp\int_{-1}^0 \Big]d\cos\thK  
   \Big[\int_0^1  -\int_{-1}^0 \Big] d\cos\thl \dsp
  \frac{d^4(\Gamma-\bar{\Gamma})}{dq^2d\cos\thl d\cos\thK d\phi}}
  {\dsp\int_0^{2\pi}d\phi\int_{-1}^1d\cos\thK \int_{-1}^1d\cos\thl\,
  \frac{d^4(\Gamma+\bar{\Gamma})}{dq^2d\cos\thl d\cos\thK d\phi}},\\
  \label{eq:A9}
  A_{9}=&\frac{\Big[\dsp\int_{0}^{\pi/2} -\dsp\int_{\pi/2}^{\pi}
  +\dsp\int_{0}^{\pi} -\dsp\int_{3\pi/2}^{2\pi}\Big] d\phi  
  \Big[\dsp\int_{-1}^1 d\cos\thK \Big] \Big[\int_{-1}^1 
  d\cos\thl\Big]\dsp 
  \frac{d^4(\Gamma-\bar{\Gamma})}{dq^2d\cos\thl d\cos\thK d\phi}} 
  {\dsp\int_0^{2\pi}d\phi\int_{-1}^1d\cos\thK \int_{-1}^1d\cos\thl\, 
  \frac{d^4(\Gamma+\bar{\Gamma})}{dq^2d\cos\thl d\cos\thK d\phi}}.
\end{align}
\end{widetext}

The well known forward--backward asymmetry $\AFB$ and the five other angular 
asymmetries,
$A_4$, $A_5$, $A_7$, $A_8$ and $A_9$ can be written directly in terms of the
transversity amplitudes as follows:
\begin{align}
&A_{\text FB}=\frac{3}{2}\frac{\Re({\cal A}_\|^L{\cal A}_\perp^{L^*}-{\cal A 
}_\|^R{\cal A}_\perp^{R^*})}{\Gamma_{\!f}}, \\
&A_4=\frac{\sqrt{2}}{\pi}\frac{\Re(\mathcal{A}_0^L\mathcal{A}_\|^{L^*}
  +\mathcal{A}_0^R\mathcal{A}_\|^{R^*})}{\Gamma_{\!f}},\\
&A_5=\frac{3}{2\sqrt{2}}\frac{\Re(\mathcal{A}_0^L\mathcal{A}_\perp^{L^*} 
-\mathcal{A}_0^R\mathcal{A}_\perp^{R^*})}{\Gamma_{\!f}},\\
&A_7=\frac{3}{2\sqrt{2}}\frac{\Im (\azeL^{}\apaL^*-\azeR^{}\apaR^*)}{\Gf}, \\
&A_8= \frac{\sqrt{2}}{\pi}\frac{\Im(\azeL^{}\apeL^* 
  +\azeR^{}\apeR^*)}{\Gf}, \\
&A_9= \frac{3}{2\pi}\frac{\Im(\apaL^{*}\apeL^{} 
  +\apaR^{*}\apeR^{})}{\Gf}.
\end{align}

The observables $A_4$, $A_5$, $\AFB$, $A_7$, $A_8$ and $A_9$ are related to the
$CP$ averaged observables $S_4$, $S_5$, $\AFB^{\sss \text{LHC}b}$, $S_7$, $S_8$
and $S_9$ measured by LHC$b$ \cite{Aaij:2013iag} as follows respectively,
\begin{align}
\label{eq:S4-S5-S9}
&A_4=-\dsp\frac{2}{\pi}S_4, ~~A_5=\dsp\frac{3}{4}S_5, ~~\AFB=-\AFB^{\sss 
\text{LHC}b}, \nn \\
&A_7=\dsp\frac{3}{4} S_7,~~A_8=-\dsp\frac{2}{\pi} 
S_8,~~A_9=\dsp\frac{3}{2\pi}S_9.
\end{align}
We emphasize that our observables $A_{4,5,7,8,9}$ are $CP$ conserving
asymmetries and in particular $A_9$ and should not be confused with the $CP$
violating asymmetry measured by LHC$b$~\cite{Aaij:2013iag} also denoted by
$A_9$. In our notation we would refer to the $CP$ violating asymmetries as
$A^{CP}_{4,5,6,7,8,9}$. The observables $F_L$ and $\AFB$ has been measured by
different experiments Babar, Belle, CDF and LHC$b$
\cite{Aaij:2013iag,:2009zv,:2008ju, Aubert:2006vb,CMS:2013, CDF,
Aaltonen:2011qs, Aaltonen:2011ja}. By doing a angular analysis in the angle
$\phi$, LHC$b$ has measured the observable $S_3$ \cite{Aaij:2013iag}. $S_3$ is
related to the transversity helicity fraction $F_\perp$ through the relation
\begin{equation}
S_3=-\frac{1-F_L-2F_\perp}{2}.
\end{equation}
The observables $F_L$, $F_\perp$, $A_4$, $A_5 $, $\AFB$, $A_7$, 
$A_8$ and $A_9$ defined in this section are not independent. In the subsequent 
sections we explore the relation between them. 

\section{The massless lepton limit. } 
\label{sec:massless_case} 

In this section we generalize the approach developed in
Refs.~\cite{Das:2012kz} to include all contribution from the SM  that
were ignored  as their effects are sub-dominant, except that we still restrict
our discussion to the limit where the lepton is massless. The corrections
arising from massive leptons will be taken into account later in
Sec~\ref{sec:massive_case}. In particular we will consider the possibility that
the amplitudes ${\cal A}_\lambda^{L,R}$ are in general complex. As already
mentioned the imaginary contribution can be totally attributed to the complex
$\widetilde{C}_9^\hel{\lambda}$ and $\widetilde{\mathcal{G}}_\lambda$. 
This would include loop contributions that are both factorizable and 
non-factorizable and all resonance contributions. We also
take into account that the non-factorizable contributions can introduce an
`effective helicity ($\lambda$) dependence' in the Wilson coefficient
$\widetilde{C}_9^\hel{\lambda}$.

In Ref.~\cite{Das:2012kz} a new variable $r_\lambda$ was introduced that led to
significant simplification. We once again introduce the same `real variable'
$r_\lambda$ defined as,
\begin{equation}
\label{eq:rlambda}
r_\lambda=\frac{\Re(\widetilde{\mathcal{G}}_\lambda)}{\mathcal{F}_\lambda}
-\Re(\widetilde{C}_9^\hel{\lambda}).
\end{equation}
Since, we now consider $\widetilde{C}_9^\hel{\lambda}$ and 
$\widetilde{\mathcal{G}}_\lambda$ 
to be complex in general,
we have modified $r_\lambda$ to include only the real contributions i.e, 
$\Re(\widetilde{C}_9^\hel{\lambda})$ and 
$\Re(\widetilde{\mathcal{G}}_\lambda)$. 
The amplitude ${\cal A}_\lambda^{L,R}$ in 
Eq.~\eqref{eq:amp-def1} can thus be written as,
\begin{eqnarray}
\mathcal{A}_\lambda^{L,R}&=&\big(\widetilde{C}_9^{\sss\lambda}\mp 
\widehat{C}_{10})\mathcal{F}_\lambda -\widetilde{\mathcal{G}}_\lambda \nn \\ 
\label{eq:amp-def2}
&=&(\mp \widehat{C}_{10}-r_\lambda)\mathcal{F}_\lambda+i\varepsilon_\lambda,
\end{eqnarray}
where $\varepsilon_\lambda\equiv
\Im(\widetilde{C}_9^\hel{\lambda})\mathcal{F}_\lambda
-\Im(\widetilde{\mathcal{G}}_\lambda)$. The use of $\varepsilon_\lambda$ is not
necessarily meant to imply that the imaginary parts are negligibly small. We
make no such assumption. It is, however, to be expected that the imaginary
contributions are sub-dominant. The presence of the $\varepsilon_\lambda$ term
introduces three extra variables in comparison to the discussion in
Ref.~\cite{Das:2012kz}. However, we now have three extra observables $A_7$,
$A_8$ and $A_9$. Hence, dealing with complex amplitude introduces only a
technical difficulty of solving for additional variables. We begin by expressing
the observables $F_L$, $F_\|$, $F_\perp$, $A_4$, $A_5$, $\AFB$, $A_7$, $A_8$ and
$A_9$ in terms of $\widehat{C}_{10}$, $r_\lambda$, $\mathcal{F}_\lambda$ and
$\epsilon_\lambda$ as follows:
\begin{align}
\label{eq2:FL}
F_L\Gf&=2\mathcal{F}_0^2\big(r_0^2+\widehat{C}_{10}^2\big)+2 \varepsilon_0^2, \\
\label{eq2:Fparallel}
F_\|\Gf&=2\mathcal{F}_\|^2\big(r_\|^2+\widehat{C}_{10}^2\big)+2 
\varepsilon_\|^2, \\
\label{eq2:Fperp}
F_{\!\perp}\Gf&=2\mathcal{F}_{\!\perp}^2\big(r_{\!\perp}^2
  +\widehat{C}_{10}^2\big)+2 \varepsilon_\perp^2, \\
\label{eq2:A4}
\sqrt{2}\pi A_4\Gf&=4\mathcal{F}_0 \mathcal{F}_\| \big(r_0 
r_\|\!+\widehat{C}_{10}^2\big)\!
  +\! 4\varepsilon_0 \varepsilon_\|, \\
\label{eq2:A5}
\sqrt{2}A_5\Gf&=3\mathcal{F}_0\mathcal{F}_{\!\perp} \widehat{C}_{10}\big(r_0+ 
r_\perp \big), \\
\label{eq2:AFB}
\AFB\Gf&=3\mathcal{F}_\|\mathcal{F}_{\!\perp} \widehat{C}_{10}\big(r_\|+r_\perp 
\big), \\
  \label{eq2:A7}
\sqrt{2}A_7\Gf&=3\widehat{C}_{10} \big(\mathcal{F}_0 
\varepsilon_\|-\mathcal{F}_\|\varepsilon_0 \big), \\
\label{eq2:A8}
\pi A_8\Gf&= 2\sqrt{2}\big(\mathcal{F}_0r_0\varepsilon_\perp 
  -\mathcal{F}_{\!\perp} r_\perp \varepsilon_0 \big), \\
\label{eq2:A9}
\pi A_9\Gf&= 3\big(\mathcal{F}_\perp r_\perp \varepsilon_\| 
  -\mathcal{F}_\| r_\| \varepsilon_\perp \big).
\end{align}
One immediately concludes that 
\begin{align}
\label{eq:eps-bound1}
2\frac{\varepsilon_0^2}{\Gf} &\le F_L, \\
2\frac{\varepsilon_\|^2}{\Gf} &\le F_\|, \\
\label{eq:eps-bound3}
2\frac{\varepsilon_\perp^2}{\Gf} &\le F_\perp. 
\end{align} 
Eqs.~\eqref{eq2:FL}--\eqref{eq2:AFB} can be easily transformed to the form in 
Ref.~\cite{Das:2012kz} by the  redefinition of the observables  $F_L$, $F_\|$, 
$F_\perp$ and $A_4$ as 
\begin{align}
\label{eq:FLprime}
F_\lambda^\prime&=F_\lambda-\frac{2 \varepsilon_\lambda^2}{\Gf}, \\
\label{eq:A4prime}
A_4^\prime&=A_4-\frac{2\sqrt{2}\varepsilon_0\varepsilon_\|}{\pi\Gf}.
\end{align}
It should be noted that $F_L^\prime+F_\|^\prime+F_\perp^\prime \le 1$. Since
only the ratios of the form-factors $\mathcal{F}_\lambda$ play a role in the
relations we wish to derive we define ratios of form-factors $ \mathsf{P_1}$, 
$ \mathsf{P_2}$ and $ \mathsf{P_3}$:
\begin{align}
  \label{eq:P_1}
  \mathsf{P_1}&=\frac{\mathcal{F}_\perp}{\mathcal{F}_\|},\\
  \label{eq:P_2}
  \mathsf{P_2}&=\dsp \frac{\mathcal{F}_\perp}{\mathcal{F}_0},\\
  \label{eq:P_3}
  \mathsf{P_3}&=\dsp\frac{\mathcal{F}_\perp}{\mathcal{F}_0+\mathcal{F}_\|} 
  \equiv  \frac{\mathsf{P_1}\mathsf{P_2}}{\mathsf{P_1}+\mathsf{P_2}}. 
\end{align}
Following these redefinitions  Eqs.~\eqref{eq2:FL}--\eqref{eq2:AFB} can be 
recast into three sets of equations just as done in Ref.~\cite{Das:2012kz}. The 
three sets of equation are:
\begin{itemize}
\renewcommand{\labelitemi}{$\diamond$}
\item \underline{\textbf{Set-I}}
\begin{align}
\label{eqI:Fparallel} 
F_\|^\prime \Gf &=2\frac{\mathcal{F}_\perp^2}{\mathsf{P_1^2}}
\big(r_\|^2+\widehat{C}_{10}^2\big)\\ 
\label{eqI:Fperp} 
F_{\!\perp}^\prime \Gf
&=2\mathcal{F}_{\!\perp}^2\big(r_{\!\perp}^2 +\widehat{C}_{10}^2\big)\\
\label{eqI:AFB} 
\AFB\Gf &=3\frac{\mathcal{F}_\perp^2}{\mathsf{P_1}}
\widehat{C}_{10}\big(r_\|+r_\perp \big)
\end{align}
\item \underline{\textbf{Set-II}}
\begin{align}
\label{eqII:FL}
F_L^\prime \Gf  &=2\frac{\mathcal{F}_\perp^2}{\mathsf{P_2^2}} 
\big(r_0^2+\widehat{C}_{10}^2\big)\\
\label{eqII:Fperp}
F_{\!\perp}^\prime \Gf &=2\mathcal{F}_{\!\perp}^2\big(r_{\!\perp}^2
+\widehat{C}_{10}^2\big)\\ 
\label{eqII:A5}
\sqrt{2}A_5\Gf&=3\frac{\mathcal{F}_\perp^2}{\mathsf{P_2}}\widehat{C}_{10} 
\big(r_0+r_\perp \big)
\end{align}
\item \underline{\textbf{Set-III}}
\begin{align}
\label{eqIII:FLp+FPp+A4p}
(F_L^\prime\!+\!F_\|^\prime\!+\!\sqrt{2}\pi A_4^\prime)\Gf &=
2 \frac{\mathcal{F}_\perp^2}{\mathsf{P_3^2}}
\big(r_{\!\wedge}^2+\widehat{C}_{10}^2\big)\\
\label{eqIII:Fperp}
F_{\!\perp}^\prime \Gf &=2\mathcal{F}_{\!\perp}^2\big(r_{\!\perp}^2
+\widehat{C}_{10}^2\big)\\ 
\label{eqIII:AFB+A5}
\big(\AFB+\!\sqrt{2}A_5\big)\Gf&=3\frac{\mathcal{F}_\perp^2}{\mathsf{P_3}}
\widehat{C}_{10}\big(r_{\!\wedge}+r_{\!\perp}\!\big)
\end{align}
\end{itemize}
In the above we have defined $r_\wedge$ as 
\begin{equation}
\label{eq:rwedge}
r_{\!\wedge}=\frac{r_\|\mathsf{P_2}+r_0 
\mathsf{P_1}}{\mathsf{P_2}+\mathsf{P_1}},
\end{equation}
Of the nine equations defined in the three Sets only six of them are
independent. These are the three equations
Eqs.~\eqref{eqI:Fparallel}--\eqref{eqI:AFB} in Set-I, two Eqs.~\eqref{eqII:FL}
and \eqref{eqII:A5} from Set-II and Eq.~\eqref{eqIII:FLp+FPp+A4p} of Set-III.
It is easy to see that Set-II and Set-III can be obtined from Set-I by the 
following replacements:
\begin{itemize}
\item \underline{\textbf{Set-II from Set-I}}\\ $F_\|^\prime\to F_L^\prime$,
$\AFB \to \sqrt{2}A_5$, $r_\|\to r_0$ and $\mathsf{P_1}\to \mathsf{P_2}$ (or
$\mathcal{F}_\| \to \mathcal{F}_0$). 
\item \underline{\textbf{Set-III from Set-I}}\\ $F_\|^\prime\to F_L^\prime+
 F_\|^\prime+\sqrt{2}\pi A_4^\prime$, $\AFB\to \AFB+\sqrt{2}A_5$, 
 $r_\|\to r_{\!\wedge}$ and $\mathsf{P_1}\to \mathsf{P_3}$ (or
 $\mathcal{F}_\| \to \mathcal{F}_\|+\mathcal{F}_0$).
\end{itemize}
It is obvious that we only need to solve Set-I to obtain $r_\|$ and $r_\perp$  
in terms of $\mathsf{P_1}$, $F_\|^\prime$,
$F_\perp^\prime$ and $\AFB$. The solutions to Set-II and Set-III can be obtained
by simple replacements.

The solution of Set-I gives (from Appendix.~\ref{sec:appendix-1}) 
\begin{align}
\label{eq:rparallel}
r_\|&=\pm \frac{\sqrt{\Gf}}{\sqrt{2}\mathcal{F}_\perp }\frac{(\mathsf{P_1^2} 
F_{\|}^\prime+\frac{1}{2}\mathsf{P_1}Z_1^\prime)}
{\sqrt{\mathsf{P_1^2}F_\|^\prime +F_\perp^\prime+ \mathsf{P_1} Z_1^\prime}\,\,},
 \\
\label{eq:rperp-1}
r_\perp&=\pm 
\frac{\sqrt{\Gf}}{\sqrt{2}\mathcal{F}_\perp}\frac{(F_{\perp}^\prime 
+\frac{1}{2}\mathsf{P_1}Z_1^\prime)} {\sqrt{\mathsf{P_1^2}F_\|^\prime
+F_\perp^\prime+ \mathsf{P_1} Z_1^\prime}\,} , 
\end{align}
where $Z_1^\prime$ is defined as,
\begin{equation}
 \label{eq:Z_1prime}
Z_1^\prime =\sqrt{4F_\|^\prime F_\perp^\prime-\frac{16}{9}\AFB^2}.
\end{equation}

The solution to Set-II is now easily seen to be
\begin{align}
\label{eq:r0}
r_0&=\pm \frac{\sqrt{\Gf}}{\sqrt{2}\mathcal{F}_\perp }
  \frac{(\mathsf{P_2^2}F_L^\prime+\frac{1}{2}\mathsf{P_2}Z_2^\prime)} 
  {\sqrt{\mathsf{P_2^2}F_L^\prime
  +F_\perp^\prime+ \mathsf{P_2} Z_2^\prime}\,}, \\
\label{eq:rperp-2}
r_\perp &=\pm \frac{\sqrt{\Gf}}{\sqrt{2}\mathcal{F}_\perp }
  \frac{(F_{\perp}^\prime+\frac{1}{2}\mathsf{P_2}Z_2^\prime)} 
  {\sqrt{\mathsf{P_2^2}F_L^\prime
  +F_\perp^\prime+ \mathsf{P_2} Z_2^\prime}\,},
\end{align}
with $Z_2^\prime$ defined as,
\begin{equation}
 \label{eq:Z_2}
Z_2^\prime =\sqrt{4F_L^\prime F_\perp^\prime-\frac{32}{9}A_5^2}.
\end{equation}
On comparing the solutions for $r_\perp$ in Eqs.~\eqref{eq:rperp-1} and
\eqref{eq:rperp-2} obtained from Set-I and Set-II respectively, we obtain a
relation for $\mathsf{P_2}$ in terms of $\mathsf{P_1}$ and observables to be
\begin{equation}
  \label{eq:P2}
\mathsf{P_2}=\frac{2\mathsf{P_1}\AFB 
F_\perp^\prime}{s\sqrt{2}A_5(2F_\perp^\prime+Z_1^\prime\mathsf{P_1})
-Z_2^\prime\mathsf{P_1}\AFB},
\end{equation}
with $s\in \{-1,+1\}$. To remove the ambiguities in the $\mathsf{P_2}$ solution 
let us divide Eq.~\eqref{eqII:A5} by Eq.~\eqref{eqI:AFB} and using 
Eqs.~\eqref{eq:rparallel} --  \eqref{eq:rperp-2} we get  
\begin{align}
\label{eq:A5byAFB}
\frac{\sqrt{2}A_5}{\AFB} &= \frac{\mathsf{P}_1}{\mathsf{P}_2}
  \frac{\sqrt{\mathsf{P_2^2}F_L^\prime +F_\perp^\prime+ \mathsf{P_2} 
  Z_2^\prime}}
  {\sqrt{\mathsf{P_1^2}F_\|^\prime +F_\perp^\prime+ \mathsf{P_1} Z_1^\prime}} \\
&= \frac{\mathsf{P}_1}{\mathsf{P}_2}\frac{(2 F_{\perp}^\prime+
  \mathsf{P_2}Z_2^\prime)}{( 2 F_{\perp}^\prime +\mathsf{P_1}Z_1^\prime)}. \nn
\end{align}
Substituting it in Eq.~\eqref{eq:P2} we have 
\begin{align*}
s ( 2 F_{\perp}^\prime + \mathsf{P_2}Z_2^\prime)- Z_2^\prime \mathsf{P_2}= 2 
F_{\perp}^\prime,
\end{align*}
which is valid for the whole $q^2$ region only for $s=1$.

Finally we write the $r_\perp$ solution obtained from Set-III:

\begin{align}
\label{eq:rperp-3}
 &\!r_\perp\!=\!\pm \frac{\sqrt{\Gf}}{\sqrt{2}\mathcal{F}_\perp }
  \frac{(F_{\perp}^\prime+\frac{1}{2}\mathsf{P_3}Z_3^\prime)} 
  {\sqrt{\mathsf{P_3^2}(F_\|^\prime\!+\!F_L^\prime\!+\!\sqrt{2}\pi A_4^\prime)
  \!+\!F_\perp^\prime\!+\! \mathsf{P_3} Z_3^\prime}\,},
\end{align}
where  $Z_3^\prime$ is defined as,
\begin{align}
&\!Z_3^\prime \!=\!\sqrt{4 (F_L^\prime \!+\!F_\|^\prime \!+\!\sqrt{2}\pi 
A_4^\prime 
)F_\perp^\prime \!-\!\frac{16}{9}(\AFB \!+\!\sqrt{2}A_5)^2}.
\end{align}
Analogous comparison of solutions for $r_\perp$ in Eqs.~\eqref{eq:rperp-1} and
\eqref{eq:rperp-3} obtained from Set-I and Set-III respectively, results in a
relation for $\mathsf{P_3}$ in terms of $\mathsf{P_1}$:
\begin{equation}
\label{eq:P3}
\mathsf{P_3}=\frac{2\mathsf{P_1}\AFB 
F_\perp^\prime}{ (\AFB+\sqrt{2}A_5)(2F_\perp^\prime+Z_1^\prime
\mathsf{P_1})-Z_3^\prime\mathsf{P_1}\AFB}.
\end{equation}
The ambiguity in the $\mathsf{P}_3$ solution is also taken to be positive
for the same reason as the $\mathsf{P}_2$ solution.
The form factor ratio $\mathsf{P}_3$ is not however independent of
$\mathsf{P}_1$ and $\mathsf{P}_2$ and is related by Eq.~\eqref{eq:P_3}.
Substituting Eqs.~(\ref{eq:P2}) and (\ref{eq:P3}) in Eq.~(\ref{eq:P_3})
we obtain the relation between the observables as:
\begin{equation}
  \label{eq:Z-constraint}
  Z_3^\prime=Z_1^\prime+Z_2^\prime.
\end{equation}

The relations derived so far involve the primed observables that depend on
$\varepsilon_\perp$,~$\varepsilon_\|$ and $\varepsilon_0$. However, the
$\varepsilon_\lambda$'s can be solved using $A_7$, $A_8$ and $A_9$ from
Eqs.~\eqref{eq2:A7}--\eqref{eq2:A9} to give
\begin{align}
\label{eq:eps_perp}
\varepsilon_\perp&=\frac{\sqrt{2}\pi\Gf}{(r_0\!-\!r_\|)\mathcal{F}_{\!\perp}} 
\Bigg[\frac{A_9 \mathsf{P_1}}{3\sqrt{2}}+\frac{A_8 \mathsf{P_2}}{4}
-\frac{A_7 \mathsf{P_1}\mathsf{P_2}r_\perp}{3\pi\widehat{C}_{10} }\Bigg], 
\\[2ex]
\label{eq:eps_parallel}
\varepsilon_\|&=\frac{\sqrt{2}\pi\Gf}{(r_0\!-\!r_\|)\mathcal{F}_{\!\perp}} 
\Bigg[\frac{A_9 r_0}{3\sqrt{2} r_\perp}+\frac{A_8 \mathsf{P_2}r_\|}
{4 \mathsf{P_1}r_\perp}-\frac{A_7\mathsf{P_2} r_\|}{3\pi\widehat{C}_{10}}
 \Bigg], \\[2ex]
\label{eq:eps_0}
\varepsilon_0&=\frac{\sqrt{2}\pi\Gf}{(r_0\!-\!r_\|)\mathcal{F}_{\!\perp}} 
\Bigg[\frac{A_9\mathsf{P_1}r_0}{3\sqrt{2} \mathsf{P_2}r_\perp}+\frac{A_8 
r_\|} {4 r_\perp} -\frac{A_7 \mathsf{P_1} r_0}{3\pi\widehat{C}_{10} }\Bigg]. 
\end{align}

A point to be noted that the $(\varepsilon_\lambda/\Gf^{\nicefrac{1}{2}})$'s are
free from the form factor $\mathcal{F}_\perp$ and $\Gf$ as can easily be seen
from the expressions for $r_\|$, $r_\perp$ and $r_0$ (Eqs.~\eqref{eq:rparallel},
\eqref{eq:rperp-1} and \eqref{eq:r0}), as well as $\widehat{C}_{10}$ derived in
Eq.~\eqref{eq:C10hat}. Indeed, since $\mathsf{P_2}$ can be expressed in terms
$\mathsf{P_1}$ and observables using Eq.~\eqref{eq:P2}, it is easy to see that
{\em each of the $\varepsilon_\lambda$'s are completely expressed in terms of
observables and the form factor ratio $\mathsf{P_1}$}. However, these solutions
are essentially iterative, since the $r_\lambda$'s and $\widehat{C}_{10}$ are
derived in terms of the primed observables that depend on $\varepsilon_\lambda$.
If the $(\varepsilon_\lambda/\Gf^{\nicefrac{1}{2}})$ are small as should be 
expected, accurate solutions for them can be found with a few iterations.

Solving for $A_4$ from Eq.~\eqref{eq:Z-constraint} the relation 
among the observables is,
\begin{widetext}
\begin{align}
  \label{eq:Obs-relation2}
  A_4\!&=\!
  \frac{2\sqrt{2}\varepsilon_\|\varepsilon_0}{\pi \Gf} \!+\! \frac{8 A_5 \AFB}
  {9 \pi \Big(F_\perp-\dsp\frac{2\varepsilon_\perp^2}{\Gf}\Big)}  
  \!+\!\sqrt{2}\,\frac{\sqrt{\Big(F_L-\dsp\frac{2\varepsilon_0^2}{\Gf}\Big) 
  \Big(F_\perp
  -\dsp\frac{2\varepsilon_\perp^2}{\Gf}\Big)
    -\frac{8}{9}A_5^2}~\sqrt{\Big(F_\|-\dsp\frac{2\varepsilon_\|^2}{\Gf}\Big)
  \Big(F_\perp-\dsp\frac{2\varepsilon_\perp^2}{\Gf}\Big)-\frac{4}{9}\AFB^2
    }}{\pi \Big(F_\perp-\dsp\frac{2\varepsilon_\perp^2}{\Gf}\Big)}.
\end{align}
\end{widetext}
This relation for $A_4$ in terms of other observables $F_L$, $F_\perp$, $A_5$,
$\AFB$, $A_7$, $A_8$ and $A_9$ is a generalization of the relation derived in
Ref.~\cite{Das:2012kz}.  A point to be noted is that while we have solved for
the observable $A_4$, we could have used Eq.~\eqref{eq:Z-constraint} to derive
an expression for any of the other observable.  However, only the solution for
$A_4$ is unique and hence the one we consider.  The validity of this relation is
a test of the consistency of the values of all measured observables.  Unlike the
expression obtained in Ref.~\cite{Das:2012kz}, we now have a relation between
observables that depends on only one hadronic parameter, the ratio of
form-factors $\mathsf{P_1}$. It is interesting to note that $\mathsf{P_1}$ does
not receive non-factorizable contributions and is uncorrected by charm loop
effects. Since, $\mathsf{P_1}$ is independent of the universal wave
functions~\cite{oliver,Beneke:2001at} in HQET, it can be reliably calculated as
an expansion in both the strong coupling constant $\alpha_s$ and
$\Lambda_{\text{QCD}}/m_b$. The dependence of $A_4$ on $\mathsf{P_1}$  is rather
weak, since the observables $A_7$, $A_8$ and $A_9$ are observed to be small and
are currently consistent with zero as expected \cite{Aaij:2013iag}. If $A_7$,
$A_8$ and $A_9$ are all observed to be zero, it is easy to see from
Eqs.~\eqref{eq:eps_perp}--\eqref{eq:eps_0} that
$\varepsilon_\perp=\varepsilon_\|=\varepsilon_0=0$ reducing the relation in
Eq.~\eqref{eq:Obs-relation2} to
\begin{equation}
  \label{eq:Obs-relation}
  A_4\!=\!
\frac{8 A_5 \AFB}{9 \pi F_\perp} +\sqrt{2}\,\frac{\sqrt{F_LF_\perp
	   \!-\!\frac{8}{9} A_5^2}\sqrt{F_\|F_\perp \!-\!\frac{4}{9}\AFB^2}}{\pi 
	   F_\perp}
\end{equation}
which was derived in Ref.~\cite{Das:2012kz}. Interestingly, in the limit of
vanishing imaginary contributions, $A_4$ can be expressed purely in terms of
observables and is free from any form factor or their ratio. In
Appendix.~\ref{sec:form-factors}  it is shown that  both $\mathsf{P}_1$ and
$\mathsf{P}_2$ are always negative.  An interesting observation that $\AFB$ and
$A_5$ always have same signs can be then made from the relation in
Eq.~\eqref{eq:A5byAFB}. Hence, we can arrive to a conclusion that, from 
Eq.~\eqref{eq:Obs-relation2} the observable $A_4$ is always positive
unless the term proportional to  $\varepsilon_\|\varepsilon_0$ is
negative and it dominates over the rest of the terms in the expression.

$A_4$ is an observable and hence must always be real. This places constraints on
the arguments of the radicals, which are directly related to the fact that
$Z_1^\prime$, $Z_2^\prime$ and $Z_3^\prime$ are all real. The constraint that
$Z_1^\prime$ is real in turn implies that
\begin{equation}
\label{eq:constraint-Z1a} 
F_\parallel F_\perp\!- \frac{4}{9} \AFB^2 \ge
F_\parallel F_\perp \Big(\frac{2\varepsilon_\parallel^2}{\Gf F_\parallel}
+\frac{2\varepsilon_\perp^2}{\Gf F_\perp} -\frac{4\varepsilon_\parallel^2
\varepsilon_\perp^2}{\Gf^2 F_\parallel F_\perp} \Big).
\end{equation}
In Eqs.~\eqref{eq:eps-bound1}-- \eqref{eq:eps-bound3}, we showed that
$0 \le \dsp\frac{2\varepsilon_\lambda^2}{\Gf F_\lambda} \le 1~$, implying that 
the R.H.S of Eq.~\eqref{eq:constraint-Z1a} must itself be greater than zero.
This imposes the following constraint:
\begin{equation}
\label{eq:constraint-Z1} 
F_\parallel F_\perp\!- \frac{4}{9} \AFB^2 \ge 0.
\end{equation}
A similar constraint arising from $Z_2^\prime$ and $Z_3^\prime$ also being real 
implies that
\begin{gather}
\label{eq:constraint-Z2} 
F_L F_\perp\!- \frac{8}{9} A_5^2 \ge 0,\\
\label{eq:constraint-Z3} 
(F_L+F_\|+\sqrt{2}\pi A_4) F_\perp\!- \frac{4}{9} (\AFB+\sqrt{2} A_5)^2 \ge 0 .
\end{gather}
The equality in the above three relations holds only when a minimum of two of
the $\varepsilon_\lambda$'s are zero. For example, $\varepsilon_\|$ and
$\varepsilon_\perp$ are zero for the equality to hold in
Eq.~\eqref{eq:constraint-Z1}, whereas $\varepsilon_0$ and $\varepsilon_\perp$
are zero for Eq.~\eqref{eq:constraint-Z2}. The three inequalities in
Eqs.~\eqref{eq:constraint-Z1}--\eqref{eq:constraint-Z3} impose constraints on
the parameter space of observables. It is obvious that non-zero
$\varepsilon_\lambda$'s will in general restrict the parameter space of
observables even further. We emphasize that this conclusion is valid without any
exception.   We will come back to this point in Sec.~\ref{sec:NPAnalysis} when
we discuss the tests of the relation for $A_4$ in Eq.~\eqref{eq:Obs-relation2}.

\section{Generalization to include lepton masses.} \label{sec:massive_case}

In this section we extend the model independent approach developed in the
previous section (Sec.~\ref{sec:massless_case}) to include the lepton mass $m$.
One of the consequences of retaining the lepton mass is the need to include an
additional amplitude in order to describe the full decay rate,
since the term proportional to $q_\mu$ in the amplitude cannot be dropped for
the massive lepton case (for a review \cite{Altmannshofer:2008dz}). In addition
to the six amplitude $\mathcal{A}_\lambda^{L,R}$ where $\lambda
\in\{0,\|,\perp\}$ the decay amplitude also depends on $\mathcal{A}_t$,
resulting in a total of seven amplitudes. These amplitudes are
given in Eqs.~\eqref{eq:trans-amp1} -- \eqref{eq:trans-ampt}. In addition, since
the massive leptons are no longer chirality eigenstates, terms involving
admixtures of heicities that are proportional to $m^2/q^2$ (see
Eqs.~\eqref{eq:I1s} and \eqref{eq:I1c}) contribute to the differential decay
rate.

These additional contributions complicate the extraction of the helicity
amplitudes. The observables $F_L$, $F_\|$,
$F_\perp$, $A_4$, $A_5$ and $\AFB$ given in Sec.~\ref{sec:massless_case} are
modified because of the presence of the new transversity amplitude
$\mathcal{A}_t$ and helicity admixture terms in the decay distribution. This in
turn results in modifying the relations in Eqs.~\eqref{eq:Obs-relation2} and
~\eqref{eq:Obs-relation}. The effect of the mass of the lepton is always
included in the measured observables and it is not possible to measure any 
observable
without the mass effects.  In order to distinguish the ``hypothetical
observables without the mass effects'' considered in
Sec.~\ref{sec:massless_case} from these true observables, we define them with a
superscript ``o'' and relate to the massless limit observables as:
\begin{subequations}
\begin{align}
    \label{eq:Gfm}
  \Gf^\o&= \beta^2 \Gf+ 3 \mathbb{T}_1,  \\
    \label{eq:FLm}
  F_L^\o &=\displaystyle\frac{1}{\Gf^\o } (\beta^2\Gf F_L + \mathbb{T}_1), \\
    \label{eq:FPam}
  F_\|^\o &= \displaystyle\frac{1}{\Gf^\o } ( \beta^2\Gf F_\|  + \mathbb{T}_1), 
  \\
    \label{eq:FPm}
  F_\perp^\o &= \displaystyle\frac{1}{\Gf^\o } (\beta^2\Gf F_\perp  + 
  \mathbb{T}_1), \\
    \label{eq:A4m}
  A_4^\o&= \displaystyle\frac{\Gf}{\Gf^\o }\beta^2  A_4,\\
    \label{eq:A5m}
  A_5^\o &= \displaystyle\frac{\Gf}{\Gf^\o } \beta A_5,  \\
    \label{eq:Afbm}
  \AFBo &= \displaystyle\frac{\Gf}{\Gf^\o }\beta \AFB, \\
   \label{eq:A7m}
  A_7^\o&= \displaystyle\frac{\Gf}{\Gf^\o }\beta A_7, \\
  \label{eq:A8m}
  A_8^\o&= \displaystyle\frac{\Gf}{\Gf^\o }\beta^2 A_8, \\
  \label{eq:A9m}
  A_9^\o&= \displaystyle\frac{\Gf}{\Gf^\o }\beta^2 A_9 .
\end{align}
\end{subequations}

In the above we have defined
\begin{align*}
\mathbb{T}_1&=\left(1+E_1\right) \dsp\frac{m^2}{q^2}\Gf ~~\text{where}\\
E_1&=\displaystyle\frac{|\at|^2}{\Gf}\!+\displaystyle\frac{2}{\Gf}
    \Re[\apaL \apaR^* + \apeL \apeR^* + \azeL \azeR^*] .
\end{align*}
Using $$2\,\Re[\alamL \alamR^*]\!=\! |\alamL + \alamR |^2-\Gf F_\lambda$$ 
and the Cauchy-Schwarz inequality, we find \\
\begin{align}
\label{eq:T1}
\mathbb{T}_1&=\big(|\at|^2+ \!\!\!\!\!\sum_{\sss\lambda=\{\|,\perp,0\}} 
\!\!\!\! 
  |\alamL + \alamR |^2 \,\big) \dsp\frac{m^2}{q^2}  \\
  & \quad 
  \le \big(|\at|^2 + 2 \Gf \big) \dsp\frac{m^2}{q^2}
\end{align}
which is always positive and bounded. This bound is important since 
$\mathbb{T}_1$ 
has not been measured so far. $\mathbb{T}_1$ can also be expressed in terms of 
angular coefficients as,
\begin{align} 
\label{eq:T1obs}
\frac{\mathbb{T}_1}{\Gf^\o}&=\frac{1}{3}-\frac{4 I_2^s-I_2^c}{3 \Gf^\o} \nn \\
&=\frac{1}{3}-\frac{16}{9} A_{10}+\frac{64}{27} A_{11}
\end{align}
and measured in terms of two new observables $A_{10}$ and $A_{11}$, defined in 
terms 
of angular asymmetries as follows:

\begin{widetext}
\begin{align}
\label{eq:A10}
A_{10}&=\dsp\frac{\dsp\int_0^{2\pi}d\phi \int_0^1 d\cos\thK
    \Big[\int_{-1}^{-1/2} -\int_{-1/2}^{1/2}  +\int_{1/2}^{1}\Big]~ d\cos\thl~ 
    \dsp\frac{d^4(\Gamma+\bar{\Gamma})}{dq^2d\cos\thl d\cos\thK d\phi}\,} 
    {\dsp\int_0^{2\pi}d\phi\int_{-1}^1d\cos\thK \int_{-1}^1d\cos\thl\,
    \frac{d^4(\Gamma+\bar{\Gamma})}{dq^2d\cos\thl d\cos\thK d\phi}}, \\
    \label{eq:A11}
A_{11}&=\dsp\frac{\dsp\int_0^{2\pi}d\phi \Big[\int_{-1}^{-1/2}-\int_{-1/2}^{1/2}
    +\int_{1/2}^{1}  \Big] ~d\cos\thK~
\Big[\int_{-1}^{-1/2}-\int_{-1/2}^{1/2} + \int_{1/2}^{1} \Big] 
    d\cos\thl~
    \dsp\frac{d^4(\Gamma+\bar{\Gamma})}{dq^2d\cos\thl d\cos\thK d\phi}\,}
     {\dsp\int_0^{2\pi}d\phi\int_{-1}^1d\cos\thK \int_{-1}^1d\cos\thl\,
    \frac{d^4(\Gamma+\bar{\Gamma})}{dq^2d\cos\thl d\cos\thK d\phi}}.
\end{align}
\end{widetext}
If the two asymmetries $A_{10}$ and $A_{11}$ are measured experimentally
then we can get the estimate of the correction term arising due to 
lepton masses. However, from Eq.~\eqref{eq:T1} it can be seen that 
$\mathbb{T}_1$
is proportional to lepton mass (square) $m^2/q^2$ which is very small and difficult to 
measure except at small $q^2$. In the 
limit of zero lepton mass $\mathbb{T}_1$ vanishes which gives a constraint on 
these two observables by,
\begin{equation}
\label{eq:A10A11}
A_{10}-\frac{4}{3} A_{11}=\frac{3}{16}.
\end{equation}
A deviation from this relation would indicate the effect of the non-zero 
lepton mass and provide an estimate of the size the mass corrections.
The observables are re-expressed in terms of the variables $r_\lambda$
(defined in Eq.~\eqref{eq:rlambda}) as follows:
\begin{align}
\label{eq2:FLo}
&F_L^\o\Gf^\o =2\beta^2\frac{\mathcal{F}_\perp^2}{\mathsf{P_2^2}} (r_0^2+ 
\widehat{C}_{10}^2)+2 \beta^2 \varepsilon_0^2 + \mathbb{T}_1,\\
\label{eq2:Fparallelo}
&F_\|^\o \Gf^\o =2\beta^2\frac{\mathcal{F}_\perp^2}{\mathsf{P_1^2}}(r_\|^2+ 
\widehat{C}_{10}^2)+2 \beta^2 \varepsilon_\|^2+ \mathbb{T}_1,\\
\label{eq2:Fperpo}
&F_\perp^\o \Gf^\o = 2\beta^2\mathcal{F}_\perp^2 (r_\perp^2+ \widehat{C}_{10}^2)
  +2 \beta^2 \varepsilon_\perp^2  + \mathbb{T}_1, \\
\label{eq2:A4o}
&\sqrt{2}\pi A_4^\o \Gf^\o =4 \beta^2 
\frac{\mathcal{F}_\perp^2}{\mathsf{P_1}\mathsf{P_2}} 
  \big(r_0 r_\|\!+\widehat{C}_{10}^2\big)\! +\! 4\beta^2 \varepsilon_0 
  \varepsilon_\|,  \\
\label{eq2:A5o}
&\sqrt{2}A_5^\o\Gf^\o =3\beta\frac{\mathcal{F}_\perp^2}{\mathsf{P_2}}  
\widehat{C}_{10}
  (r_0+r_\perp), \\ 
\label{eq2:AFBo}
&\AFBo\Gf^\o = 3\beta\frac{\mathcal{F}_\perp^2}{\mathsf{P_1}}  
 \widehat{C}_{10} 
(r_\|+r_\perp ), \\
  \label{eq2:A7o}
&\sqrt{2}A_7^\o \Gf^\o =3\beta \widehat{C}_{10} \big(\mathcal{F}_0 
\varepsilon_\|-\mathcal{F}_\|\varepsilon_0 \big), \\
\label{eq2:A8o}
&\pi A_8^\o \Gf^\o = \dsp 2\sqrt{2}\beta^2 
\big(\mathcal{F}_0r_0\varepsilon_\perp 
  -\mathcal{F}_{\!\perp} r_\perp \varepsilon_0 \big), \\ 
\label{eq2:A9o}
&\pi A_9^\o \Gf^\o = 3\beta^2 \big(\mathcal{F}_\perp r_\perp \varepsilon_\| 
  -\mathcal{F}_\| r_\| \varepsilon_\perp \big). 
\end{align}

In analogy with the previous solutions of $\mathsf{P_2}$ and $\mathsf{P_3}$ in
Eqs.~\eqref{eq:P2} and \eqref{eq:P3} using the three sets (Set-I, II, III)  we 
can solve for $\mathsf{P_2}$ and $\mathsf{P_3}$ once again in terms
of $\mathsf{P_1}$  and ``true observables'' as,
\begin{align}
  \label{eq:P2m}
\mathsf{P_2}&=\frac{2\mathsf{P_1}\AFBo 
\big(F_\perp^\o-\frac{\mathcal{T}_\perp}{\Gf^\o}\big)}
{\sqrt{2}A_5^\o\Big(2\big(F_\perp^\o- 
\frac{\mathcal{T}_\perp}{\Gf^\o}\big)+Z_1^\o\mathsf{P_1}\Big)
-Z_2^\o\mathsf{P_1}\AFBo},
   \\
\label{eq:P3m}
\mathsf{P_3}\!&=\!\frac{2\mathsf{P_1}\AFBo 
\big(F_\perp^\o- \frac{\mathcal{T}_\perp}{\Gf^\o}\big)}{ 
(\AFBo+\sqrt{2}A_5^\o)\Big(2\big(F_\perp^\o \!-\! 
  \frac{\mathcal{T}_\perp}{\Gf^\o}\big) \!+\!Z_1^\o
\mathsf{P_1}\Big)\!-\!Z_3^\o\mathsf{P_1}\AFBo},
\end{align} 
where positive sign ambiguity is chosen for $\mathsf{P_2}$ and $\mathsf{P_3}$
solutions because of the same reason discussed in the massless case. The definitions
of $Z_1^\o$, $Z_2^\o$ and $Z_3^\o$ are given by
\begin{gather}
Z_1^\o=\sqrt{4\big(F^\o_\|- \frac{\mathcal{T}_\|}{\Gf^\o}\big)\big(F^\o_\perp
- \frac{\mathcal{T}_\perp}{\Gf^\o}\big)- 
\dsp\frac{16}{9}\beta^2{\AFBo}^{\!\!\!\!\!2}}~ ,\\
Z_2^\o=\sqrt{4\big(F^\o_L- \frac{\mathcal{T}_0}{\Gf^\o}\big)\big(F^\o_\perp
    - \frac{\mathcal{T}_\perp}{\Gf^\o}\big)-\dsp\frac{32}{9}\beta^2{A_5^\o}^2} ,
\end{gather}
\begin{widetext}
\begin{equation}
Z_3^\o=\sqrt{4\Big((F^\o_L- \frac{\mathcal{T}_0}{\Gf^\o})+(F^\o_\|
  - \frac{\mathcal{T}_\|}{\Gf^\o})+ \sqrt{2}\pi A^\o_4 
-\frac{4\beta^2 \varepsilon_0 \varepsilon_\| }{\Gf^\o }\Big)\big(F^\o _\perp
    - \frac{\mathcal{T}_\perp}{\Gf^\o}\big)-\dsp\frac{16}{9}\beta^2\big(\AFBo 
    +\sqrt{2}A_5^\o \big)^2} .
\end{equation}
\end{widetext}
To simplify notation we have defined
\begin{align}
\label{eq:Tlambda}
\mathcal{T}_\lambda= \mathbb{T}_1+2 \beta^2 
\varepsilon_\lambda^2~;\qquad\lambda \in \{0,\perp,\| \}
\end{align}
Substituting Eqs.~\eqref{eq:P2m} and \eqref{eq:P3m} in Eq.\eqref{eq:P_3}
we can get the condition valid over whole $q^2$ range as:
\begin{align}
Z_3^\o=Z_1^\o+Z_2^\o \label{eq:Z3oZ2oZ1o}.
\end{align}

The $\varepsilon_\lambda$'s can be solved as was done in the previous section using 
Eqs.~\eqref{eq2:A7o}--\eqref{eq2:A9o} to give
\begin{align}
\label{eq:eps_perpo}
&\!\varepsilon_\perp\!=\!\frac{\sqrt{2}\pi\Gf^\o}{\beta^2(r_0\!-\!r_\|) 
\mathcal{F}_{\!\perp}}\! 
\Bigg[ \! \frac{A_9^\o \mathsf{P_1}}{3\sqrt{2}}\!+ \!\frac{A_8^\o 
\mathsf{P_2}}{4}\! 
-\!\frac{A_7^\o\beta \mathsf{P_1}\mathsf{P_2}r_\perp} {3\pi\widehat{C}_{10} } 
\! \Bigg], \\
\label{eq:eps_parallelo}
&\!\!\varepsilon_\|\!=\!\frac{\sqrt{2}\pi\Gf^\o}{\beta^2(r_0\!-\!r_\|) 
\mathcal{F}_{\!\perp}}\! 
\Bigg[ \! \frac{A_9^\o r_0}{3\sqrt{2} r_\perp}\!+\!\frac{A_8^\o 
\mathsf{P_2}r_\|}
{4 \mathsf{P_1}r_\perp}\!-\!\frac{A_7^\o \beta \mathsf{P_2} 
r_\|}{3\pi\widehat{C}_{10}}
 \! \Bigg], \\[4ex]
\label{eq:eps_0o}
&\!\!\varepsilon_0\!=\!\frac{\sqrt{2}\pi\Gf^\o}{\beta^2(r_0\!-\!r_\|)
\mathcal{F}_{\!\perp}}\! 
\Bigg[\frac{A_9^\o\mathsf{P_1}r_0}{3\sqrt{2} 
\mathsf{P_2}r_\perp}\!+\!\frac{A_8^\o 
r_\|} {4 r_\perp}\! -\!\frac{A_7^\o \beta \mathsf{P_1} 
r_0}{3\pi\widehat{C}_{10} }\Bigg].
\end{align}

From Eqs.~\eqref{eq:22} -- \eqref{eq:C10hat} it can be easily seen that the 
$(\varepsilon_\lambda/{\Gf^\o}^{\nicefrac{1}{2}})$'s are free from the form-factor 
$\mathcal{F}_\perp$ and $\Gf^\o$ {\em and are completely expressed in terms of
observables and the form factor ratio $\mathsf{P_1}$}. However the accurate 
solutions of $(\varepsilon_\lambda/{\Gf^\o}^{\nicefrac{1}{2}})$'s 
can be found with a few iterations as described in the previous massless case.

Solving for $A_4^\o$ from Eq.~\eqref{eq:Z3oZ2oZ1o} the relation among 
the observables including lepton masses turns out
\begin{widetext}
\begin{align}
\label{eq:Obs-relationnew} A_4^\o\!=\!
\dsp\frac{2\sqrt{2}\beta^2\varepsilon_\|\varepsilon_0}{\pi \Gf^\o} \!+\! \dsp
\frac{8 \beta^2A_5^\o \AFBo}{9 \pi \big(F_\perp^\o-\dsp\frac{\mathcal{T}_\perp}
{\Gf^\o}\big)} \!+\!\sqrt{2}\,\frac{\sqrt{\big(F_L^\o-\dsp
\frac{\mathcal{T}_0}{\Gf^\o}\big) \big(F_\perp^\o-
\dsp\frac{\mathcal{T}_\perp}{\Gf^\o}\big) \!-\frac{8}{9}\beta^2 {A_5^\o}^2}
\sqrt{\big(F_\|^\o-\dsp\frac{\mathcal{T}_\|}{\Gf^\o}\big)
\big(F_\perp^\o-\dsp\frac{\mathcal{T}_\perp}{\Gf^\o}\big)
\!-\frac{4}{9}\beta^2{\AFBo}^{\!\!\!\!\!2}}} {\pi
\big(F_\perp^\o-\dsp\frac{\mathcal{T}_\perp}{\Gf^\o}\big)}.
\end{align}

In analogy to the massless case, each of $Z_1^\o$, $Z_2^\o$ and $Z_3^\o$ are 
also real. A real $Z_1^\o$ implies that 
\begin{equation}
\label{eq:constraint-Z1oa}
F_\|^\o F_\perp^\o\!- \frac{4}{9} {\AFBo}^{\!\!\!\!\!2}\, \ge F_\|^\o 
F_\perp^\o 
\Big(\frac{\mathcal{T}_\|}{\Gf^\o F_\|^\o}
  +\frac{\mathcal{T}_\perp}{\Gf^\o F_\perp^\o}
  -\frac{\mathcal{T}_\| \mathcal{T}_\perp}{{\Gf^\o}^2 F_\|^\o F_\perp^\o}\Big)
  -\frac{16 m^2 {\AFBo}^{\!\!\!\!\!2}}{9\,q^2}.
\end{equation}
Since, $\dsp 0\leq\frac{\mathcal{T}_\lambda}{\Gf F_\lambda^\o}\leq 1$ as can
be seen from Eqs.~\eqref{eq2:FLo}--\eqref{eq2:Fperpo}, we can obtain a bound on
the L.H.S. of Eq.~\eqref{eq:constraint-Z1oa}. The bounds arising from real
$Z_1^\o$, $Z_2^\o$ and $Z_3^\o$ are
\begin{subequations}
\begin{gather}
\label{eq:constraint-Z1o}
F_\|^\o F_\perp^\o\!- \frac{4}{9} {\AFBo}^{\!\!\!\!\!2} \, \geq- \frac{16 
m^2{\AFBo}^{\!\!\!\!\!2}}{9\,q^2} , \\ 
\label{eq:constraint-Z2o}
F_L^\o F_\perp^\o\!- \frac{8}{9} {A_5^\o}^2 
\geq-\frac{32 m^2 {A_5^\o}^2}{9\,q^2} , \\ 
\label{eq:constraint-Z3o}
(F_L^\o+F_\|^\o+\sqrt{2}\pi A_4^\o)
F_\perp^\o - \frac{4}{9} (\AFBo+\sqrt{2} A_5^\o)^2  \geq-\frac{16 
m^2{(\AFBo+\sqrt{2}
A_5^\o)}^2}{9\,q^2}
\end{gather}
\end{subequations}
respectively. Clearly the L.H.S. of the above inequalities can, in the worst
case, be a small negative number. Comparing this with the massless case we note
that while the effect of the imaginary contributions is to restrict the
parameter space further the effect of mass dependent terms is to oppose this
restriction. The mass term should have the maximum effect at $q^2$ close to $4
m^2$, but as we will see in the next section (Sec.~\ref{sec:observables}) in the
limit  $q^2\to 4\,m^2$ all the asymmetries approach zero. The contribution from
the mass term should hence be insignificant, indicating that in practice the
allowed parameter space of observables is not noticeably altered.  This
conclusion is borne out to be true in numerical estimates as we will see in
Sec.~\ref{sec:NPAnalysis}. We conclude, therefore, that the most conservative
allowed parameter space remains unaltered even if the small lepton mass term is
dropped compared to $q^2$ and the imaginary contributions to the amplitudes are
completely ignored.

The zero crossings of angular asymmetries $\AFBo$, $A_5^\o$ and
$\AFBo+\sqrt{2}A_5^\o$  provide interesting limits where the relation in
Eq.~\eqref{eq:Obs-relationnew} simplifies to three independent relations with
each of them  providing an interesting test for NP. At the zero
crossing of $\AFBo$, $A_5^\o$ and $\AFBo+\sqrt{2}A_5^\o$,
Eq.~\eqref{eq:Obs-relationnew} reduces to
\begin{subequations}
\begin{gather}
\frac{8 {A_5^\o}^2}{9 \big(F_L^\o-\dsp\frac{\mathcal{T}_0}{\Gf^\o}\big)
  \big(F_\perp^\o-\dsp\frac{\mathcal{T}_\perp}{\Gf^\o}\big)}
  +\frac{\pi^2 
  {\Big(A_4^\o-\dsp\frac{2\sqrt{2}\beta^2\varepsilon_\|\varepsilon_0}
  {\pi \Gf^\o}\Big)}^{\!\!\!2}}{2 
  \big(F_L^\o-\dsp\frac{\mathcal{T}_0}{\Gf^\o}\big)
  \big(F_\|^\o-\dsp\frac{\mathcal{T}_\|}{\Gf^\o}\big)} =1 \\
\frac{4 {\AFBo}^{\!\!\!\!\!2}}{9 
\big(F_\|^\o-\dsp\frac{\mathcal{T}_\|}{\Gf^\o}\big)
  \big(F_\perp^\o-\dsp\frac{\mathcal{T}_\perp}{\Gf^\o}\big)}
  +\frac{\pi^2 
  {\Big(A_4^\o-\dsp\frac{2\sqrt{2}\beta^2\varepsilon_\|\varepsilon_0}
  {\pi \Gf^\o}\Big)}^{\!\!\!2}}{2 
  \big(F_L^\o-\dsp\frac{\mathcal{T}_0}{\Gf^\o}\big)
  \big(F_\|^\o-\dsp\frac{\mathcal{T}_\|}{\Gf^\o}\big)} =1 \\
\frac{2 ({\AFBo}^{\!\!\!\!\!2}+2{A_5^\o}^2)\Big((F^\o_L- 
\dsp\frac{\mathcal{T}_0}{\Gf^\o})+(F^\o_\|
  - \dsp\frac{\mathcal{T}_\|}{\Gf^\o})+ \sqrt{2}\pi A^\o_4 
  -\dsp\frac{4\beta^2 \varepsilon_0 \varepsilon_\| }{\Gf^\o }\Big)}{9 
\big(F_\|^\o-\dsp\frac{\mathcal{T}_\|}{\Gf^\o}\big)
  \big(F_L^\o-\dsp\frac{\mathcal{T}_0}{\Gf^\o}\big)
  \big(F_\perp^\o-\dsp\frac{\mathcal{T}_\perp}{\Gf^\o}\big)}
  +\frac{\pi^2 
  {\Big(A_4^\o-\dsp\frac{2\sqrt{2}\beta^2\varepsilon_\|\varepsilon_0}
  {\pi \Gf^\o}\Big)}^{\!\!\!2}}{2 
  \big(F_L^\o-\dsp\frac{\mathcal{T}_0}{\Gf^\o}\big)
  \big(F_\|^\o-\dsp\frac{\mathcal{T}_\|}{\Gf^\o}\big)} =1
\end{gather}
\end{subequations}
respectively. In the limit where both the mass effect and the imaginary
contributions to the Wilson coefficients $\widehat{C}_7$ and $\widehat{C}_9$ 
can be ignored these relations simplify to depend only on observables
\begin{equation}
\label{eq:zeroassym0}
\begin{array}{cl}
\dsp\frac{8A_5^2}{9 F_L F_\perp}+\frac{\pi^2 A_4^2}{2 F_L F_\|}=1 &\qquad 
\text{if} 
~\AFB=0\\ [3.5ex]
\dsp\frac{4\AFB^2}{9 F_\| F_\perp}+\frac{\pi^2 A_4^2}{2 F_L F_\|}=1 &\qquad 
\text{if}~A_5=0\\ [3.5ex]
\dsp\frac{2(\AFB^2+2 A_5^2)(F_L+F_\|+\sqrt{2}\pi A_4)}{9 F_\|F_L F_\perp}
  +\frac{\pi^2 A_4^2}{2 F_L F_\|}=1 &\qquad \text{if}~\AFB+\sqrt{2}A_5=0
\end{array}
\end{equation}
The zero-crossings of these observables are also interesting as the form factor
ratios $\mathsf{P}_1$, $\mathsf{P}_2$ and $\mathsf{P}_3$ can be related to the
helicity fractions at those $q^2$ points. Eq.~\eqref{eq2:AFBo} implies that when
$\AFBo=0$, $r_\|+r_\perp $ must be zero. Then, the expression for
$r_\|+r_\perp $ (see Eq.~\eqref{eq:22} for the massive case in Appendix
\ref{sec:appendix-1}) gives,
\begin{align}
  \label{eq:u+v}
r_\|+r_\perp\big|_{
\AFBo=0}&=\pm\frac{\sqrt{\Gf^\o}}{\sqrt{2}\mathcal{F}_\perp}
\Big(\sqrt{F_\perp^\o
  -\dsp\frac{\mathcal{T}_\perp}{\Gf^\o}}+ \mathsf{P_1}\sqrt{F_\|^\o
  -\dsp\frac{\mathcal{T}_\|}{\Gf^\o}}~\Big)=0 \nn \\ 
  &\Longrightarrow \mathsf{P_1}\big|_{\AFBo=0} =-\frac{\sqrt{F_\perp^\o
  -\dsp\frac{\mathcal{T}_\perp}{\Gf^\o}}}
  {\sqrt{F_\|^\o -\dsp\frac{\mathcal{T}_\|}{\Gf^\o}}}
\end{align}
$\mathsf{P_1}$ can be iteratively solved from the above equation. We note that
in order one has real positive form-factors by definition
(Eq.~\eqref{eq:P_1}) $\mathsf{P_1}$ is always negative. The zero 
crossing of $\AFBo$ is observed at $q^2=4.9^{+1.1}_{-1.3}\gev^2$ 
~\cite{Aaij:2013iag} which is in the large recoil region where it is believed 
that reliable calculations can be done in HQET.  Hence, we can check the 
predictability of HQET in large recoil region, when enough data for all 
observables are available at this $q^2$ point.

Eqs.~\eqref{eq:P2m} and \eqref{eq:P3m} can now be used to obtain 
$\mathsf{P_2}$  and $\mathsf{P_3}$ at the zero crossings $A_5^\o=0$ and 
$\AFB^\o+\sqrt{2}A_5^\o=0$ respectively, 
\begin{align}
  \label{eq:P2limit}
\mathsf{P_2}\big|_{A_5^\o=0}&=-\frac{\sqrt{F_\perp^\o
  -\dsp\frac{\mathcal{T}_\perp}{\Gf^\o}}}
  {\sqrt{F_L^\o - \dsp\frac{\mathcal{T}_0}{\Gf^\o}}} ,\\
  \label{eq:P3limit}
\mathsf{P_3}\big|_{\AFB^\o+\sqrt{2}A_5^\o=0}& =-\frac{\sqrt{F_\perp^\o
  -\dsp\frac{\mathcal{T}_\perp}{\Gf^\o}}}
  {\sqrt{\Big((F^\o_L- \dsp\frac{\mathcal{T}_0}{\Gf^\o})+(F^\o_\|
  - \dsp\frac{\mathcal{T}_\|}{\Gf^\o})+ \sqrt{2}\pi A^\o_4 
  -\dsp\frac{4\beta^2 \varepsilon_0 \varepsilon_\| }{\Gf^\o }\Big)}}.
\end{align}
\end{widetext}

The relation derived in Eq.~\eqref{eq:Obs-relationnew} incorporates all the
possible effects within SM. It includes a finite lepton mass, electromagnetic
correction to hadronic operators at all orders and all factorizable and
non-factorizable contributions including resonances to the decay. It can be seen
from the Eq.~\eqref{eq:Tlambda} the term $\mathcal{T}_\lambda/\Gf^\o$ contains
$\mathbb{T}_1/\Gf^\o$ which is expressed in Eq.~\eqref{eq:T1obs} in terms of the
asymmetries $A_{10}$ and $A_{11}$ which can be measured experimentally and the
other term $(\varepsilon_\lambda/{\Gf^\o}^{\nicefrac{1}{2}})$ depends only on
the observables and one form-factor ratio $\mathsf{P}_1$. Thus, the relation in
Eq.~\eqref{eq:Obs-relationnew} is complete and exact in the sense that it
involves all the eleven observables and only one hadronic input which can be 
reliably estimated using HQET.

\section{Observables at kinematic extreme points} 
\label{sec:observables}

In this section we will briefly discuss the limiting value of the observables at
the two kinematic extremities of $q^2$, the dilepton invariant mass squared. The
minimum $q^2$ value, $q^2\!=\!\qmin \!=\!4 m^2$ and the endpoint $q^2\!=\!\qmax
\!=\left(m_B-m_\kstar\right)^{\!\,2}$. The values of the observables we obtain
below can be experimentally verified and any exception must imply NP.

\begin{itemize}
\item \underline{\textbf{Case-I:}}  $q^2\!=\!4 m^2$
\end{itemize}
It is easy to see that at $\qmin$ the two lepton carry equal momentum and recoil
against the $\kstar$. In the dilepton rest frame the two leptons carry zero
momentum. Hence, angles $\theta_\ell$ and $\phi$ cannot be defined. The angular
distribution in Eq.~\eqref{eq:helicity} thus implies that all asymmetries
i.e $A_4$, $A_5$, $\AFB$, $A_7$, $A_8$ and $A_9$ must
vanish in this limit. This implies that there is no preferred direction, leading
to the conclusion that all helicities are equally probable.

Using the expressions of the observables derived in the previous
section (Eqs.~\eqref{eq:Gfm} and \eqref{eq:FLm}) we can write
\begin{align}
F_L^\o &=\dsp\frac{1}{\Gf^\o } \big(\beta^2\Gf F_L 
  + \frac{1}{3}(\Gf^\o-\beta^2 \Gf)\big) \nn \\
  &\mathop{=}_{\mathrm{\beta\to 0}} ~\frac{1}{3} 
\end{align}
This limiting value holds for the other two helicity fractions as well.
Hence, at the kinematic starting point we can write
\begin{align}
F_\lambda^\o \mathop{=}_{q^2\to 4 m^2} 
  ~\dsp\frac{1}{3},\qquad \lambda \in \{L,\perp,\|\}.
\end{align}
We conclude that each observed helicity fraction should be $1/3$ at 
$q^2_{\min}$, which can be easily verified experimentally.
The asymmetries defined in Eq.~\eqref{eq:A10} and \eqref{eq:A11} also
vanish at $q^2\!=\!\qmin$ implying $(\mathbb{T}_1/\Gf^\o) \to \frac{1}{3}$
(from Eq.~\eqref{eq:T1obs}).
Thus the observable $A_4^\o$ from Eq.\eqref{eq:Obs-relationnew} 
at $q^2\!=\!\qmin$ is given by,
\begin{align}
   A_4^\o& \mathop{=}_{\beta \to 0}
\frac{\sqrt{2}}{\pi}\sqrt{F_L^\o-\frac{\mathbb{T}_1}{\Gf^\o}}
  \sqrt{F_\|^\o-\frac{\mathbb{T}_1}{\Gf^\o}} 
\mathop{\mathop{=}_{F_\lambda^\o \to \frac{1}{3}}}_{\frac{\mathbb{T}_1}{\Gf^\o}\to \frac{1}{3}}
0
\end{align}
as it was expected above.

\begin{center}
\begin{figure*}[htbp]
 \begin{center}
	\includegraphics*[width=0.25\textwidth]{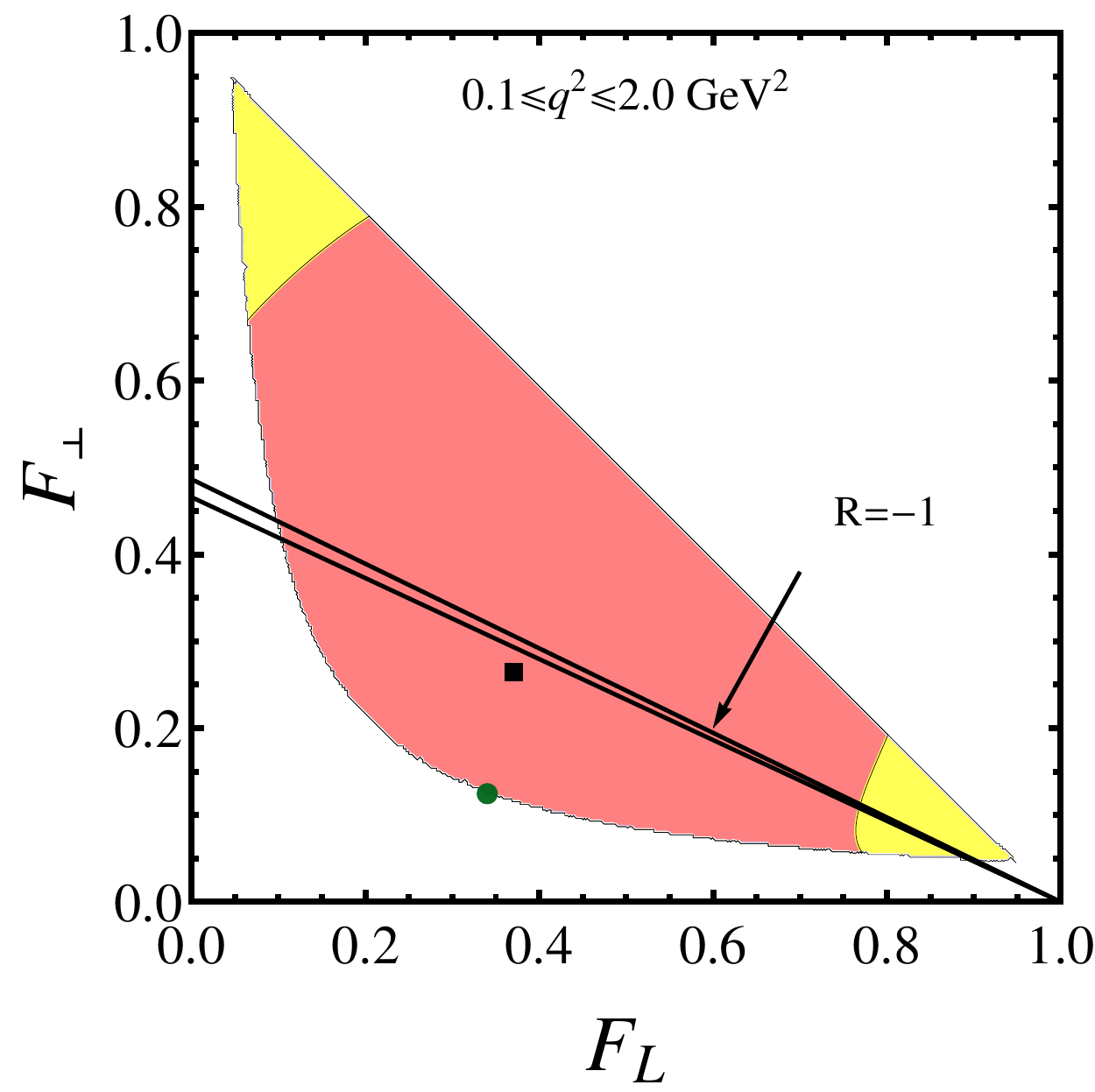}%
	\includegraphics*[width=0.25\textwidth]{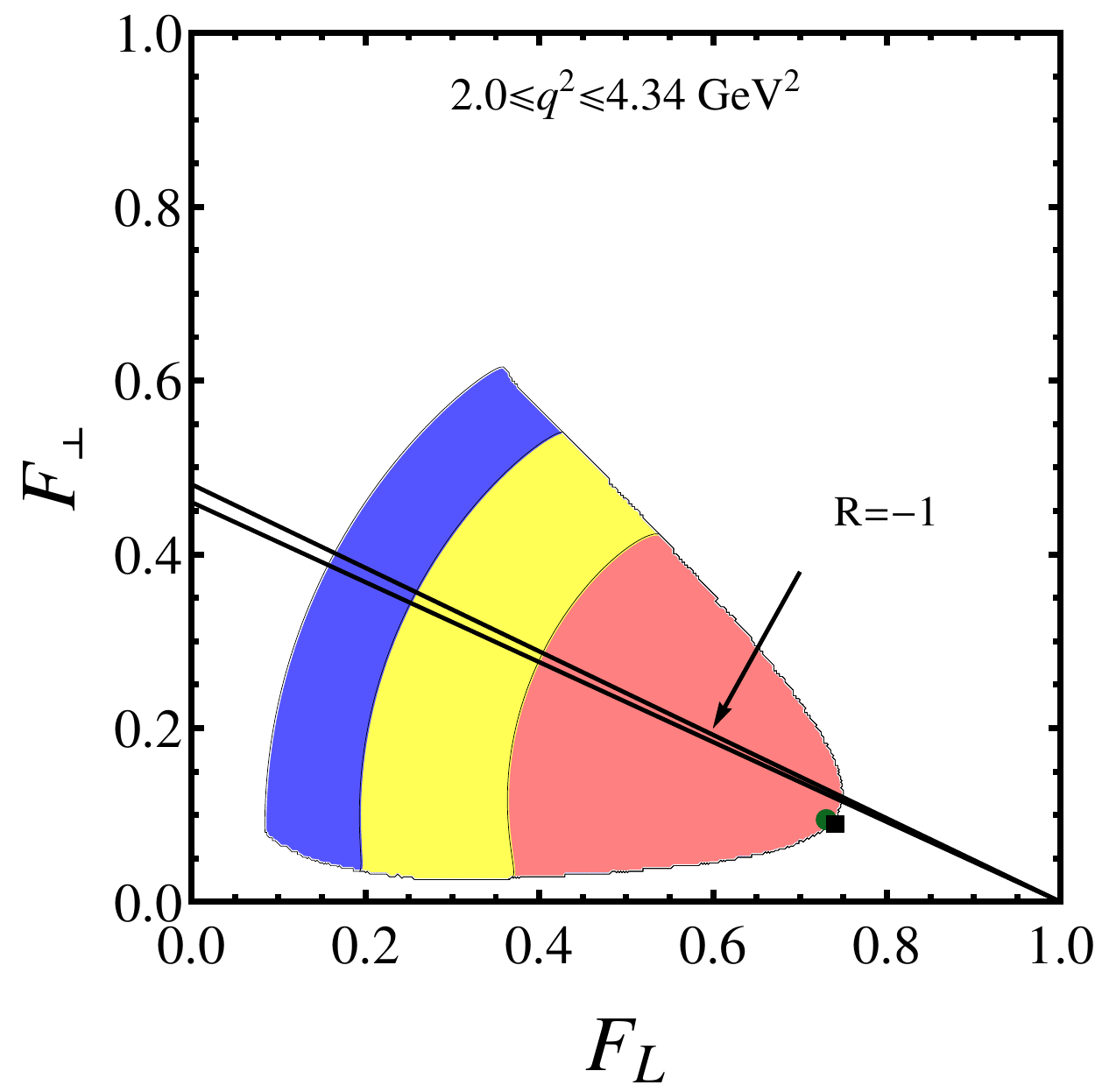}
	\includegraphics*[width=0.25\textwidth]{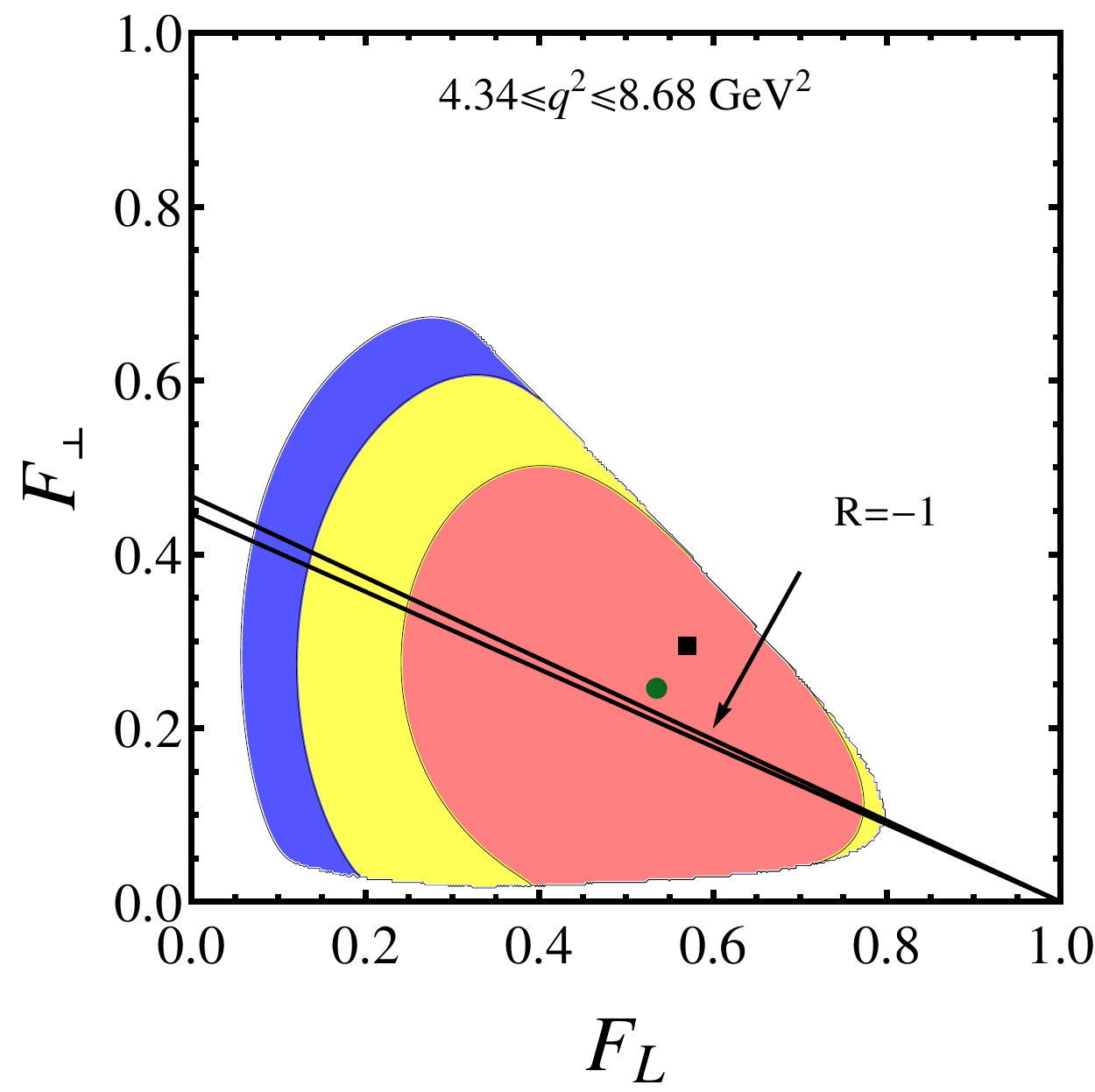}
	\includegraphics*[width=0.25\textwidth]{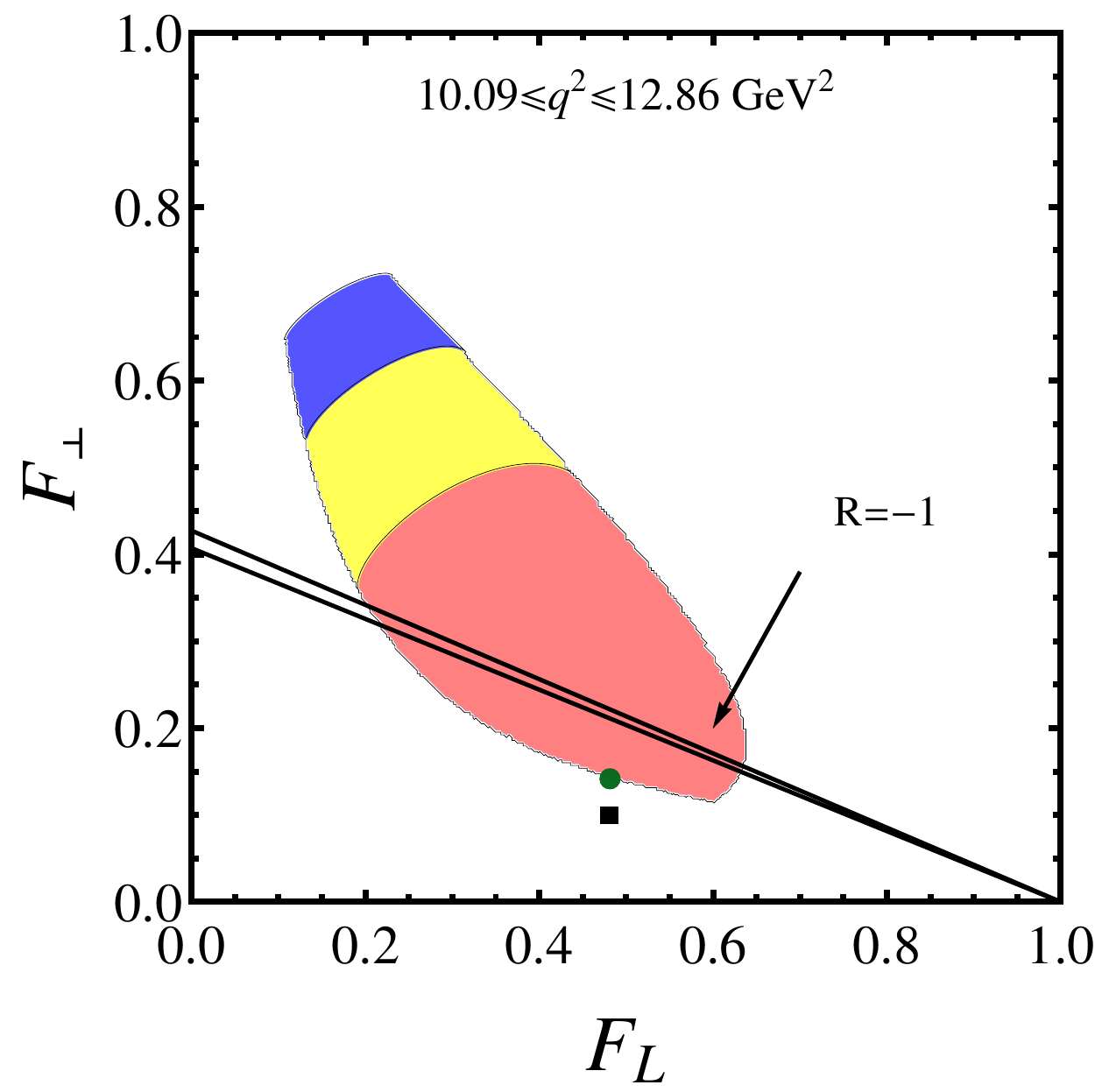}
	\includegraphics*[width=0.25\textwidth]{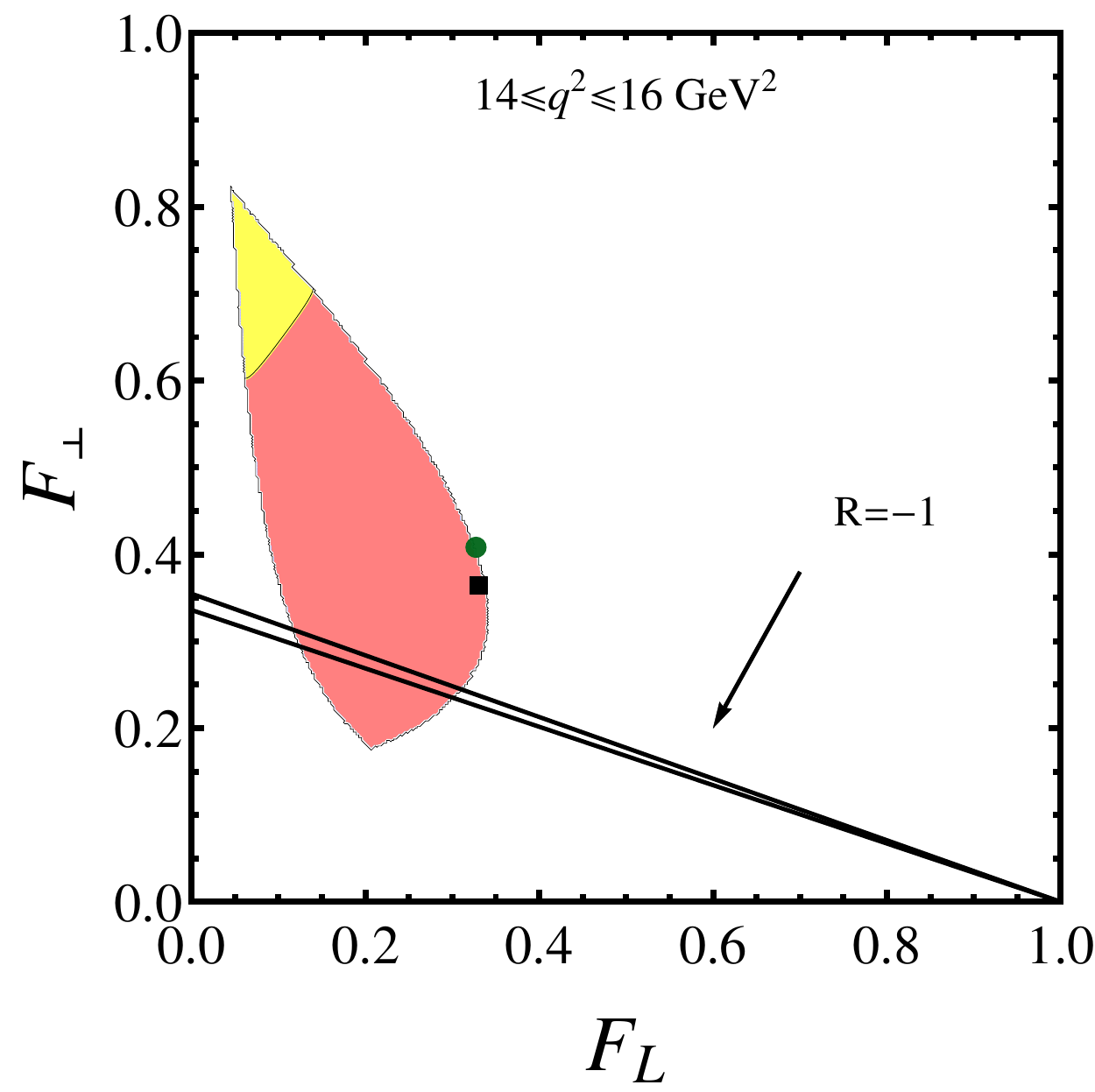}
	\includegraphics*[width=0.25\textwidth]{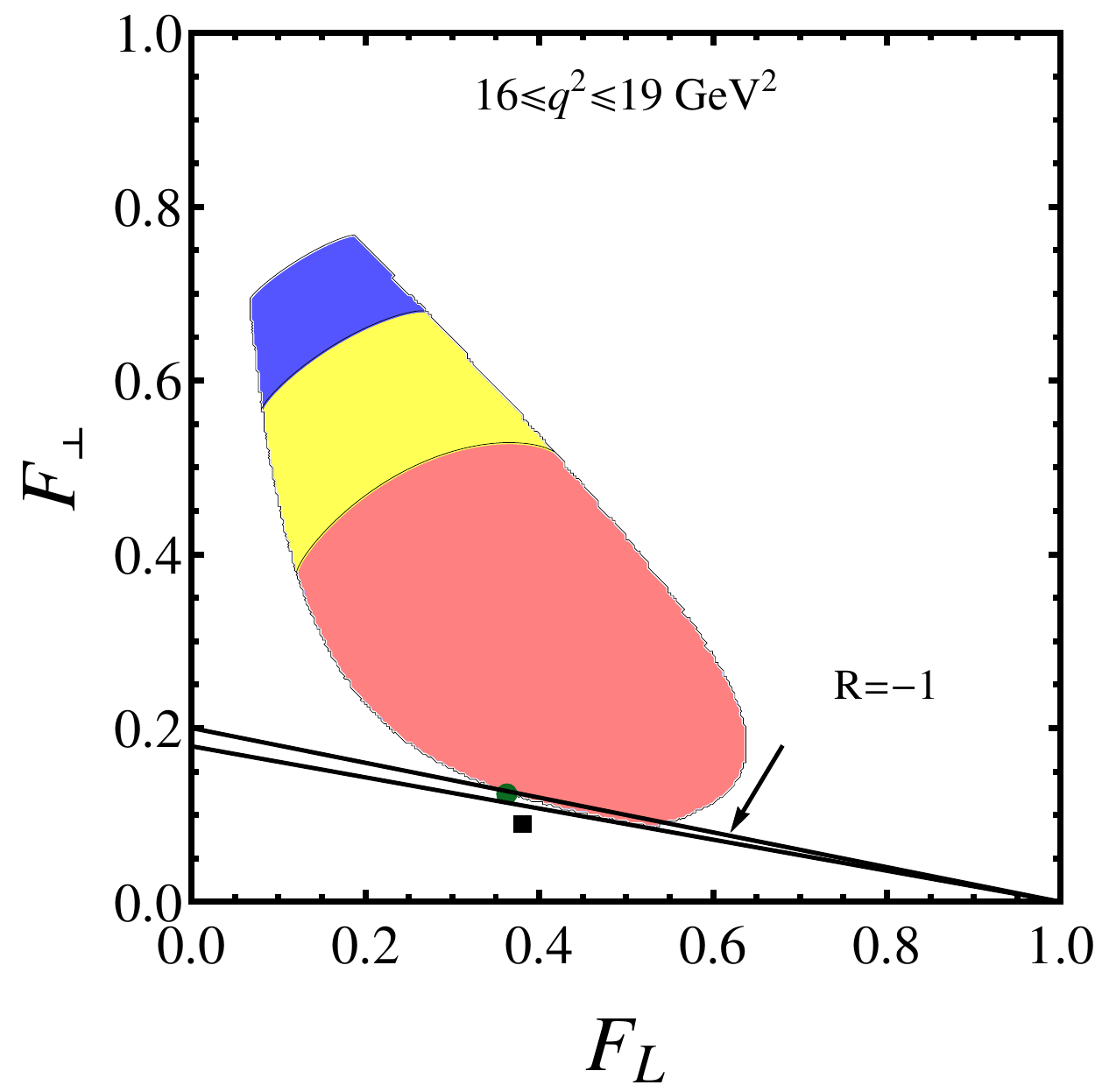} \caption{The 
	$\chi^2$ projection onto the plane of observables $F_L$ and $F_\perp$. The 
	experimental values of all the observables are taken from $1\invfb$ LHC$b$ 
	measurements Ref.\cite{Aaij:2013iag}. The green dots corresponds to best 
	fit value from $\chi^2$ minimization and the black squares corresponds to 
	the measured central value. The pink (dark), yellow (light) and blue 
	(darkest)  correspond to the $1\sigma$, $2\sigma$ and $3\sigma$
	confidence level regions respectively. If the amplitudes are real, 
	non-factorizable contributions vanish and the form-factors were reliably 
	evaluated at leading order in HQET then using SM estimated values of Wilson 
	coefficients we find $F_L-F_\perp$ are constrained to lie in the narrow 
	region between the two solid black lines. See text for details.} 
	\label{fig:FLFP}
 \end{center}
\end{figure*}
%
\begin{figure*}[hbtp]
 \begin{center}
	\includegraphics*[width=0.25\textwidth]{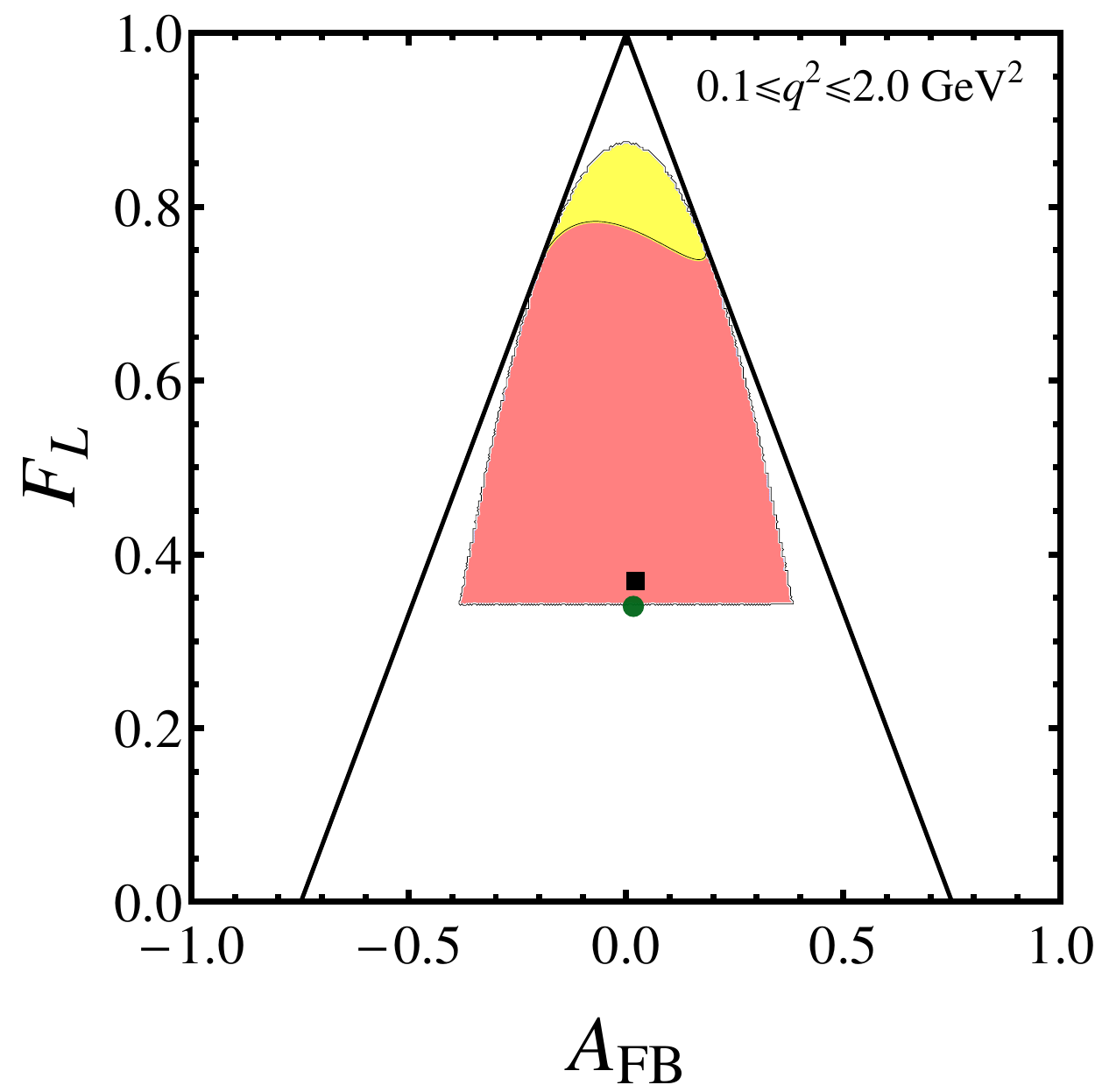}
	\includegraphics*[width=0.25\textwidth]{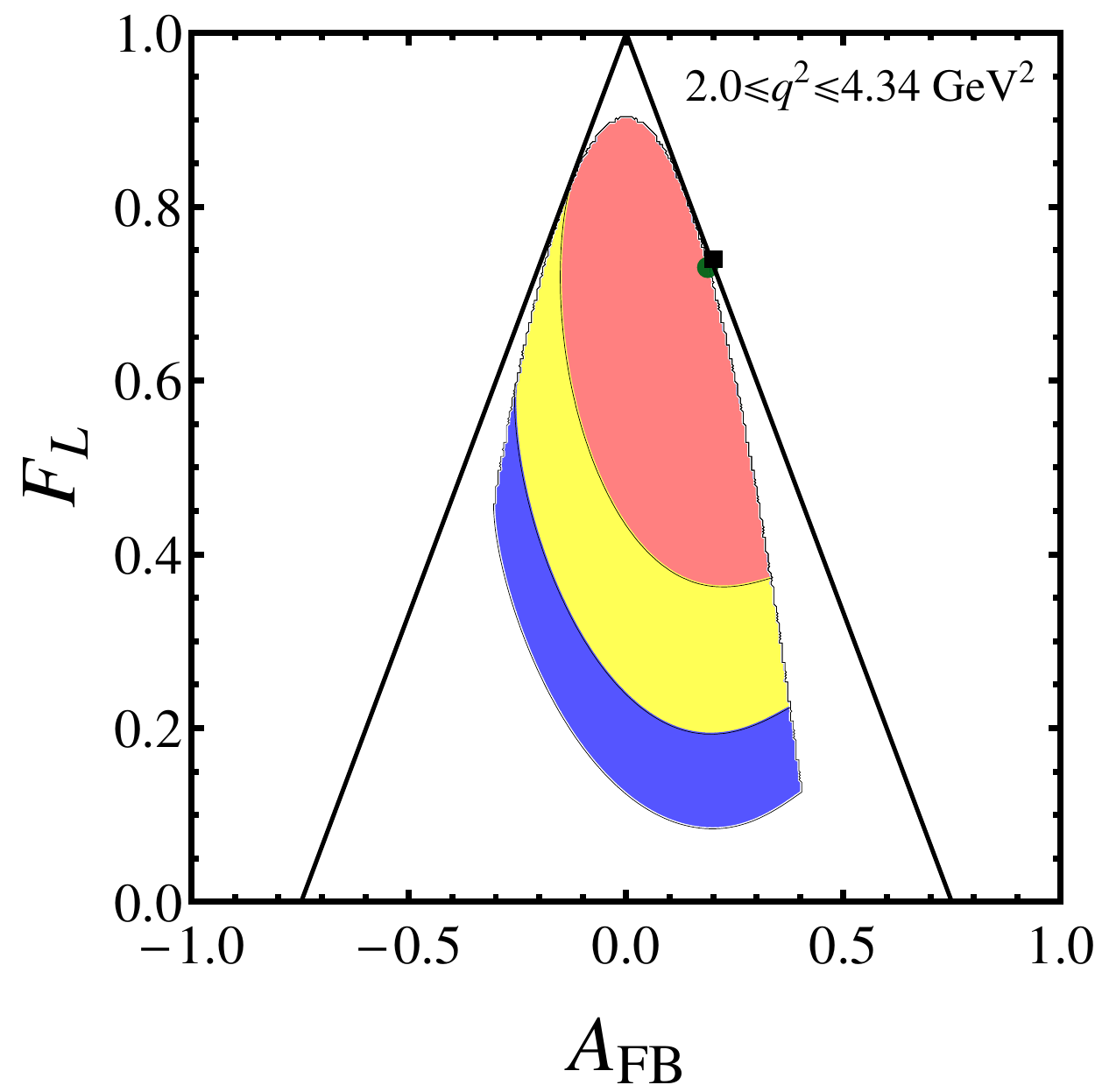}
	\includegraphics*[width=0.25\textwidth]{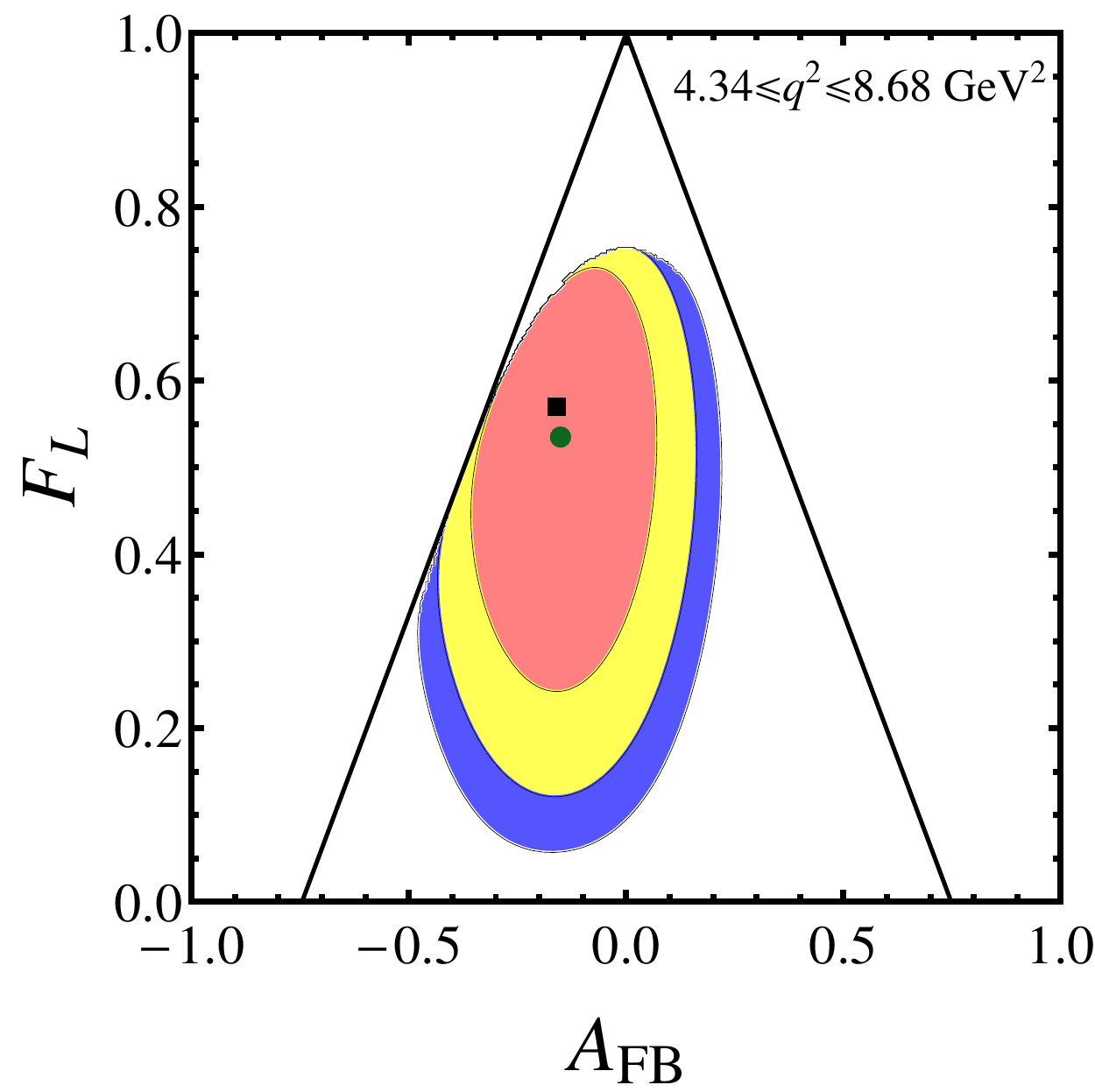}
	\includegraphics*[width=0.25\textwidth]{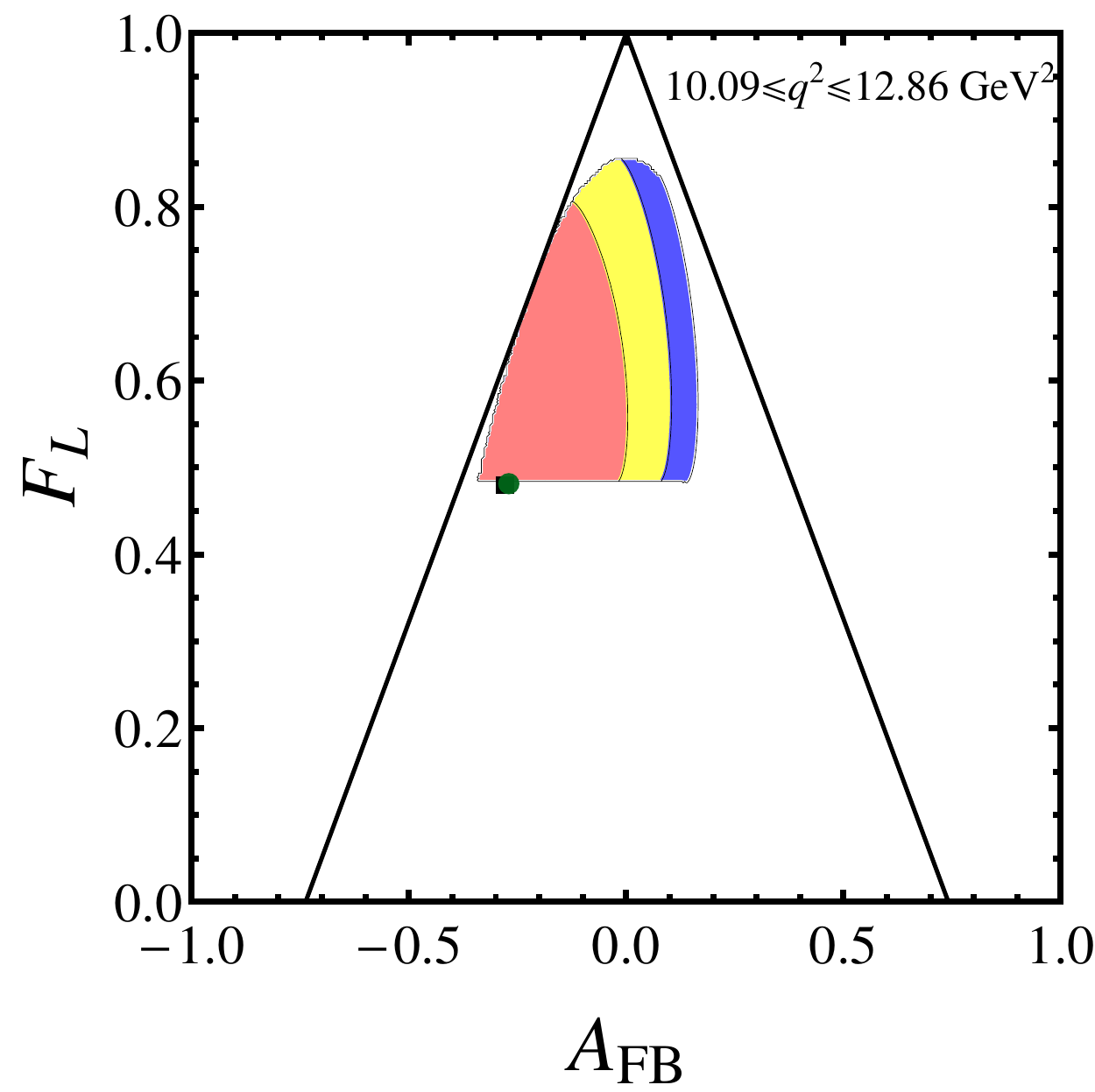}
	\includegraphics*[width=0.25\textwidth]{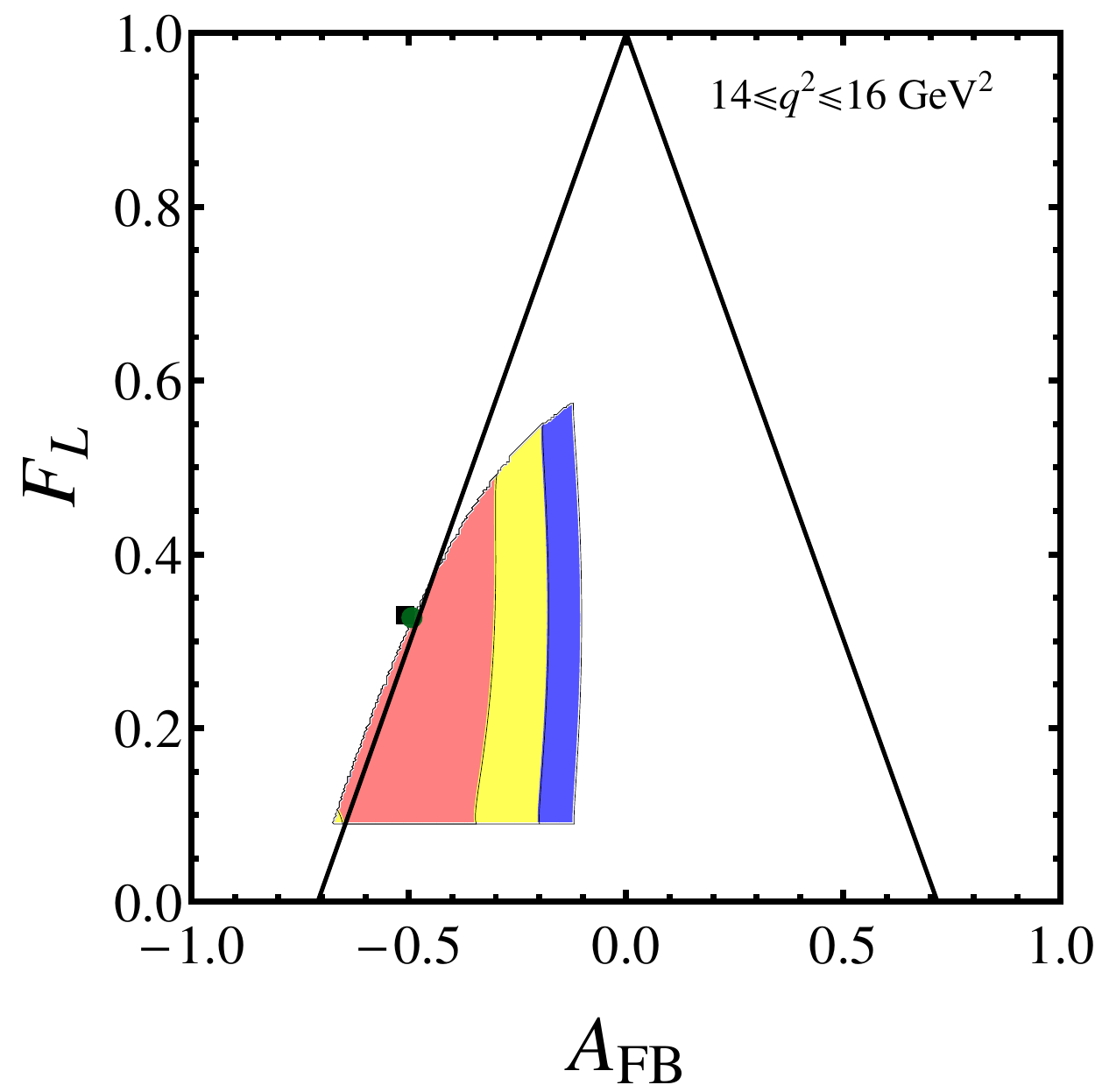}
	\includegraphics*[width=0.25\textwidth]{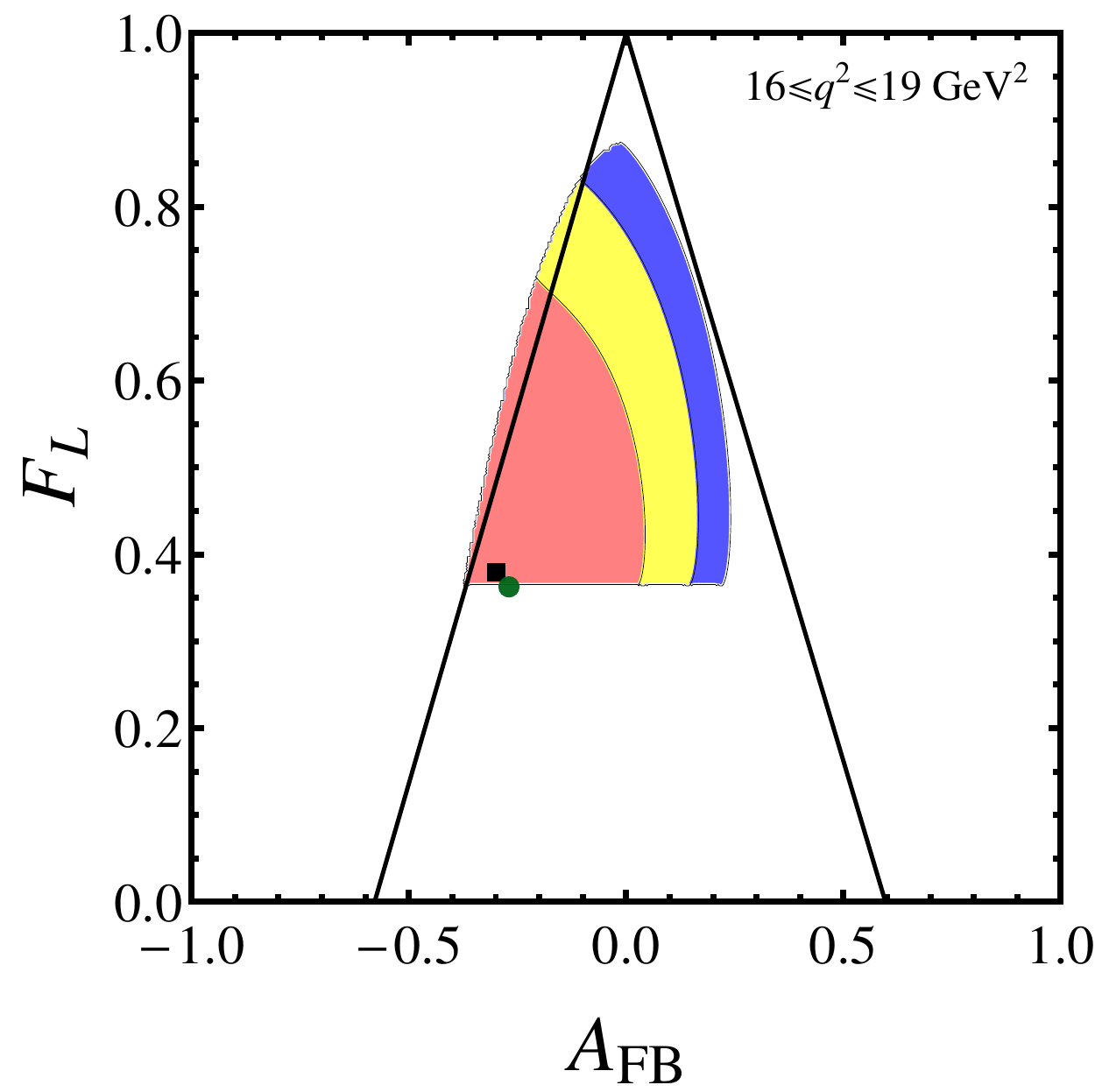}
	\caption{The $\chi^2$ projection onto the plane of observables $F_L$ and 
	$\AFB$. The experimental values of all the observables are taken from 
	Refs.~\cite{Aaij:2013iag}. The color codes are the same as in Fig. 
	\ref{fig:FLFP}. If the amplitudes are real, non-factorizable contributions 
	vanish and the form-factors were reliably evaluated at leading order in 
	HQET then using SM estimated values of Wilson coefficients we find 
	$\AFB-F_L$ are constrained to lie in the two solid black triangular region. 
	See text for details.}
	\label{fig:FLAFB}
 \end{center}
\end{figure*}
\end{center}

\begin{itemize}
\item  \underline{\textbf{Case-II:}} $q^2\!=\!\left(m_B-m_\kstar\right)^{\!\,2}$
\end{itemize}
In this kinematic limit the $\kstar$ is at rest and the two leptons go back to
back in the $B$ meson rest frame. Therefore, we can always choose the angle
$\phi$ to be zero. The entire decay takes place in one plane, resulting in
vanishing $F_\perp$. Also, the left and right chirality of the leptons
contribute equally. These together results in only the angular asymmetry $A_4$
being finite with all other asymmetries vanishing. The relations among the
various angular coefficients at this kinematical endpoint are derived in
Ref.~\cite{Hiller:2013} where it is explicitly shown that
\begin{eqnarray}
F_L(\qmax)=\frac{1}{3},\qquad \AFB(\qmax)=0.
\end{eqnarray}
Solving for the other observables from Eq.~(3.2) of Ref.~\cite{Hiller:2013}
we can write
\begin{align}
& F_\perp(\qmax)=0, & F_\|(\qmax)=\frac{2}{3},\\
\label{eq:A4max}
& A_4(\qmax)= \frac{2}{3\pi}, & 
A_{5,7,8,9}(\qmax)=0.
\end{align}
These limiting values of the observables imply that  $\varepsilon_\lambda \to 0$
at the extremum $q^2\!=\!\qmax$ as can be seen from
Eqs.~\eqref{eq:eps_perpo} -- \eqref{eq:eps_0o}. The lepton mass can be safely
ignored at $\qmax$ as it would have almost no effect at this endpoint hence,
we have dropped the `o' index from all the observables for this discussion only.
Thus, in the limit $\varepsilon_\lambda\to 0$, we find that
Eq.~\eqref{eq:Obs-relationnew} reduces to Eq.~\eqref{eq:Obs-relation}. Hence,
the observable $A_4$ at $q^2\!=\!\qmax$ turns out to be
\begin{align*}
   A_4&=
\frac{8 A_5 \AFB}{9 \pi F_\perp}+\sqrt{2}\,\frac{\sqrt{F_LF_\perp
-\frac{8}{9} A_5^2}\sqrt{F_\|F_\perp -\frac{4}{9}\AFB^2}}{\pi F_\perp} \\ 
& \mathop{\mathop{=}_{\AFB\to 0}}_{A_5\to 0}
\frac{\sqrt{2}\sqrt{F_LF_\|}}{\pi}  
\mathop{\mathop{=}_{F_L\to \frac{1}{3}}}_{F_\|\to \frac{2}{3}}
\frac{2}{3\pi}
\end{align*}
which exactly matches with the limit predicted in Eq.~\eqref{eq:A4max}.
\begin{center}
\begin{figure*}[htbp]
 \begin{center}
	\includegraphics*[width=0.21\textwidth]{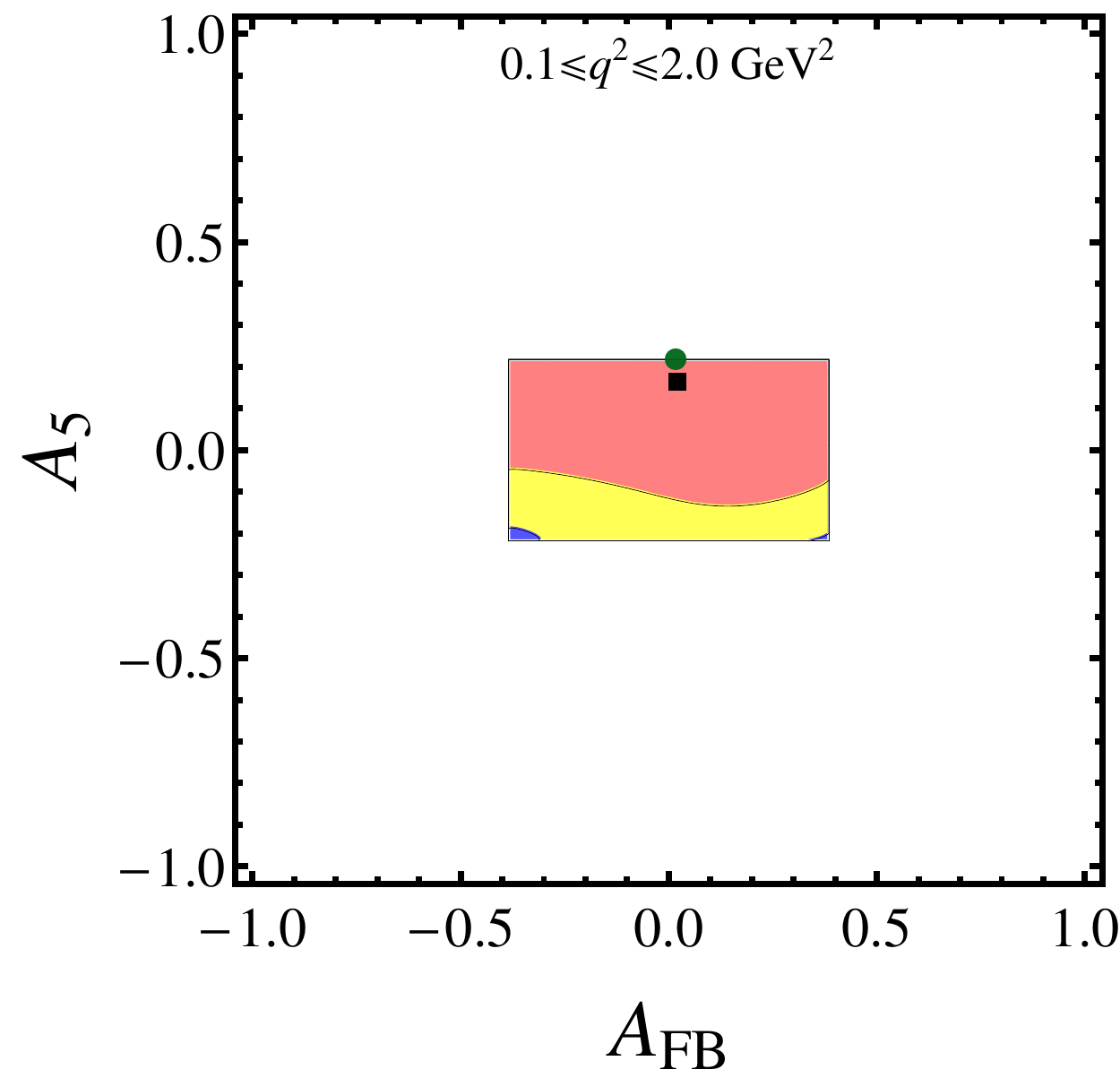}
	\includegraphics*[width=0.2\textwidth]{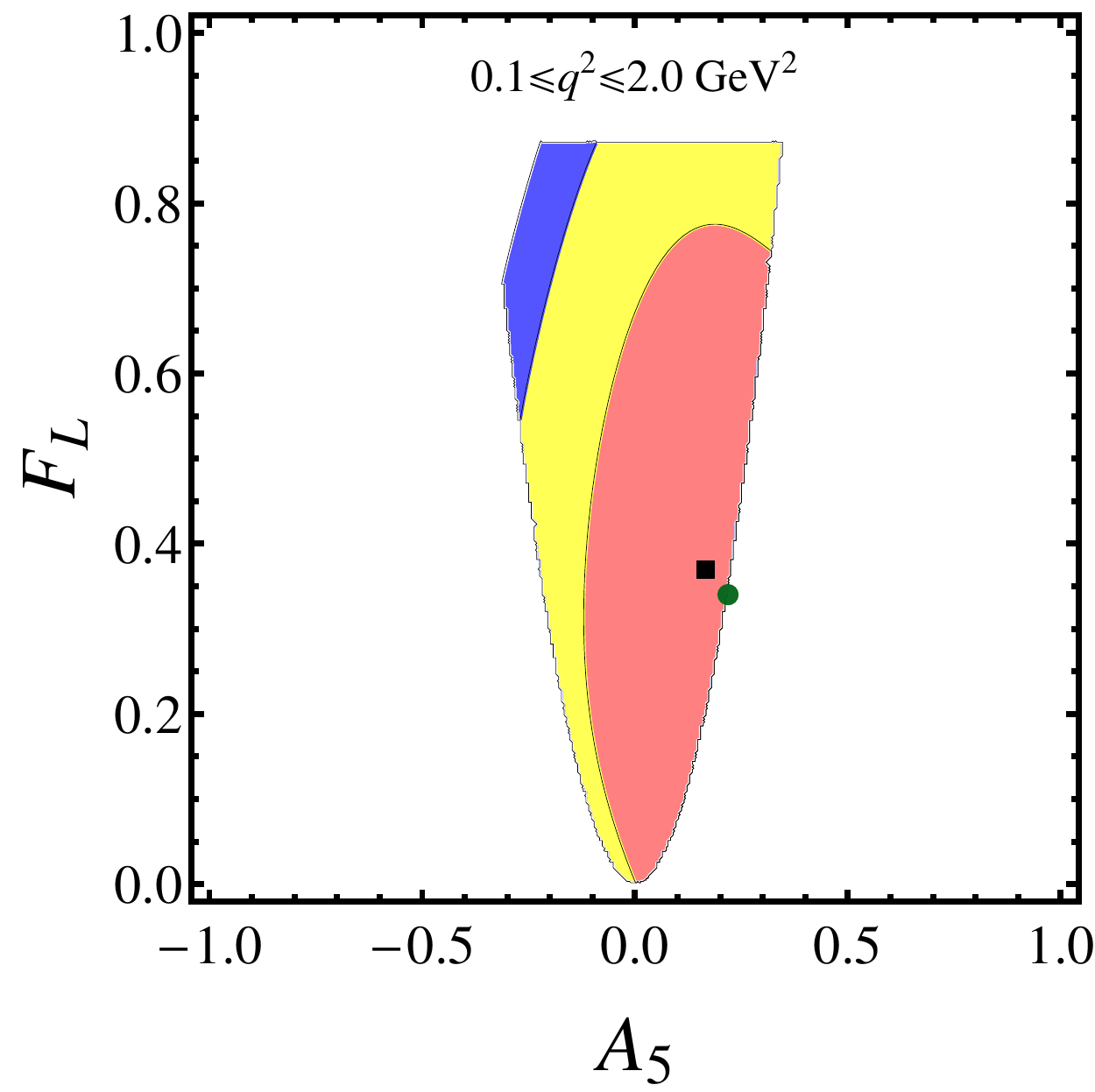}
	\includegraphics*[width=0.2\textwidth]{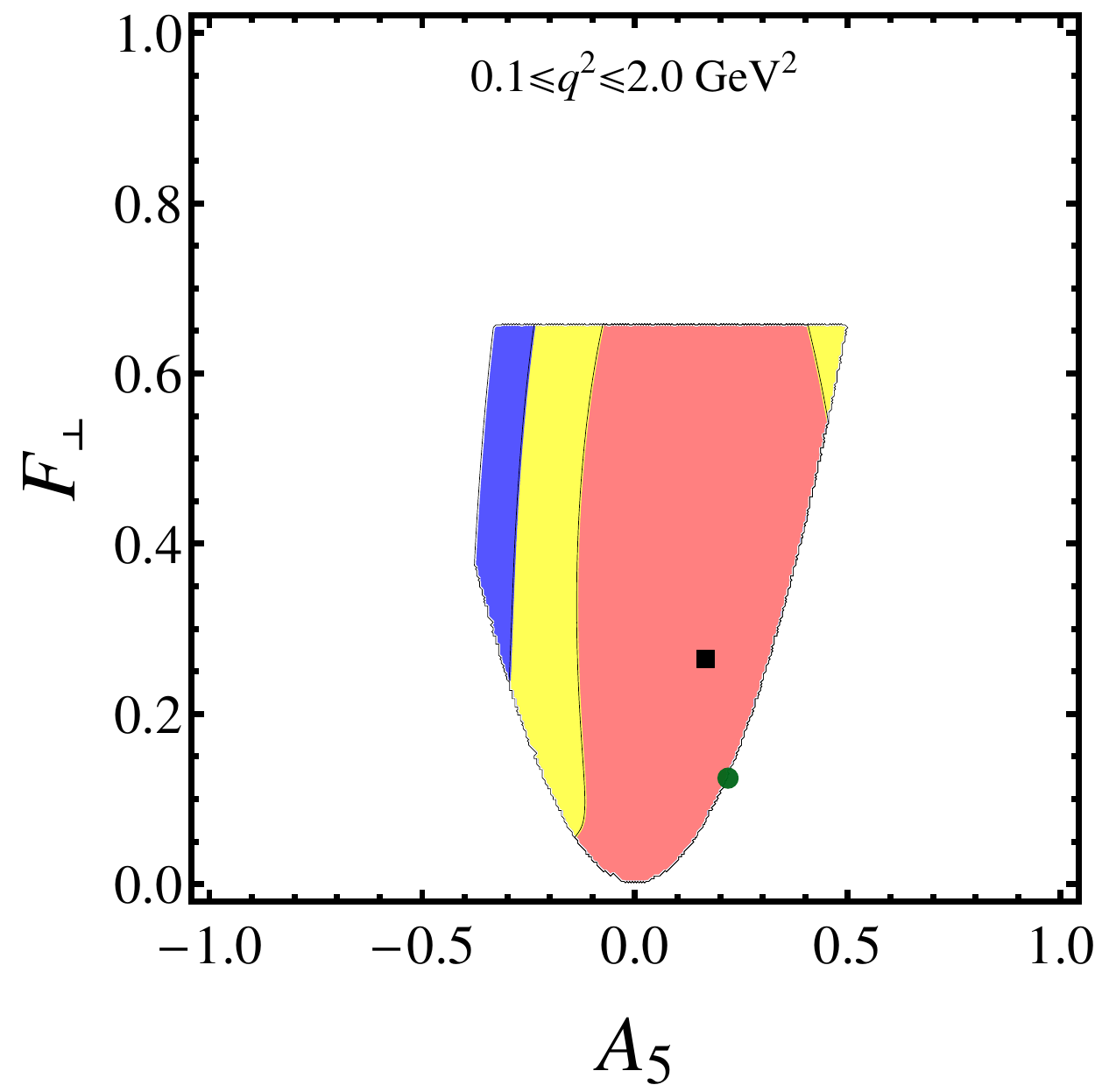}
	\includegraphics*[width=0.2\textwidth]{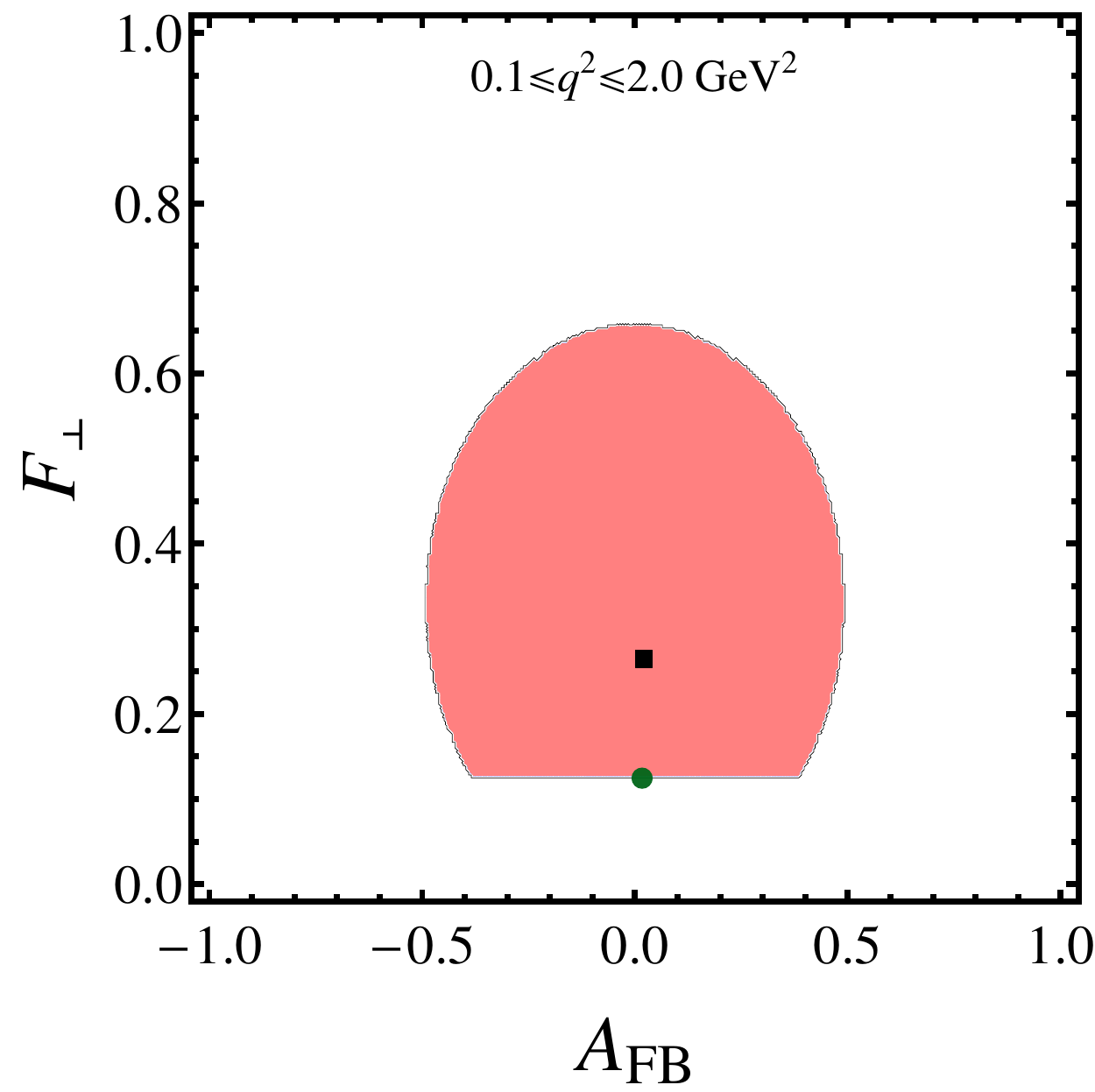}
	\includegraphics*[width=0.21\textwidth]{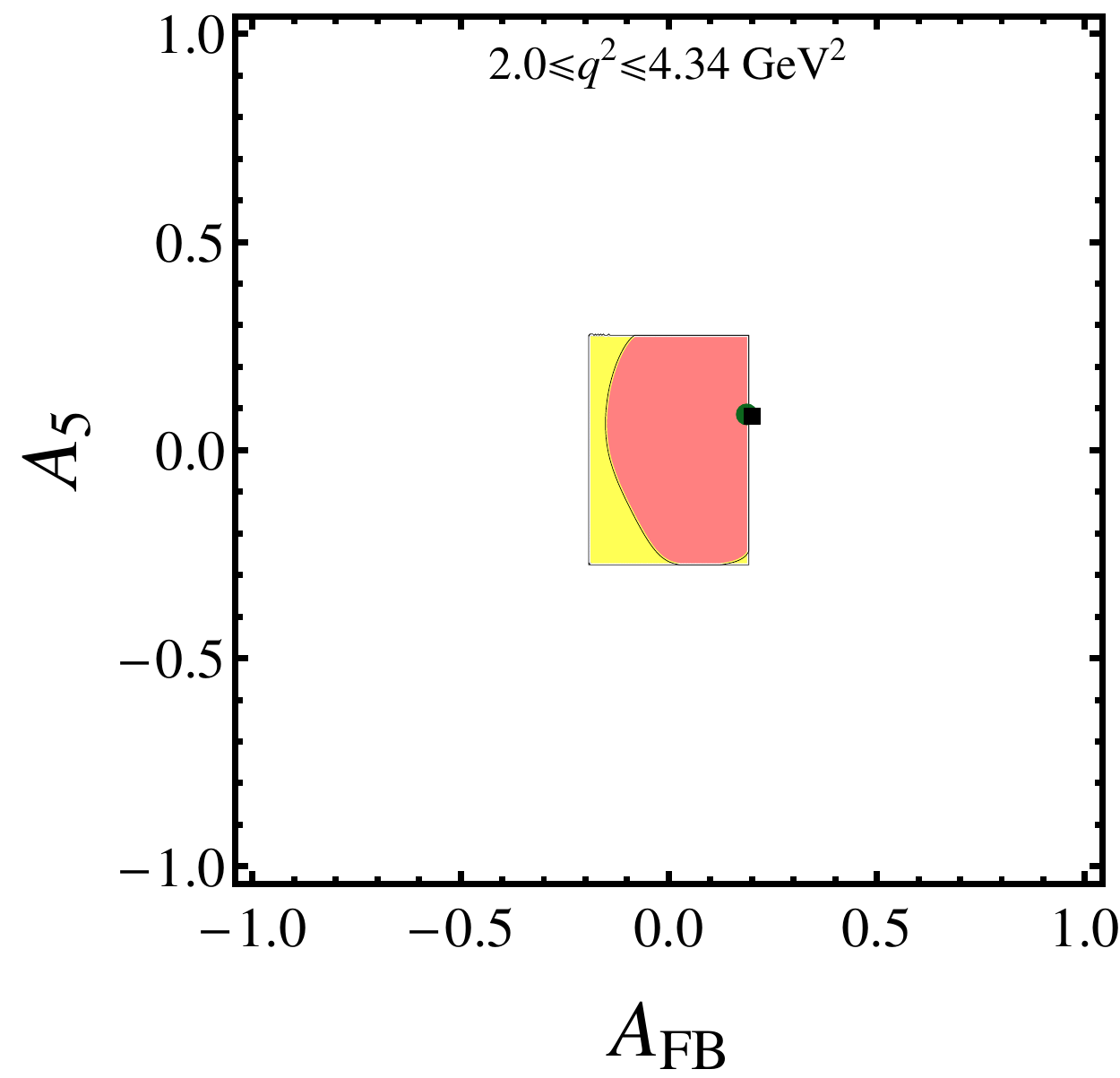}
	\includegraphics*[width=0.2\textwidth]{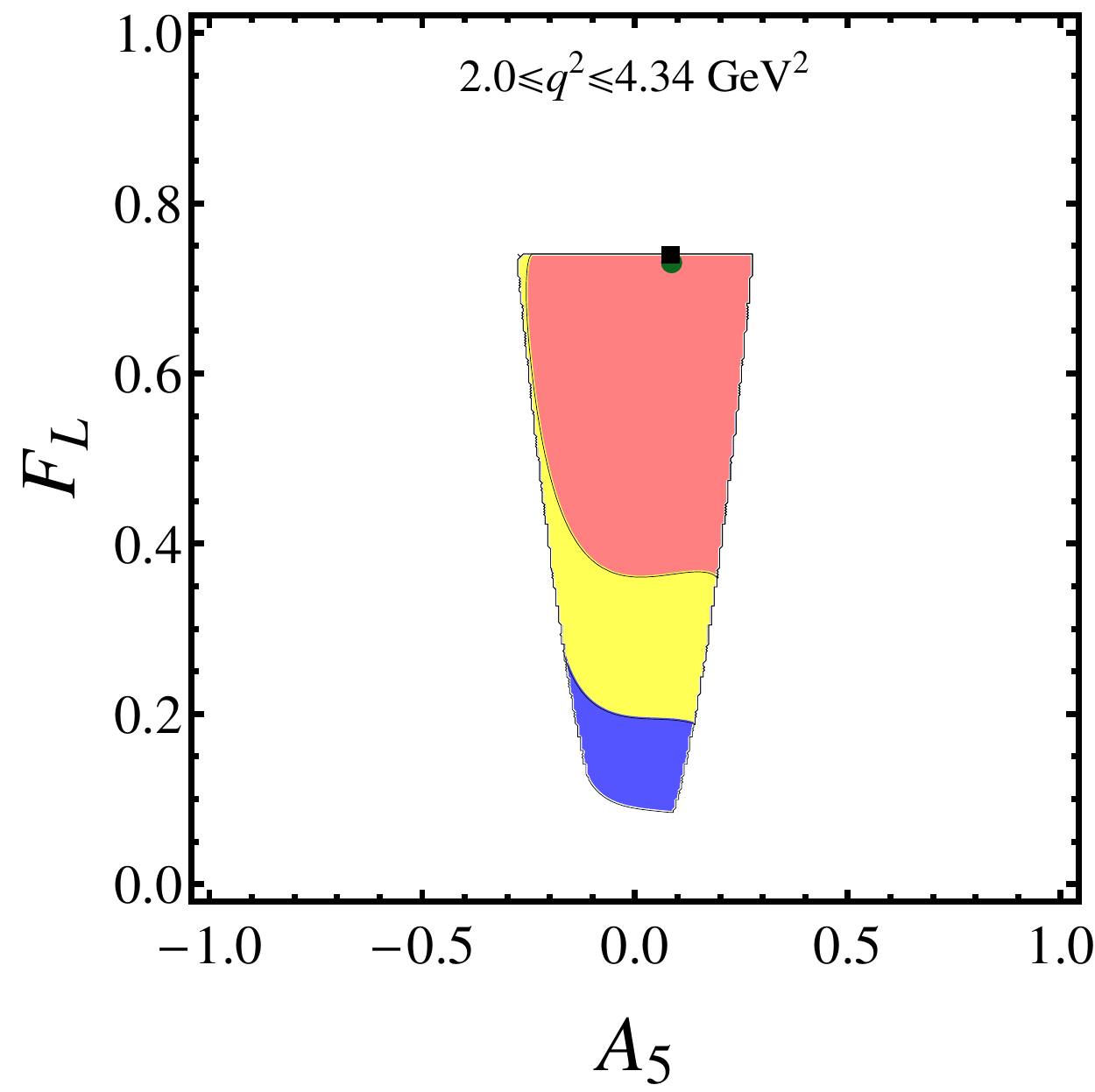}
	\includegraphics*[width=0.2\textwidth]{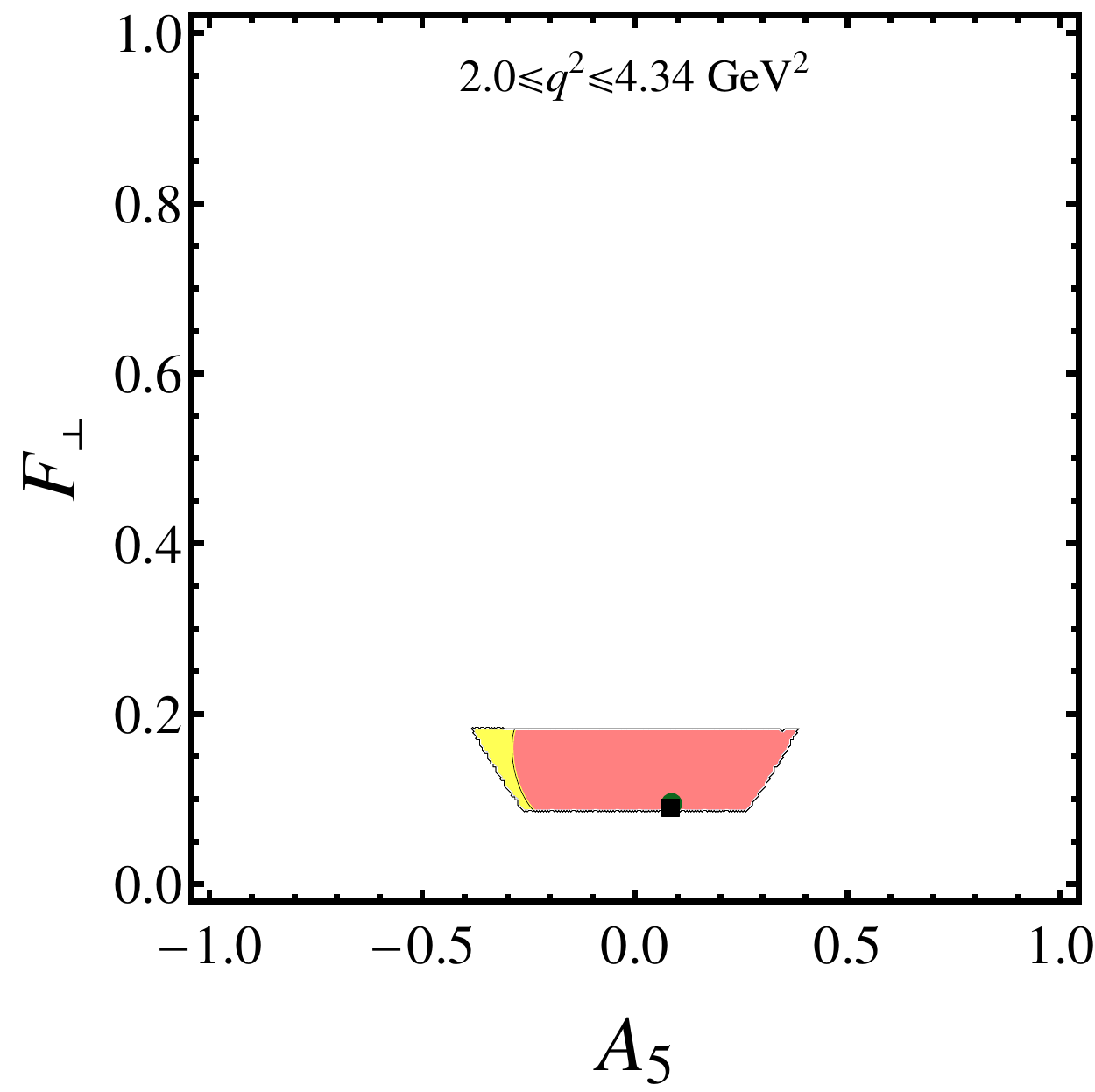}
	\includegraphics*[width=0.2\textwidth]{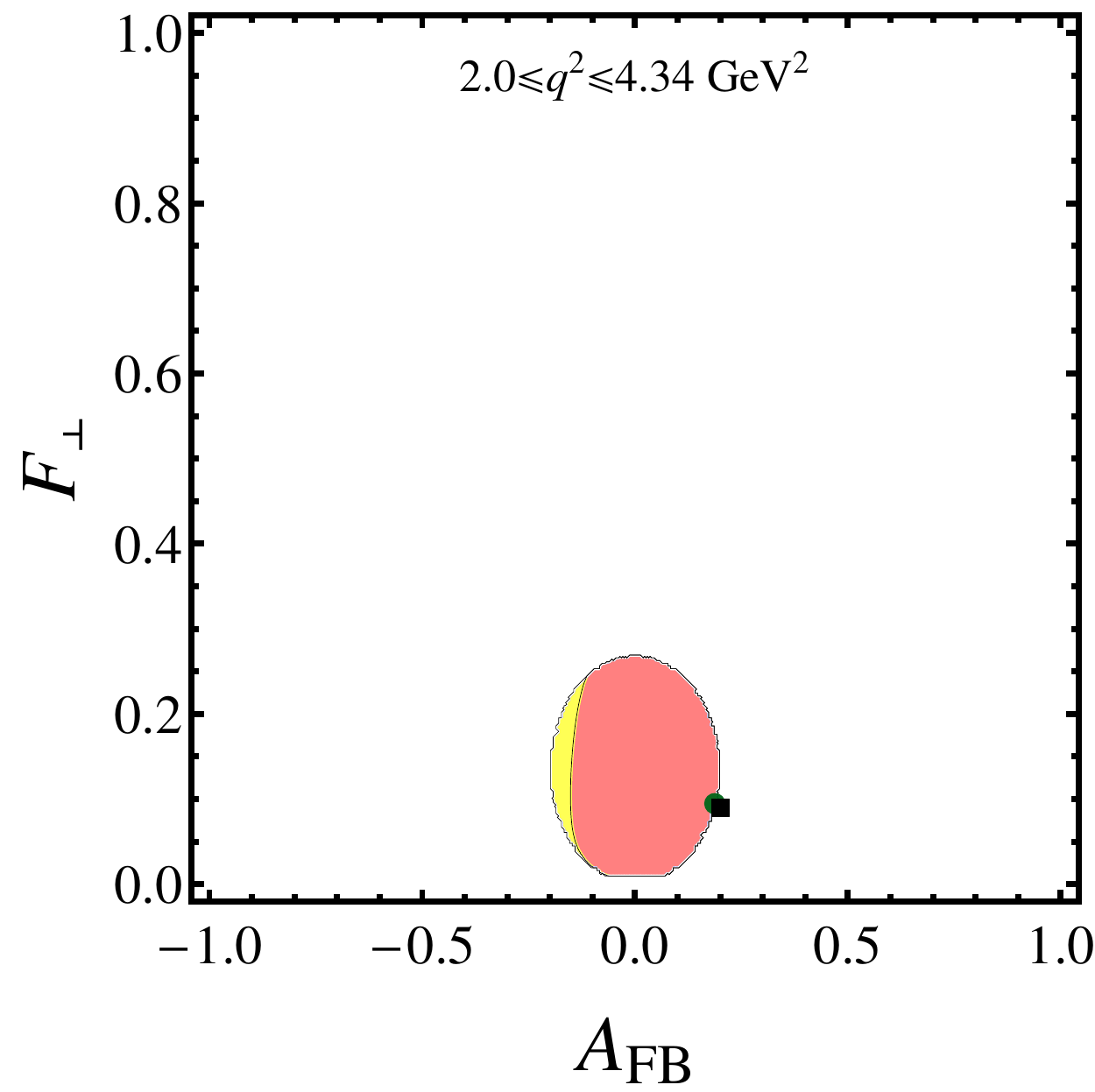}
	\includegraphics*[width=0.21\textwidth]{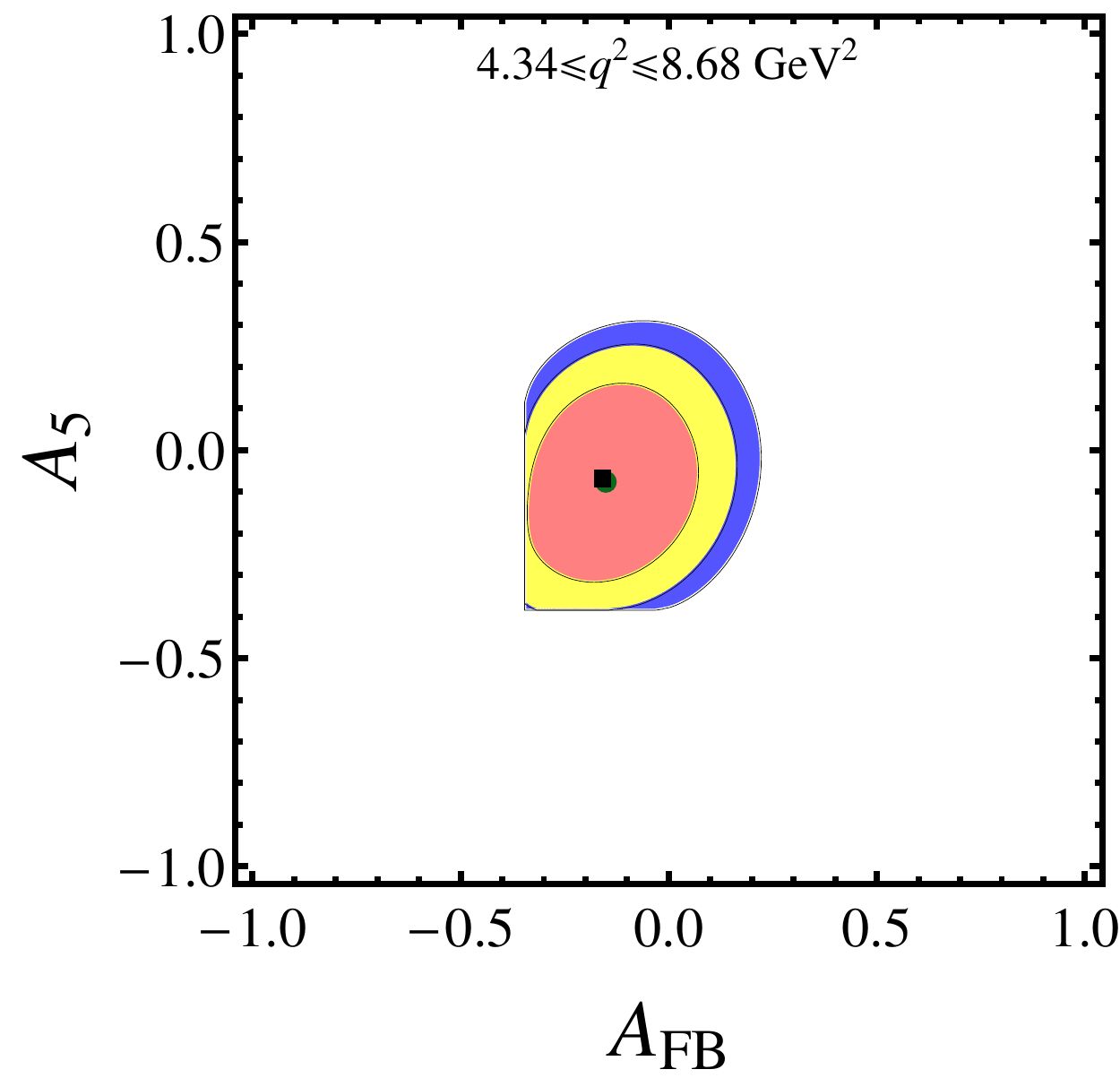}
	\includegraphics*[width=0.2\textwidth]{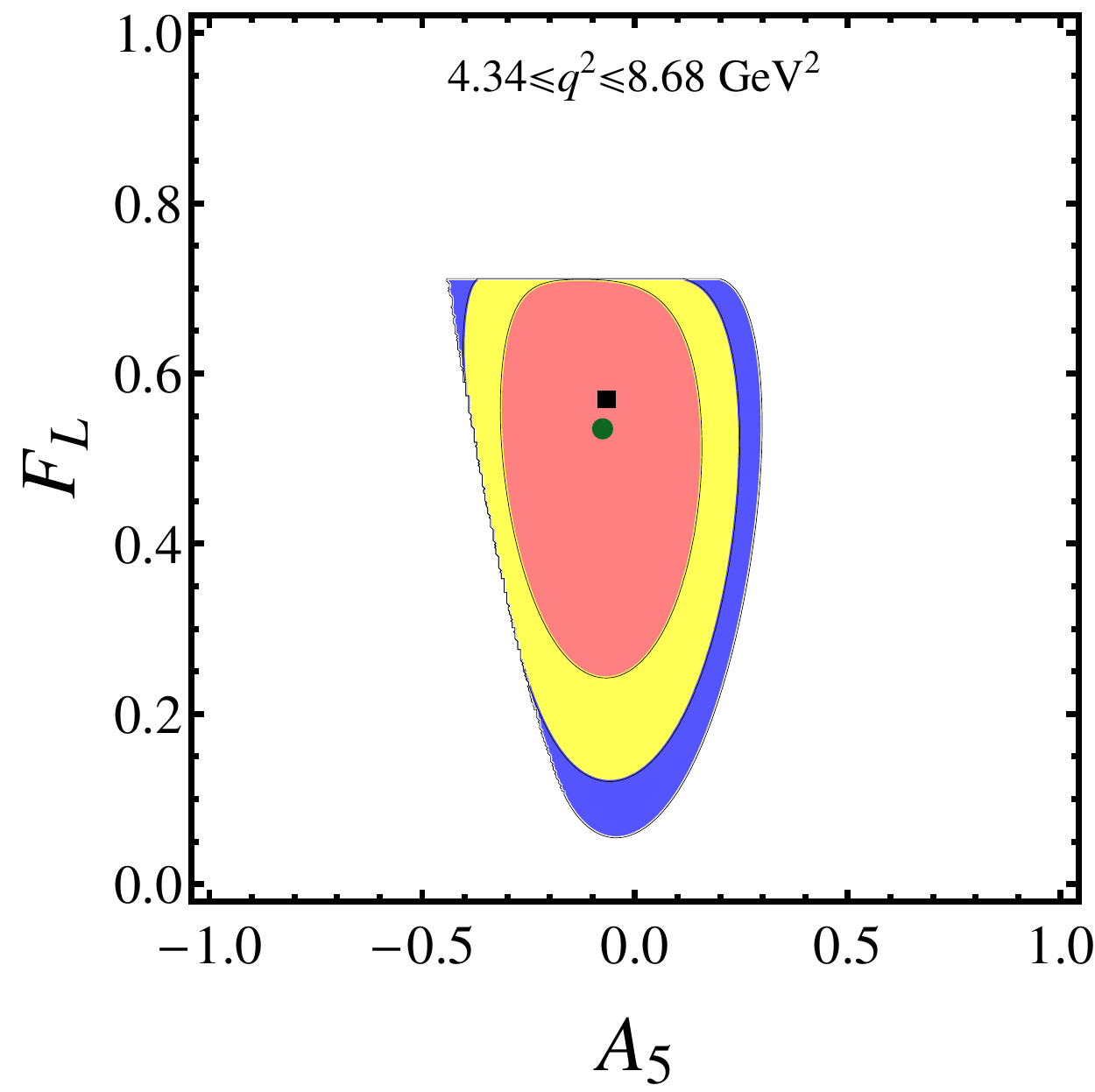}
	\includegraphics*[width=0.2\textwidth]{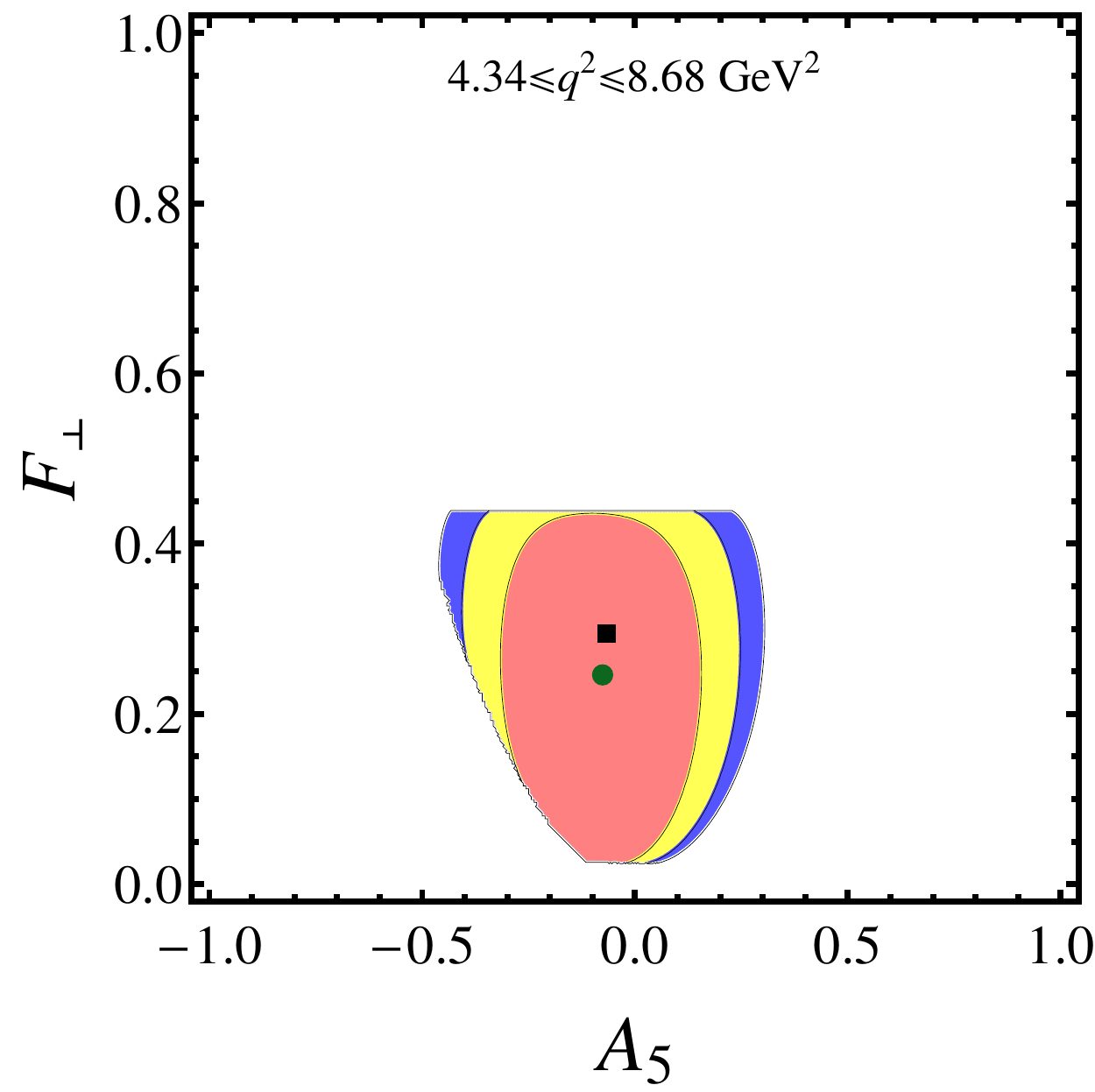}
	\includegraphics*[width=0.2\textwidth]{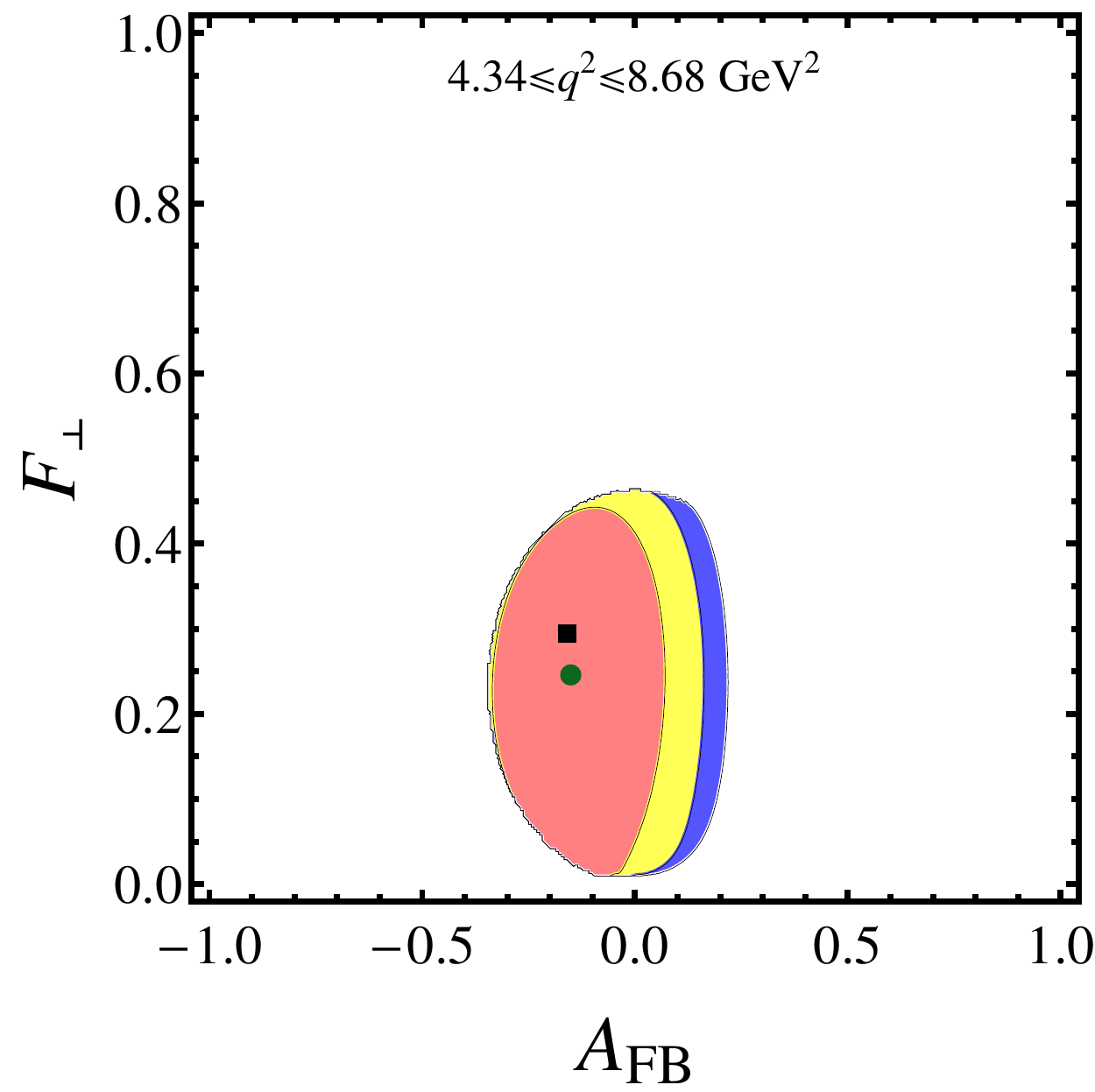}
	\includegraphics*[width=0.21\textwidth]{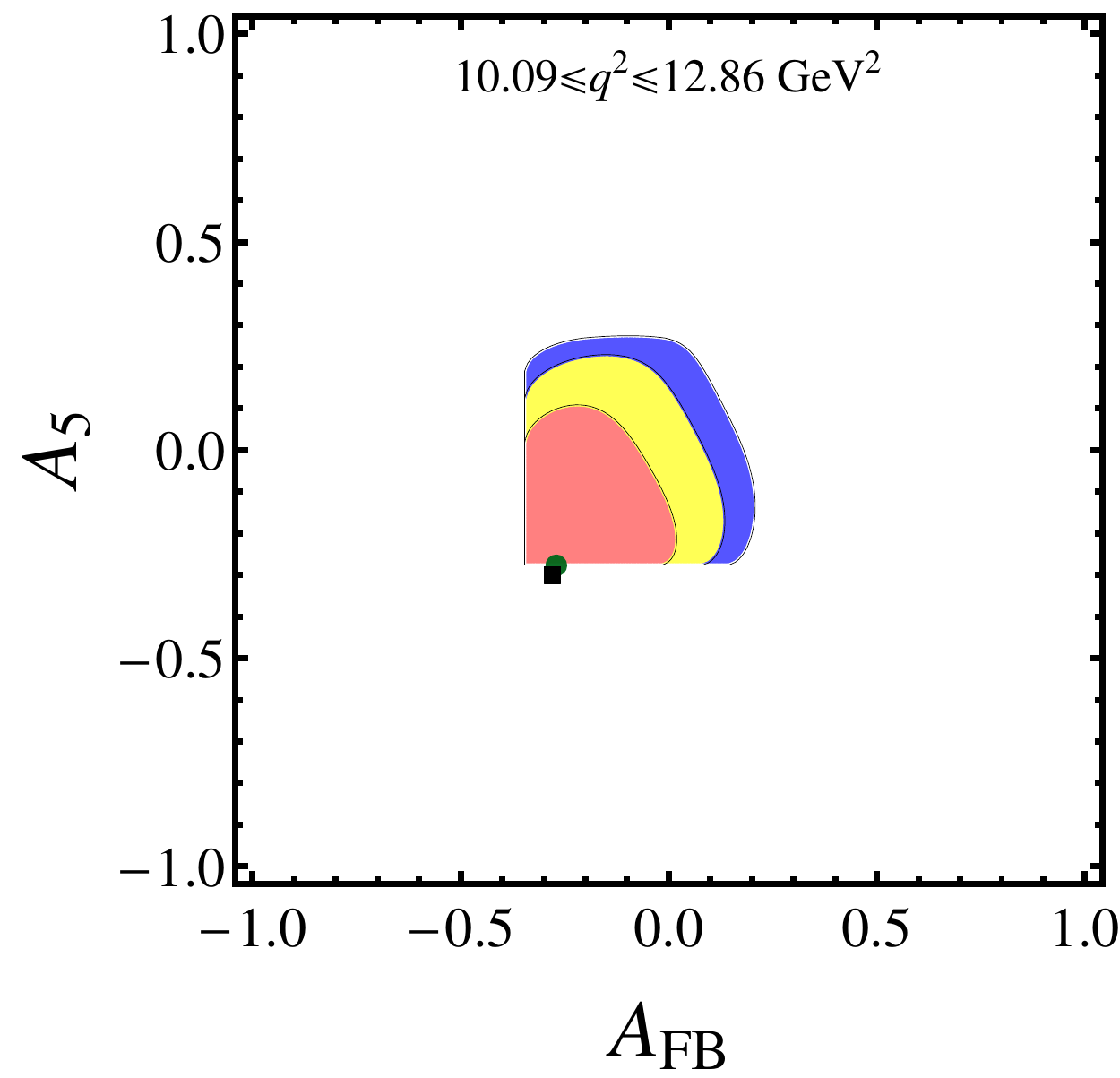}
	\includegraphics*[width=0.2\textwidth]{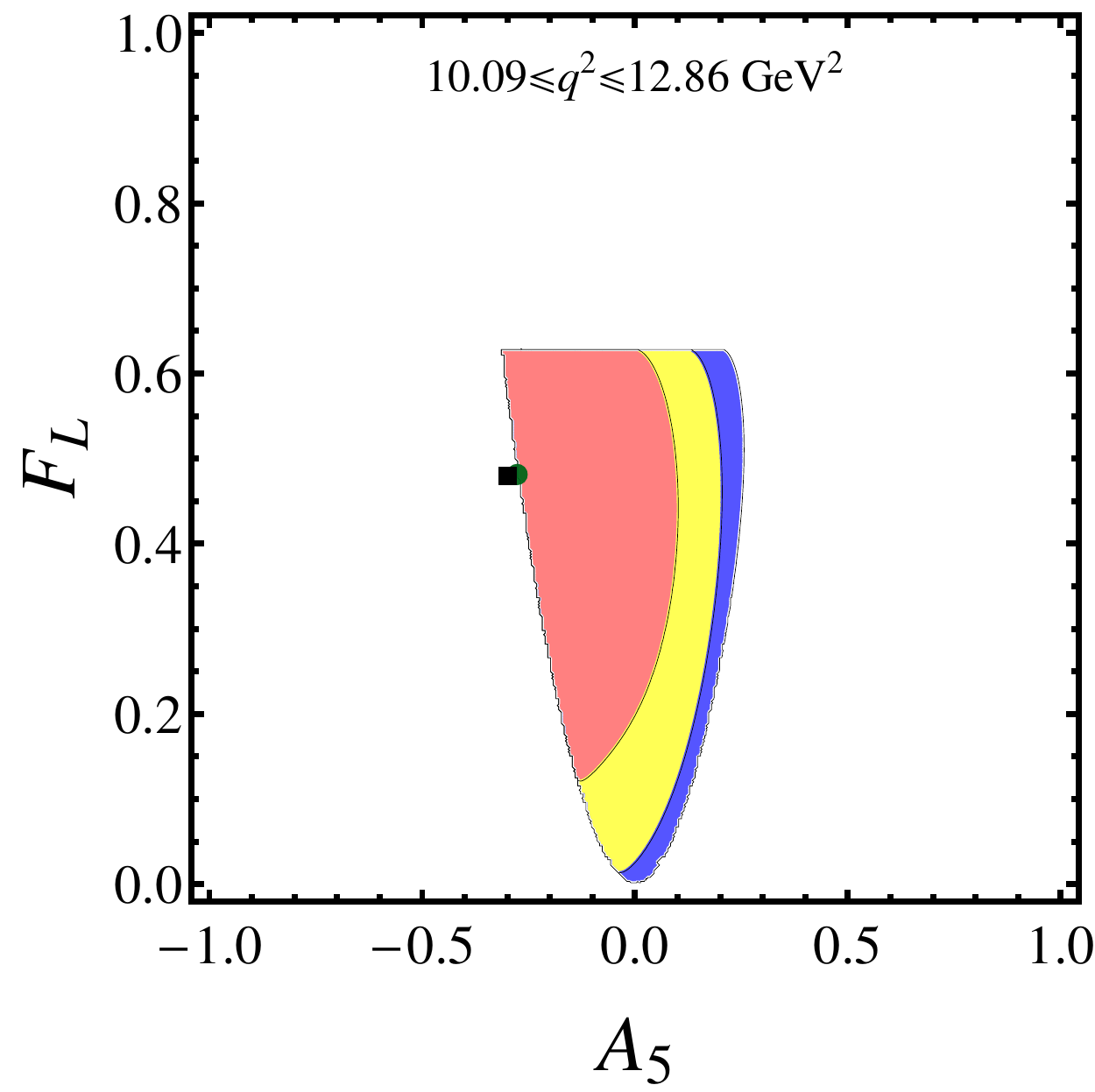}
	\includegraphics*[width=0.2\textwidth]{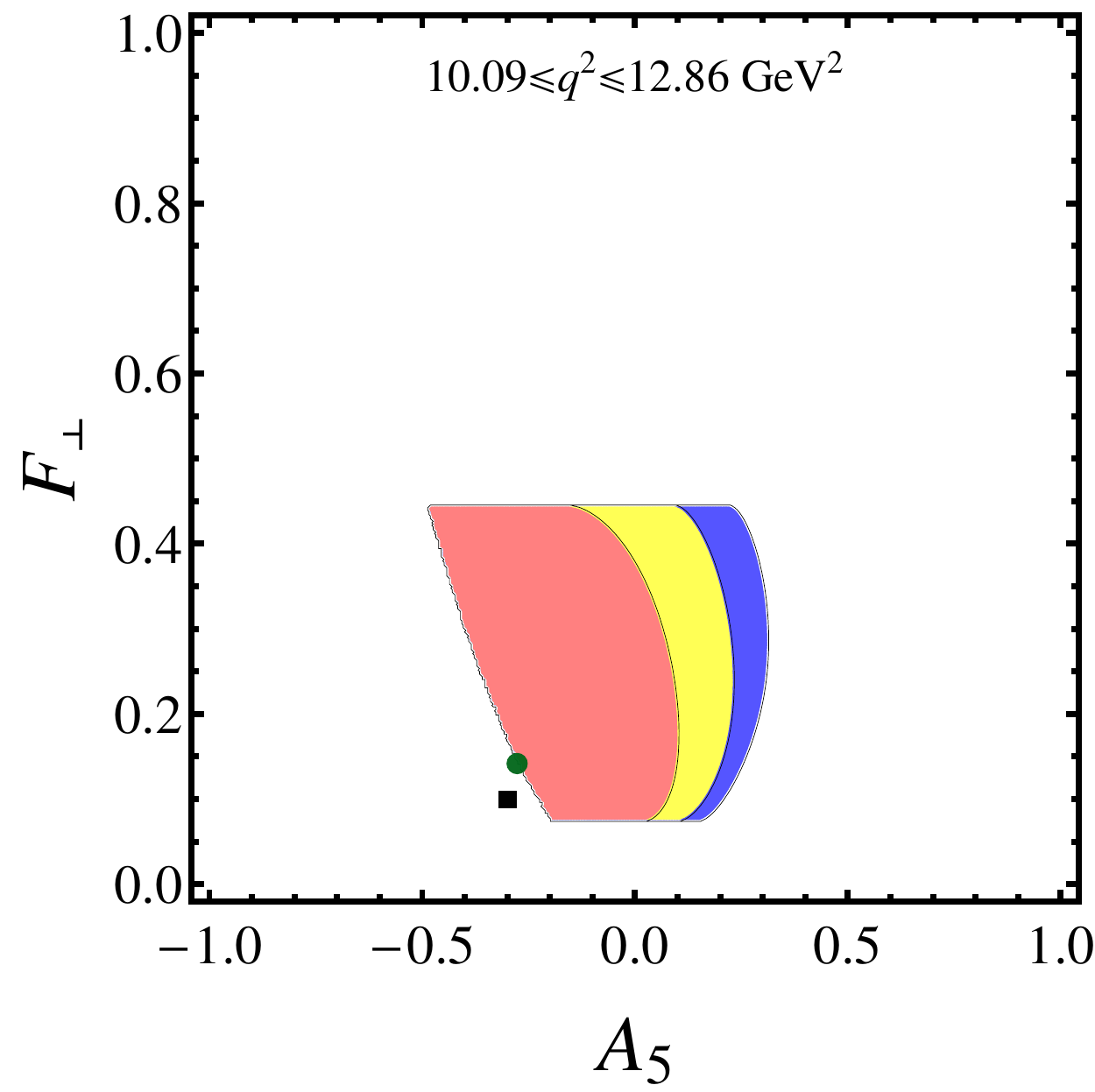}
	\includegraphics*[width=0.2\textwidth]{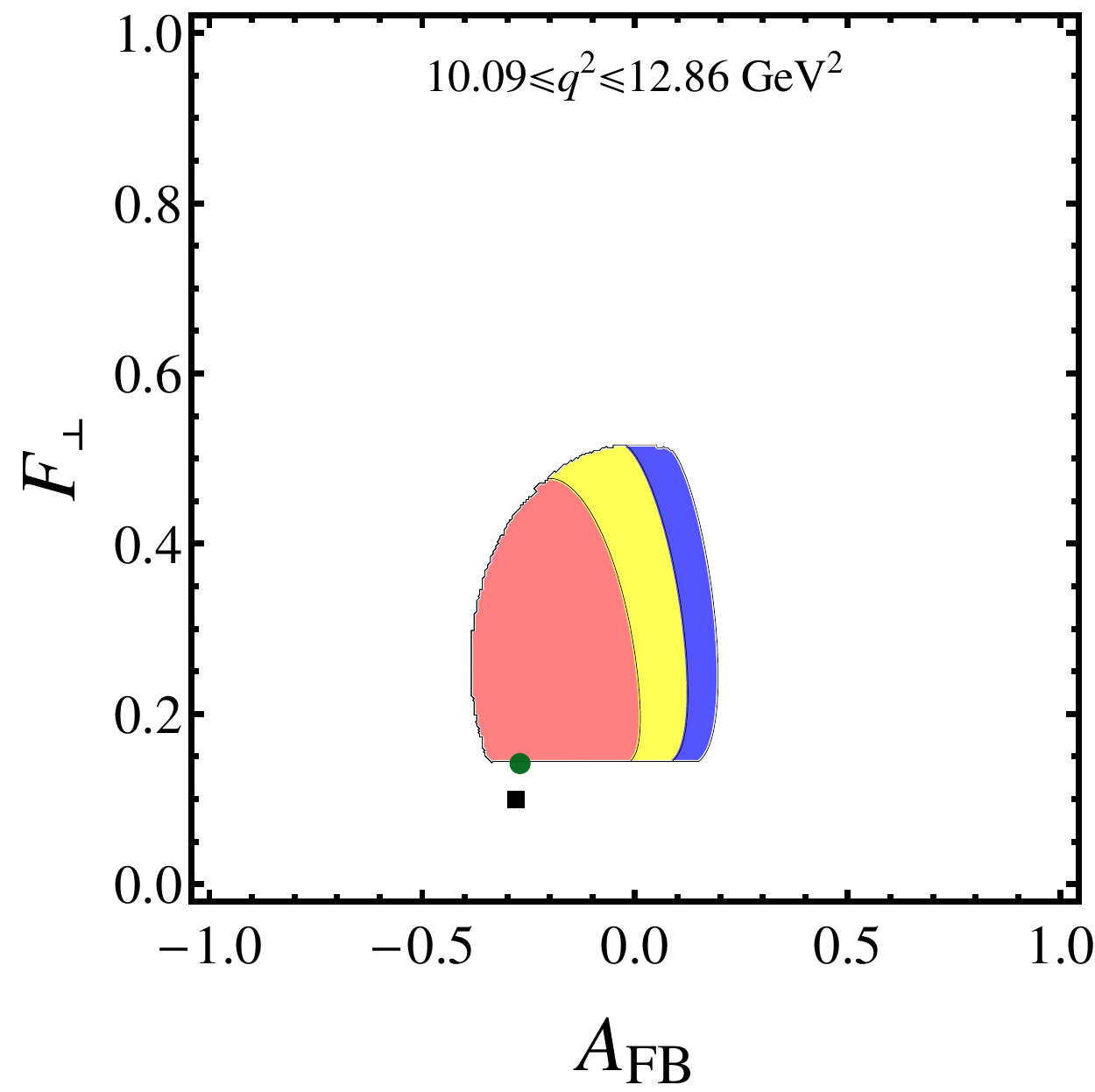}
	\includegraphics*[width=0.21\textwidth]{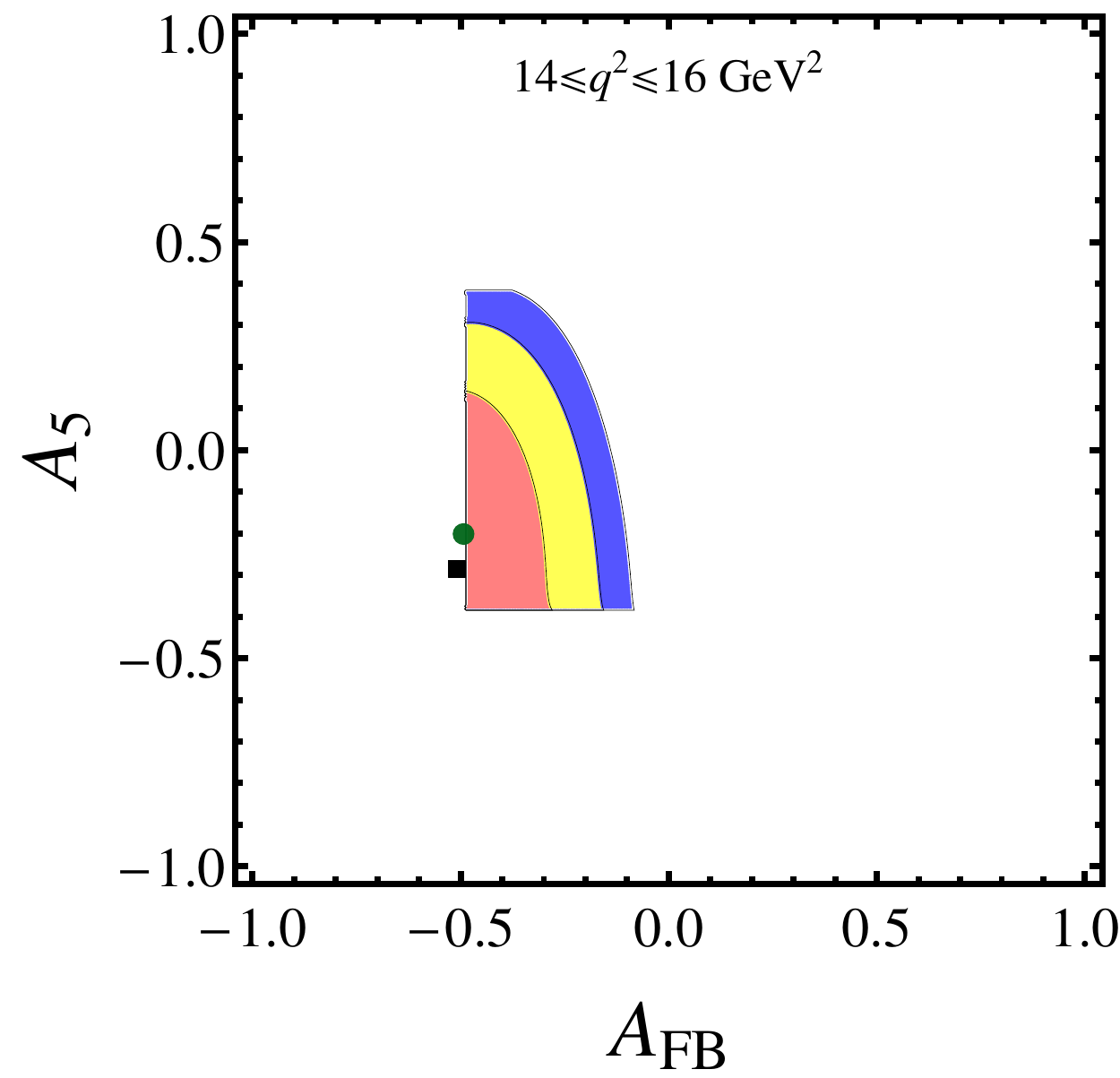}
	\includegraphics*[width=0.2\textwidth]{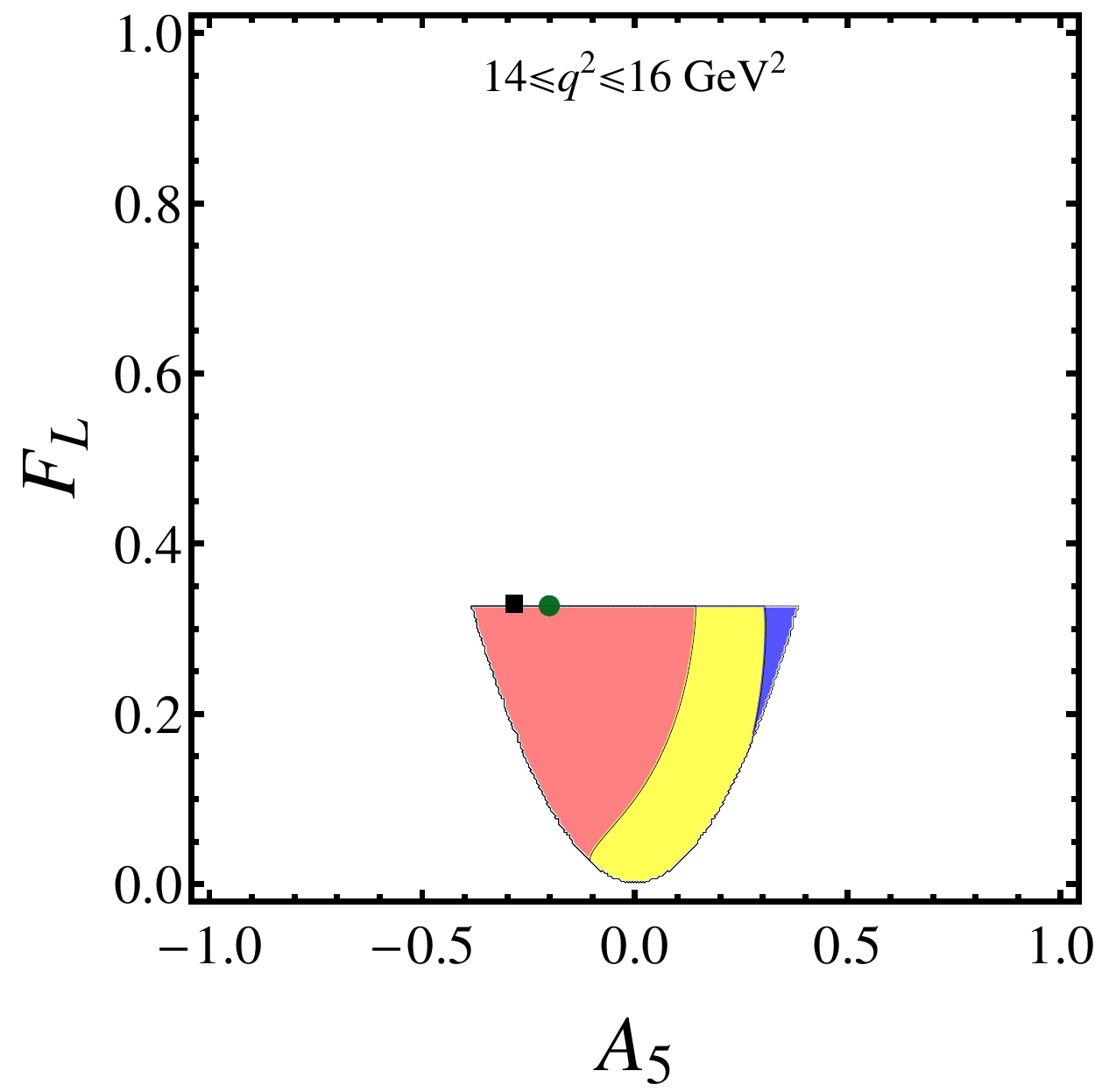}
	\includegraphics*[width=0.2\textwidth]{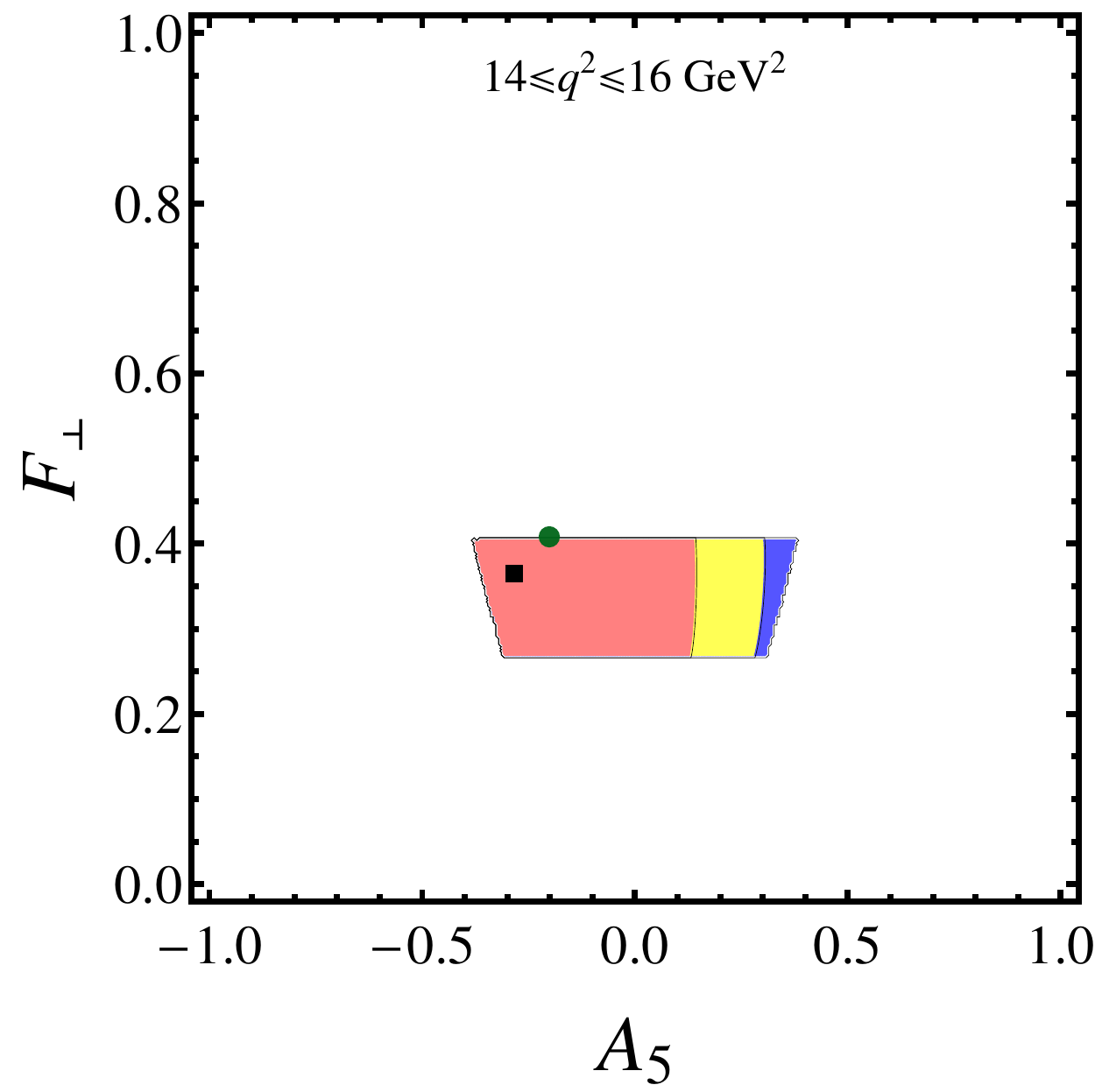}
	\includegraphics*[width=0.2\textwidth]{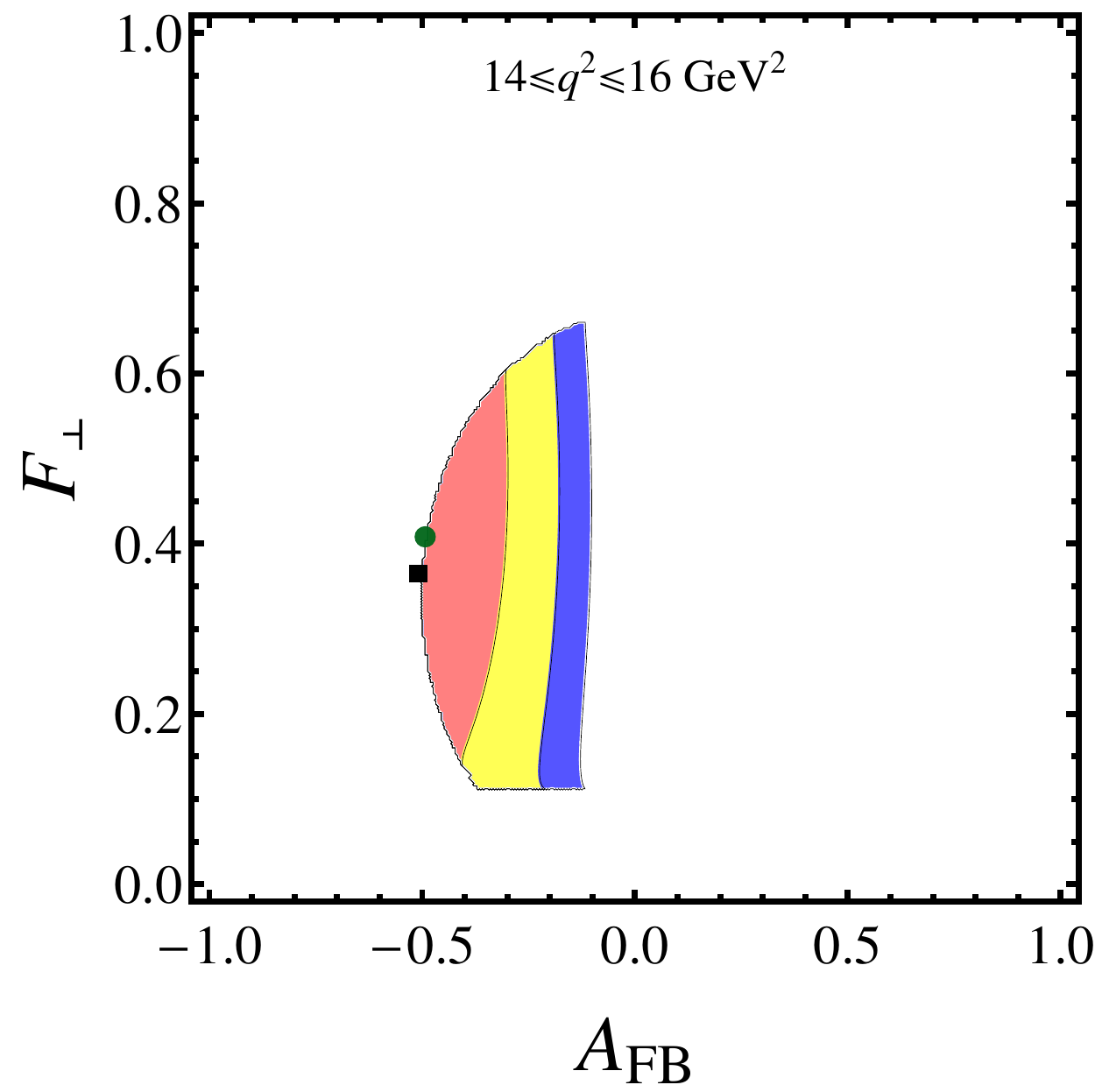}
	\includegraphics*[width=0.21\textwidth]{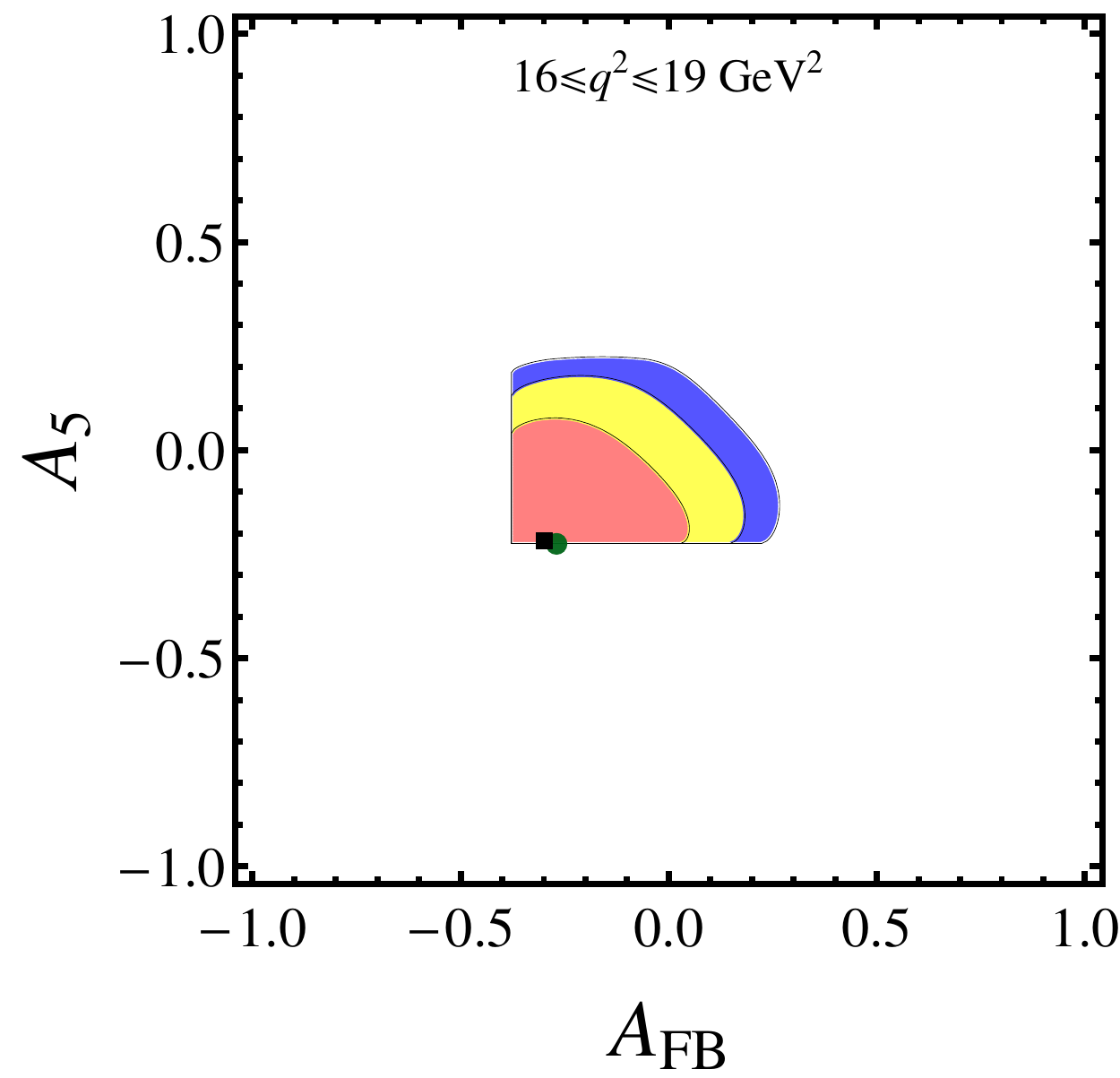}
	\includegraphics*[width=0.2\textwidth]{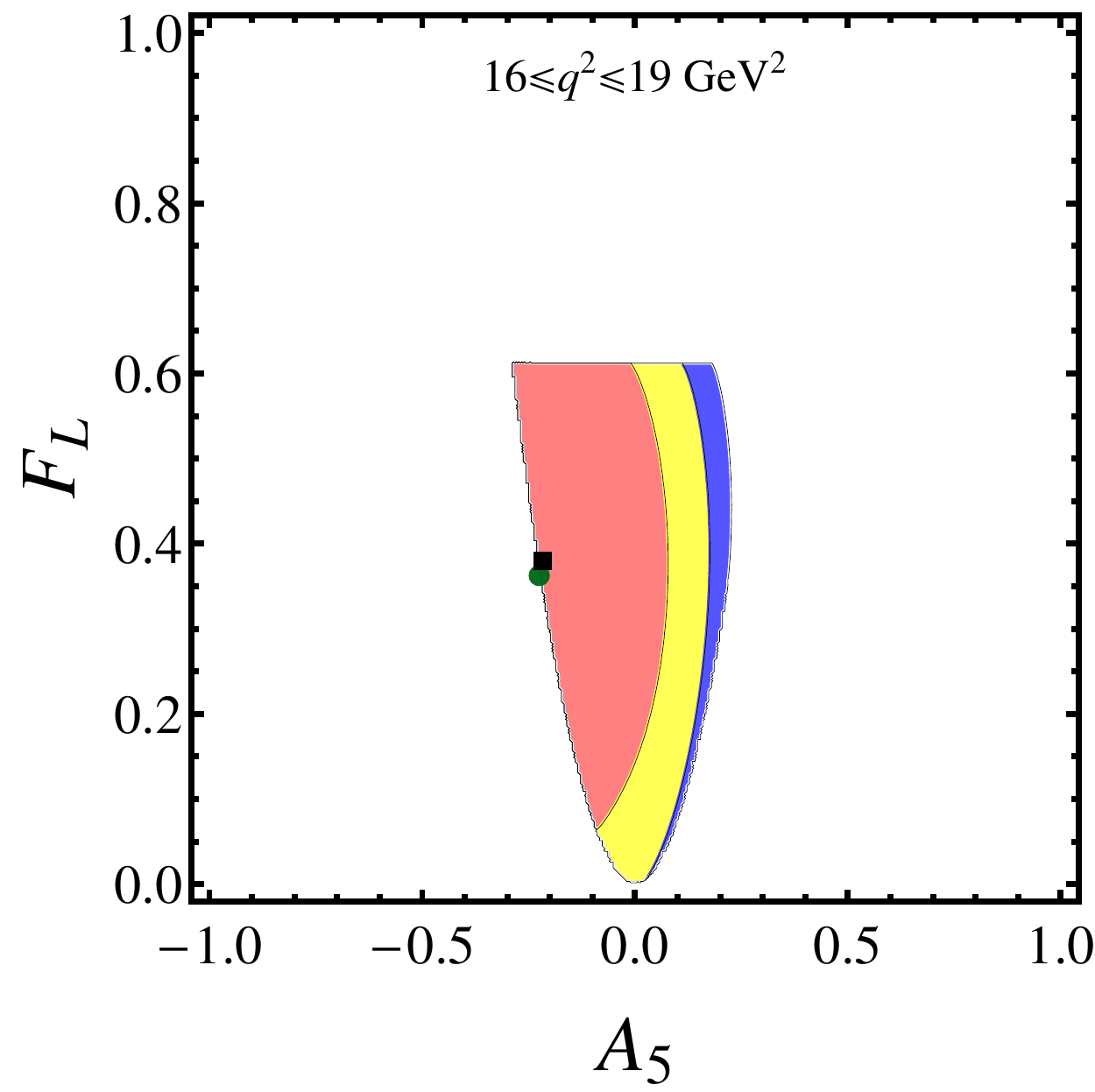}
	\includegraphics*[width=0.2\textwidth]{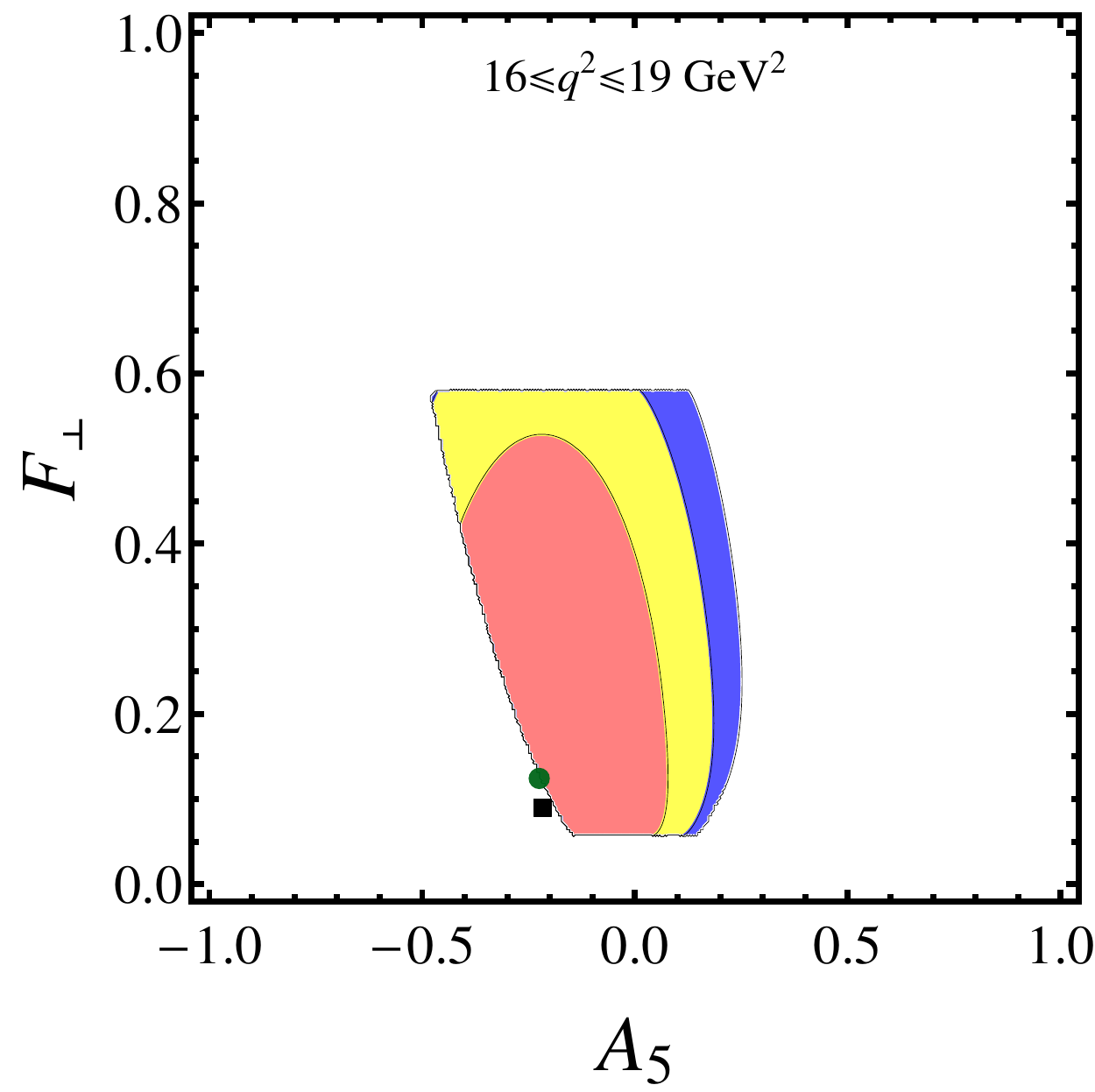}
	\includegraphics*[width=0.2\textwidth]{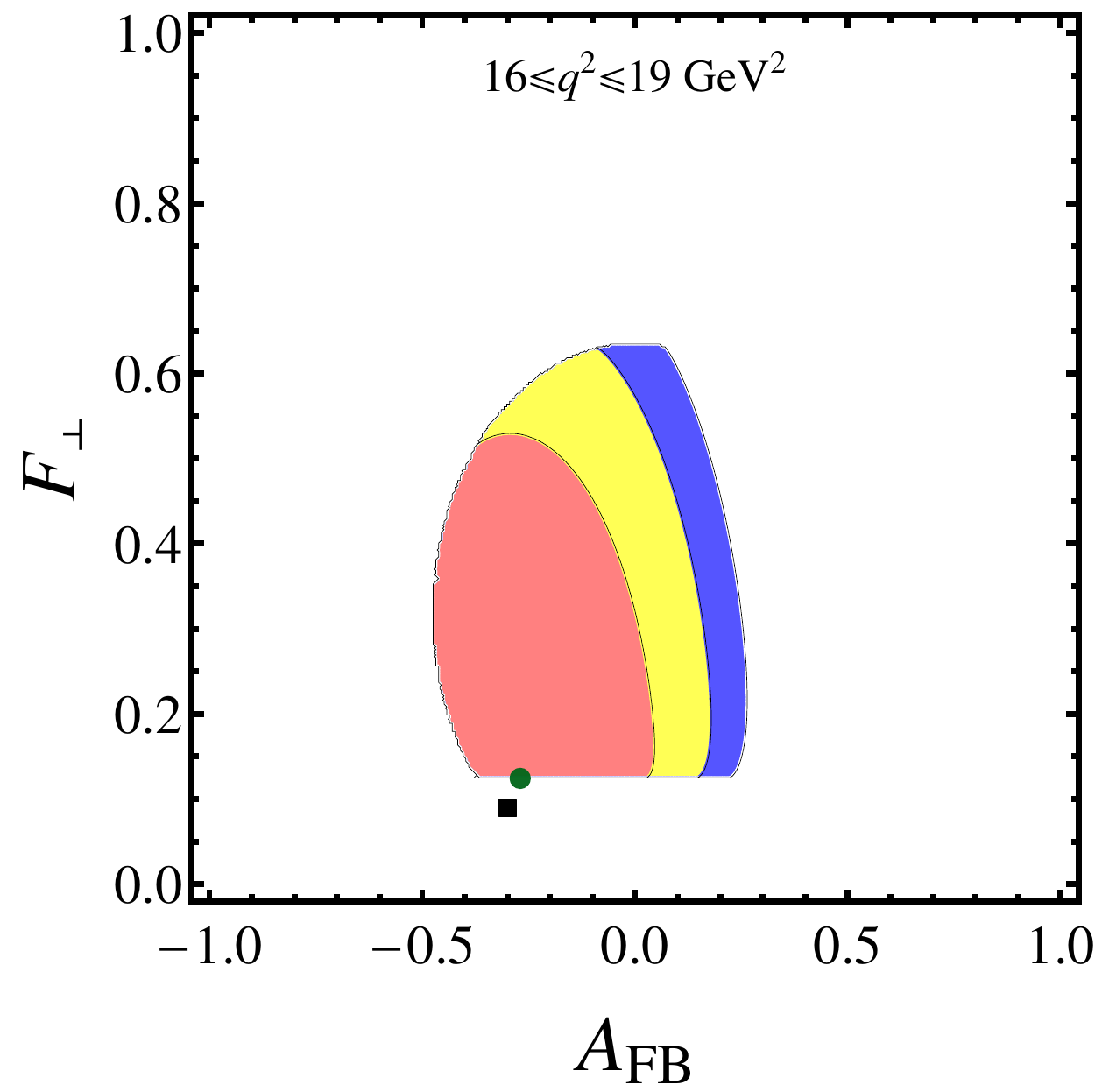}
	\caption{The $\chi^2$ projection onto the sets observables $A_5-\AFB$,
	$A_5-F_L$,  $A_5-F_\perp$ and $\AFB-F_\perp$ for various $q^2$ 
	bins going vertically from first to the sixth bin. The experimental values 
	of all the observables are taken from Refs.~\cite{Aaij:2013iag}. The color 
	codes are same as in Fig. \ref{fig:FLFP}.}
	\label{fig:A5AFB}
 \end{center}
\end{figure*}
\end{center}

\section{New Physics Analysis} 
\label{sec:NPAnalysis}

In this section, we demonstrate the possibility of how new physics could  be
tested using the relations derived in this paper. The basis of our analysis is
the relation, which involves all the nine observables $F_L$, $F_\|$, $F_\perp$,
$\AFB$, $A_4$, $A_5$, $A_7$, $A_8$, $A_9$ and a single form factor ratio
$\mathsf{P}_1$ derived in Eq.~(\ref{eq:Obs-relation2}). Since the helicity
fractions are related by $F_L+F_\|+F_\perp=1$, we eliminate $F_\|$. All the
observables have been measured by LHC$b$ collaboration using $1\invfb$ data.
However, currently the observables $A_7$, $A_8$ and $A_9$ are measured to be
consistent with zero. Eqs.~\eqref{eq:eps_perpo}--\eqref{eq:eps_0o} therefore
implies that $\varepsilon_\lambda$ are all consistent with zero.  In
Sec.~\ref{sec:massive_case} we have shown that the most conservative allowed
parameter space remains unaltered even if the small lepton mass term is dropped
compared to $q^2$ and the imaginary contributions to the amplitudes are
completely ignored. Since the inclusion of $\varepsilon_\lambda$ reduces the
parameter space of observables, in order to check the consistency of measured
observables we take a conservative approach and set all the
$\varepsilon_\lambda$'s to be equal to zero for the numerical analysis. Thus,
the relation among the observables reduces to Eq.~\eqref{eq:Obs-relation} which
is in terms of six observables $F_L$, $F_\|$, $F_\perp$, $\AFB$, $A_4$, $A_5$
and is completely free from any form factor dependence. If $A_7$, $A_8$ and
$A_9$ are measured to be non zero in future experiments with reduced
uncertainties, $\varepsilon_\lambda$ can be solved iteratively using
Eqs.~\eqref{eq:eps_perpo}--\eqref{eq:eps_0o} and an exact numerical analysis can
always be done. We, emphasize that a non-zero $\varepsilon_\lambda$ would only
restrict the allowed parameter space depicted in Figs.~\ref{fig:FLFP},
~\ref{fig:FLAFB} and \ref{fig:A5AFB} further as was already pointed out in
Sec.~\ref{sec:massive_case}. Later in this section we will, nevertheless, solve
for $\varepsilon_\lambda$ in terms of $A_7$, $A_8$ and $A_9$ since the predicted
value $A_4^{\text{pred}}$ depends on the values of $\varepsilon_\lambda$.

We use the SM relation  derived in Eq.~\eqref{eq:Obs-relation}, for
$\varepsilon_\lambda=0$ and $4m^2/q^2\to 0$  instead of 
Eq.~\eqref{eq:Obs-relationnew},
to check for consistency between measurements of all the observables. As noted
above a finite value for $\varepsilon_\lambda$ would provide a stronger
constraint and since $\varepsilon_\lambda$'s are consistent with zero,
Eq.~\eqref{eq:Obs-relation} provides a more conservative test.  In order to
preform the test we define a $\chi^2$ function
\begin{eqnarray}
\label{eq:chisq}
\chi^2&=&\frac{1}{4}\Bigg[\Bigg(\frac{A_4^\text{exp}-A_4^\text{pred}}{\Delta 
A_4^\text{exp}}\Bigg)^{\!2}
+\Bigg(\frac{F_L^\text{exp}-F_L}{\Delta F_L^\text{exp}}\Bigg)^{\!2}\nn\\&+&
\Bigg(\frac{F_\perp^\text{exp}-F_\perp}{\Delta F_\perp^\text{exp}}\Bigg)^{\!2}+
\Bigg(\frac{\AFB^\text{exp}-\AFB}{\Delta \AFB^\text{exp}}\Bigg)^{\!2}\nn\\&+&
\Bigg(\frac{A_5^\text{exp}-A_5}{\Delta A_5^\text{exp}}\Bigg)^{\!2\,}\Bigg],
\end{eqnarray}
where $A_4^\text{exp}, F_L^\text{exp}, F_\perp^\text{exp}, \AFB^\text{exp},
A_5^\text{exp}$ indicate experimental central values of the observables and
$\Delta A_4^\text{exp}, \Delta F_L^\text{exp}, \Delta F_\perp^\text{exp},
\Delta\AFB^\text{exp}, \Delta A_5^\text{exp}$ are the experimental
uncertainties. The statistical and systematic uncertainties are added in
quadrature for all the numerical analysis presented. We used
Mathematica~\cite{mathematica} to do all the numerical calculations presented in
this paper. The $\chi^2$ function in Eq.~\eqref{eq:chisq} is minimized in the
$4$-dimensional parameter space of the observables by varying each of them
simultaneously within the allowed region i.e $0\le F_L \le 1$, $0 \le F_\perp
\le 1$, $-1\le \AFB \le 1$, $-1 \le A_5\le 1$, while $A_4^\text{pred}$ is taken
to be the theoretically calculated value for $A_4$ using
Eq.~\eqref{eq:Obs-relation}. The minimized $\chi^2$ function is projected in
different sets of planes of the observables, $(F_L, F_\perp)$, $(\AFB, F_L)$,
$(\AFB,A_5)$ $(A_5, F_L)$, $(A_5, F_\perp)$ and $(\AFB, F_\perp)$ for the
contour plots. In Fig.~\ref{fig:FLFP} we show the allowed domain of
$F_L-F_\perp$ values for all the six $q^2$ bins corresponding to the $q^2$
values in the range $(0.1-2)~\text{GeV}^2$, $(2-4.34)~\gev^2$, $(4.34-8.68)~
\gev^2$, $(10.09-12.86)~\gev^2$, $(14.0-16.0)~\gev^2$ and $(16.0-19.0)~\gev^2$.
The pink, yellow and blue correspond to $1\sigma$, $2\sigma$ and $3\sigma$ %
confidence level regions. The black squares correspond to the experimentally
measured  central value and the green points correspond to best fit values
obtained  by minimizing $\chi^2$ using Eq.~\eqref{eq:chisq}. As can be seen form
Fig.~\ref{fig:FLFP} the bounds derived in this paper, involving only
observables, have resulted in very significantly constraining the allowed
parameters range of observables.

If it were true that there are no significant non-factorizable contributions to
the decay mode, rendering $\widetilde{C}_9^\hel{\lambda}$ independent of the helicity 
index `$\lambda$', we can solve for $\widetilde{C}_9$ as was shown in
Ref.\cite{Das:2012kz}. The ratio of $\widetilde{C}_9/\widehat{C}_{10}$ so
obtained could be inverted to solve for $\AFB$ resulting in the constraint
between $F_L$ and $F_\perp$ given in Eq.(55) of Ref.\cite{Das:2012kz}. The
narrow constraint region between the two solid black lines depicted in
$F_L-F_\perp$ plane in Fig.~\ref{fig:FLFP} is derived assuming real transversity
amplitudes, form-factors are calculated at leading order in
$\Lambda_{\text{QCD}}/m_b$ using HQET and the estimate that
$\widetilde{C}_9/\widehat{C}_{10}=-1$ is used. We emphasize that except for the
two solid black lines for each of the $q^2$ bins all other information in
Fig.~\ref{fig:FLFP} is completely free from any theoretical assumption. As can
be seen from Fig.~\ref{fig:FLFP} the best fit values as well as the
experimentally measured central values are largely not inside the narrow
constraint region within two solid black lines. This indicates that there could
exist any or all of the possibilities: imaginary contributions to the transversity
amplitudes or sizable non-factorizable contributions or higher order corrections
in HQET could also be relevant.

\begin{center}
\begin{figure*}[htbp]
 \begin{center}
	\includegraphics*[width=0.3\textwidth]{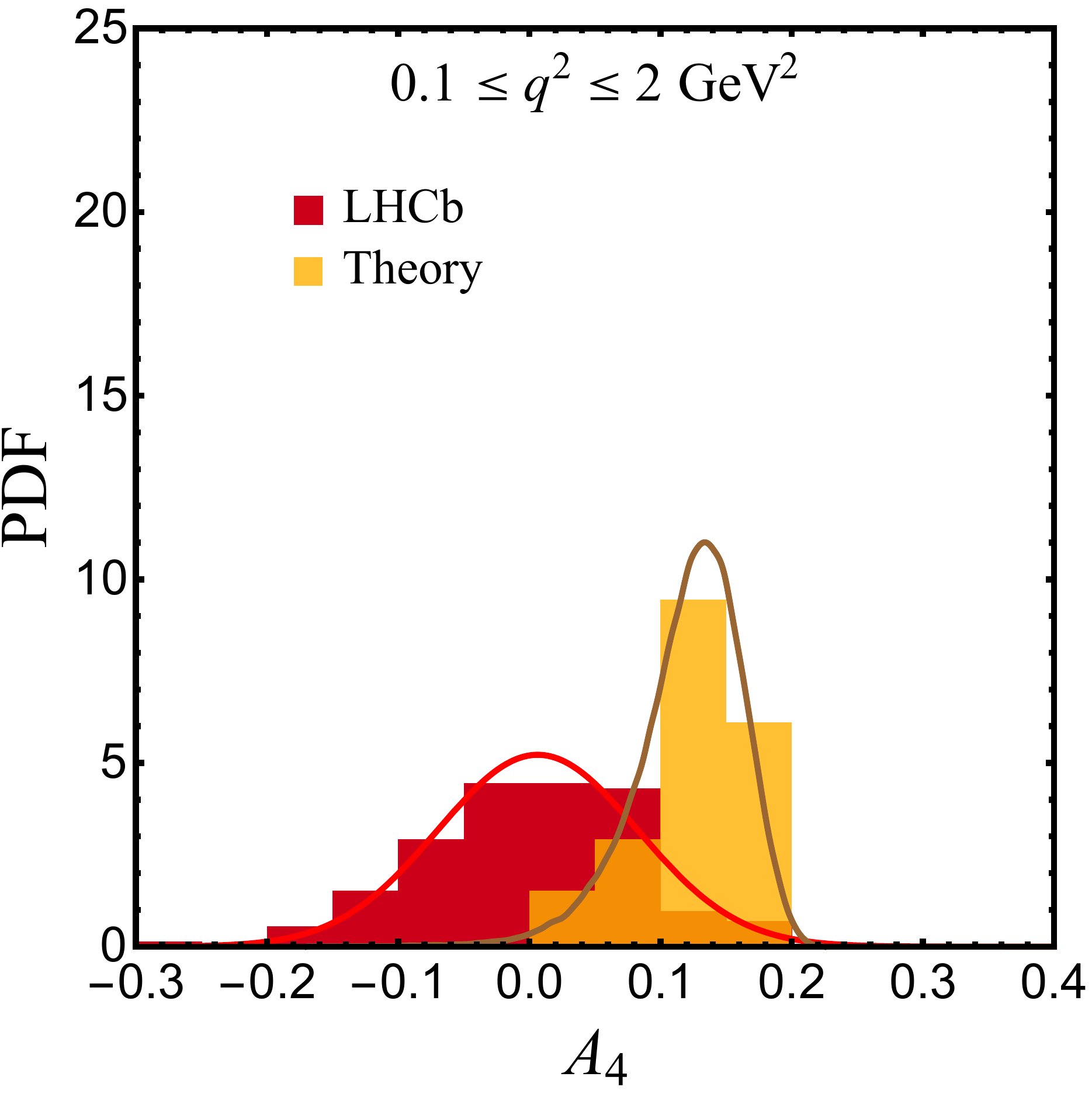}%
	 \includegraphics*[width=0.3\textwidth]{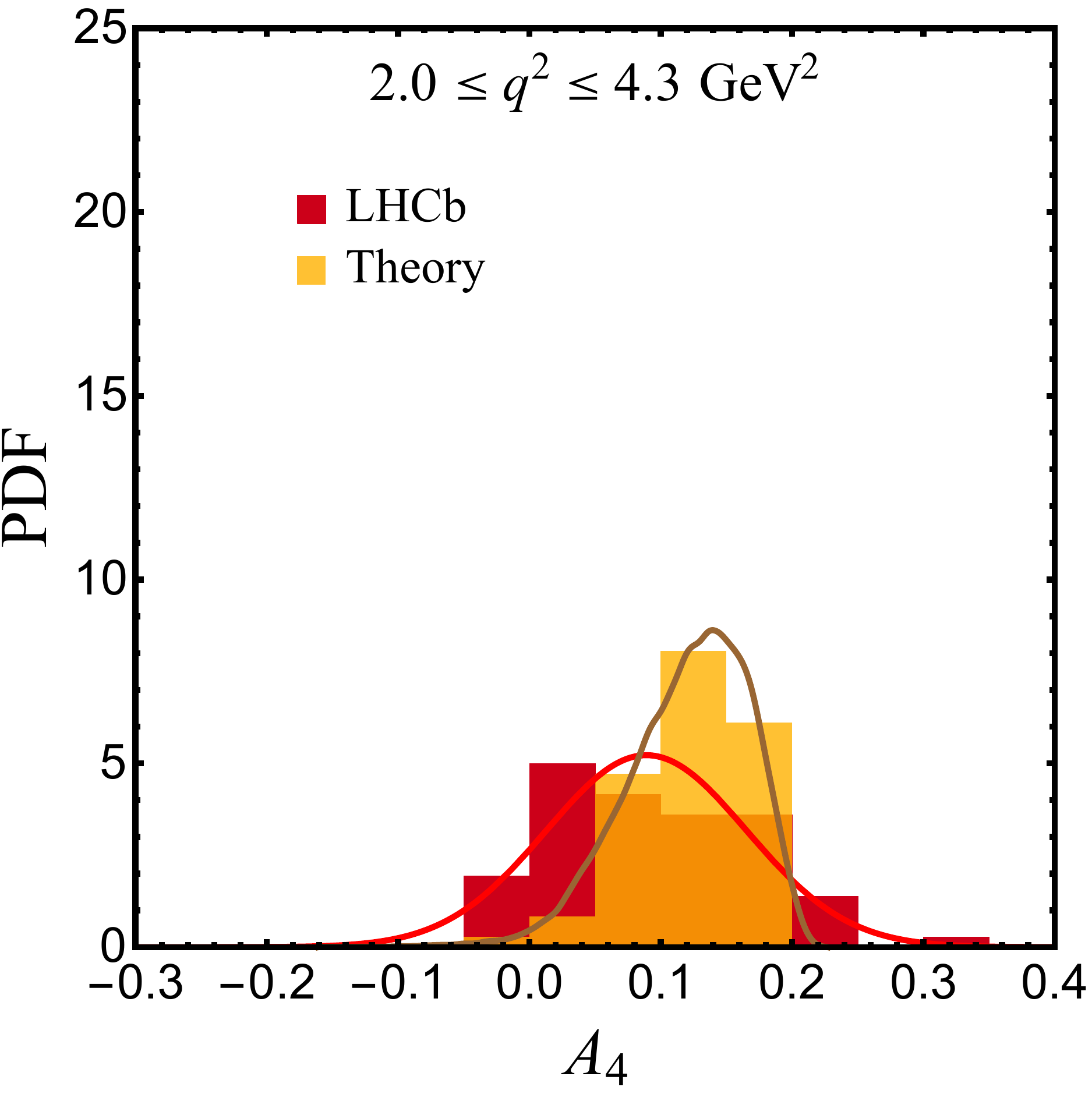}
	 \includegraphics*[width=0.3\textwidth]{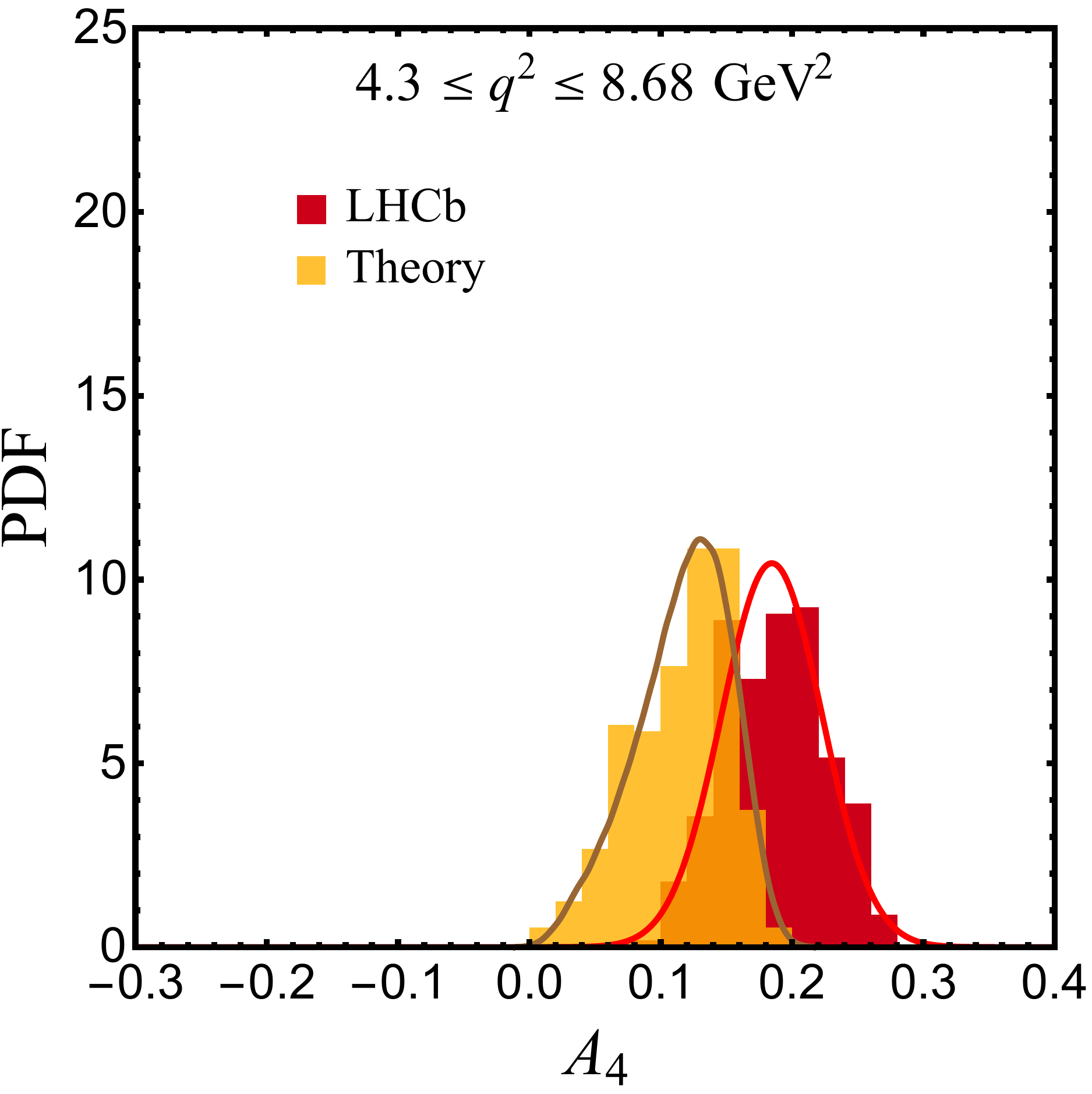}
	 \includegraphics*[width=0.3\textwidth]{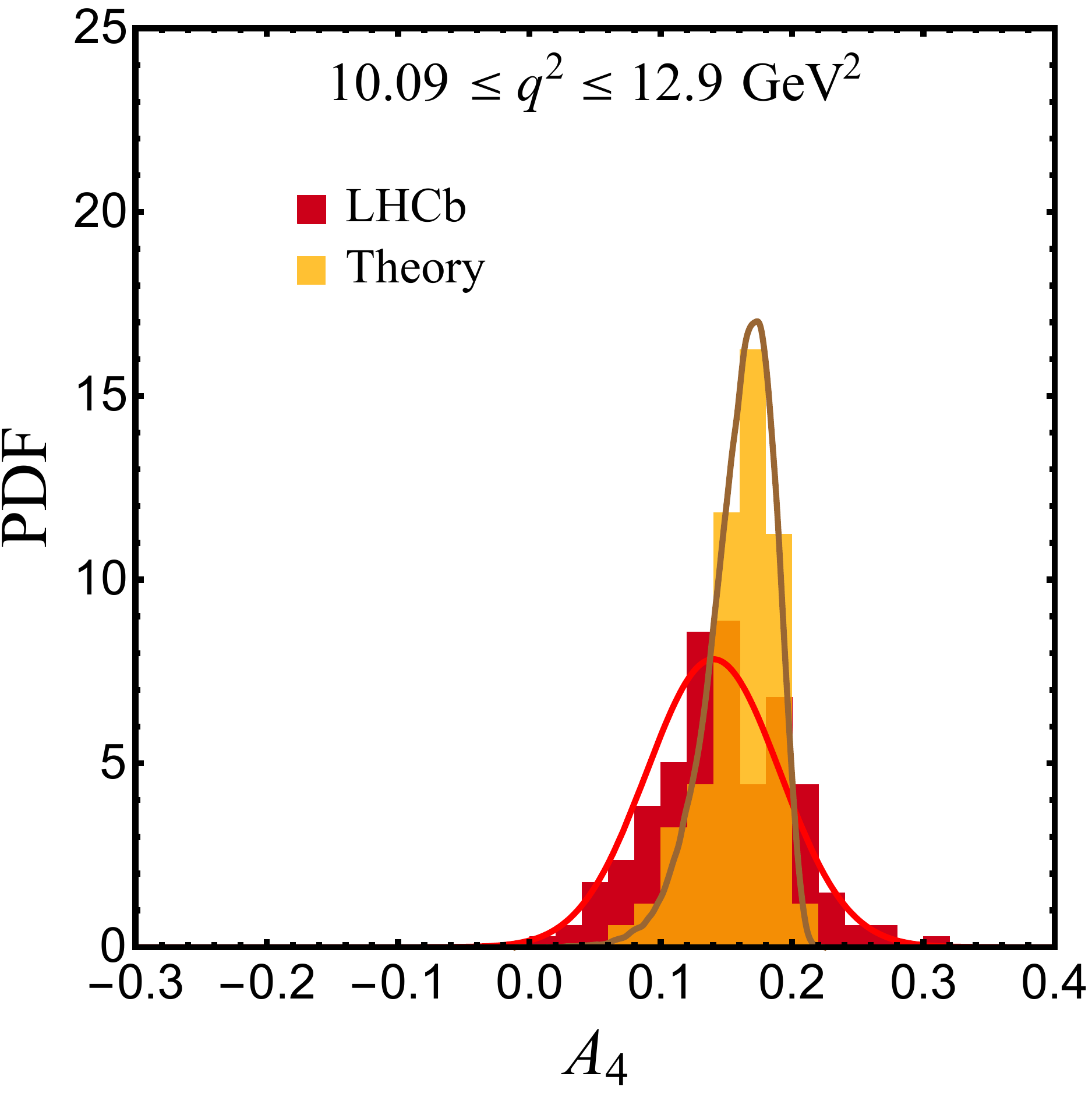}%
	 \includegraphics*[width=0.3\textwidth]{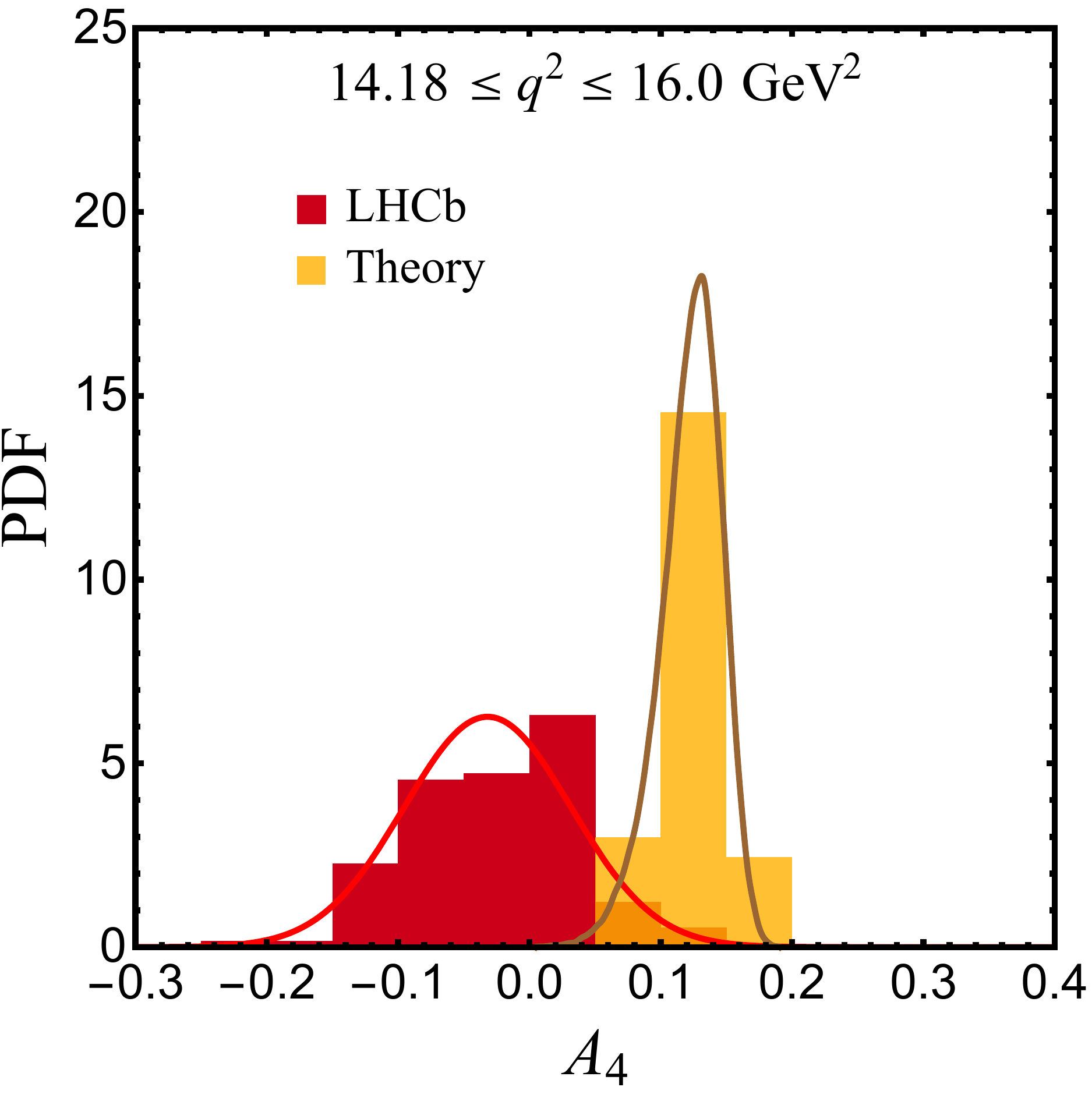}
	 \includegraphics*[width=0.3\textwidth]{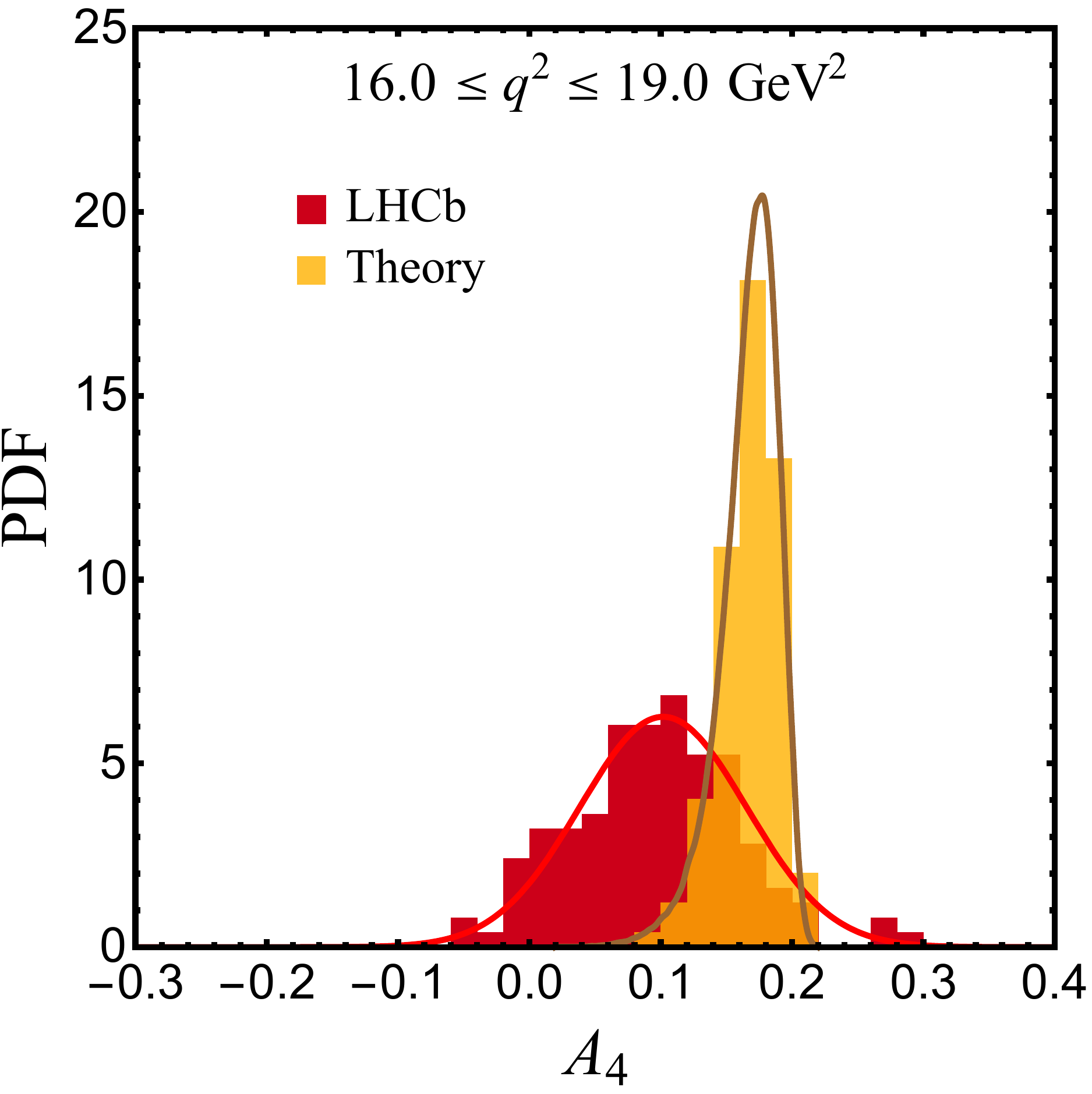}
	 \caption{A comparison of the measured and the predicted $A_4$ values for 
	 the six $q^2$ bins assuming that $A_7$, $A_8$ and $A_9$ are all zero. The 
	 simulated values of $A_4$ assuming Gaussian error in the LHC$b$ data are 
	 shown in red (dark), whereas the yellow (light) distributions referred to 
	 as ``Theory'' correspond to the values of $A_4^\text{pred}$ computed using 
	 Eq.~\eqref{eq:Obs-relation}. The plots correspond to a simulated theory 
	 (LHC$b$ $1\invfb$ data~\cite{Aaij:2013iag})  sample of  $144$ ($140 $), 
	 $76$ ($73$), $281$ ($271$), $169$ ($168$), $114$ ($115$) and $124$ ($116$) 
	 events corresponding to first through sixth $q^2$ bins as depicted in the 
	 figure. We have randomly chosen the events to be statistically 
	 consistent with the LHC$b$ observation in each bin for this decay 
	 mode. For a comparison, the probability distribution function (PDF) curves 
	 corresponding to $1000$ times more events are also shown for theory using 
	 brown (light) curve and data using red (dark) curve. We compare the two 
	 simulated distributions shown in the Histograms using the Mathematica 
	 routine ``DistributionFitTest''~\cite{fittest}.  The $P$-values obtained 
	 by comparing the two are found to be less than $10^{-9}$ 
	 for each of the bins, except the second and fourth bins, where the 
	 $P$-values obtained are $2.54\times 10^{-5}$ and $6.47\times 10^{-6}$ 
	 respectively.}
	\label{fig:A4large}
  \end{center}
\end{figure*}
\end{center}

The allowed range for observables $\AFB-F_L$ is depicted in Fig.~\ref{fig:FLAFB}
for all the six bins. The color code and markers follow the same convention used
in Fig.~\ref{fig:FLFP}. The constraint of the allowed triangular region between
two solid black line comes from Eq.(53) of Ref.~\cite{Das:2012kz}. Once again
the constraint region within the solid black triangular depicted in $\AFB-F_L$
plane is derived assuming real transersity amplitudes, form-factors calculated at
leading order in $\Lambda_{\text{QCD}}/m_b$ using HQET and the estimate that
$\widetilde{C}_9/\widehat{C}_{10}=-1$. However, note that the constraints
depicted by the contour plots are completely free from any theoretical
assumptions. The allowed region in the other four planes of observables  i.e
$\AFB-A_5$, $A_5-F_L$, $A_5-F_\perp$ and $\AFB-F_\perp$ are shown in
Fig.~\ref{fig:A5AFB}. We emphasize once again that the plots are free from any
theoretical uncertainty. In most of the contour plots depicted in
Figs.~\ref{fig:FLFP}, \ref{fig:FLAFB} and \ref{fig:A5AFB} the best fit points
(green point) lie at the edge of the boundaries except for the third bin. The
experimental measured central values (black squares) are mostly overlapping with
the best fit points except for fourth and sixth bin. In the fourth bin the black
squares stay outside the physically allowed region. In third bin both the best
fit and experimental measurement are very consistent with the allowed region and
sit almost at the center of it. It is interesting to note that the best fits are
always in the $1\sigma$ region perhaps validating the LHC$b$ data set.

\begin{center}
\begin{figure}[!htb]
 \begin{center}
	\includegraphics*[width=3.3in]{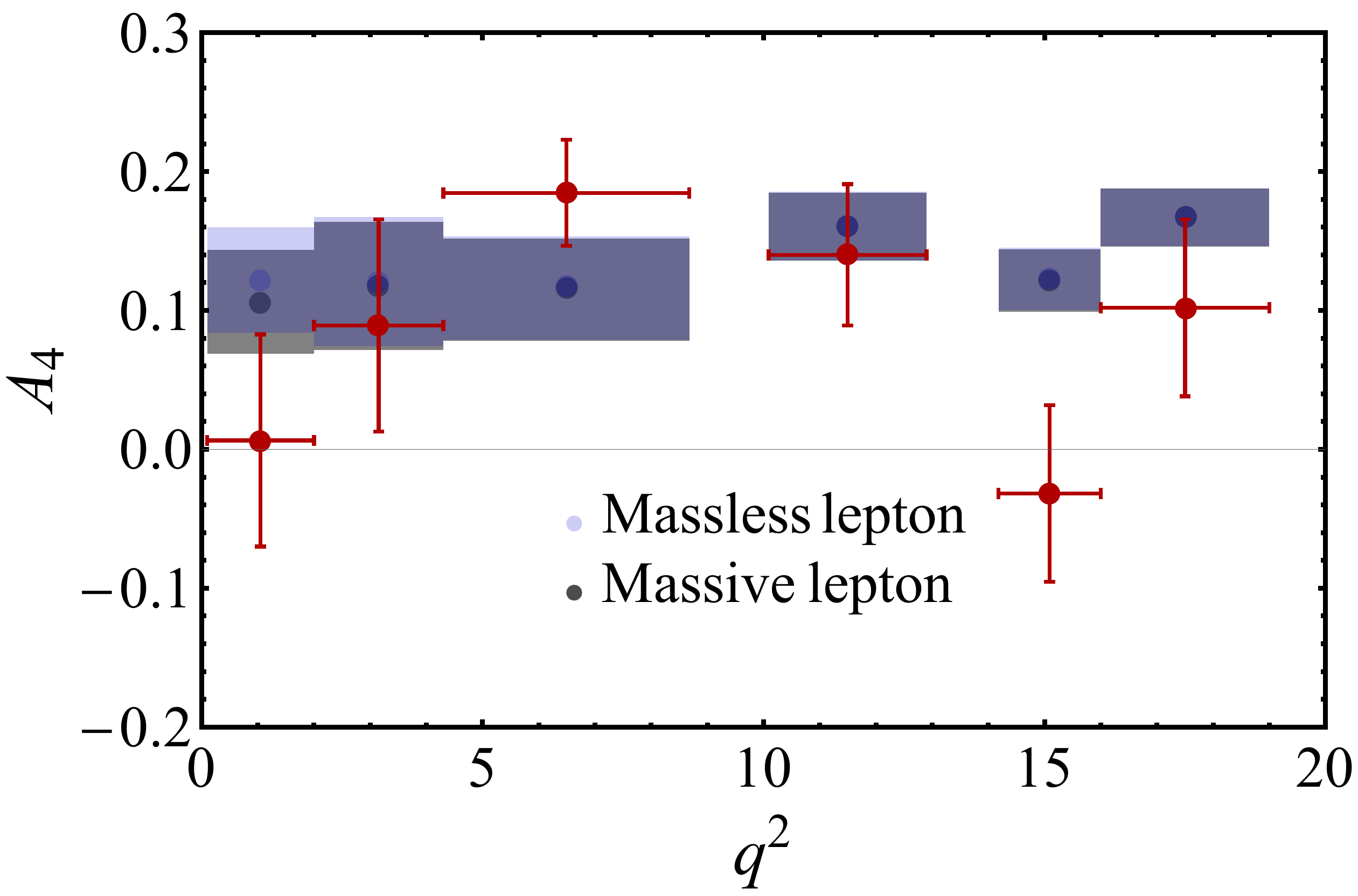}
	 \caption{The mean values and 1$\sigma$ regions for theoretically
	 calculated $A_4$ distributions excluding lepton masses 
	 (Eq.~\eqref{eq:Obs-relation}) and with massive leptons 
	 (Eq.~\eqref{eq:Obs-relationnew}) are shown in purple (light) and 
	 gray (dark) bands respectively. The simulated samples consist of $50,000$ 
	 events to start with, for each bin. The observables $A_7$, $A_8$ and $A_9$ 
	 are assumed to be zero. The error bars in red correspond to the 
	 experimentally measured \cite{Aaij:2013iag} central values and errors in 
	 $A_4$ for the respective $q^2$ bins.}
	\label{fig:comp}
  \end{center}
\end{figure}
\end{center}
\begin{center}
\begin{figure}[!bht]
 \begin{center}
	\includegraphics*[width=03.4in]{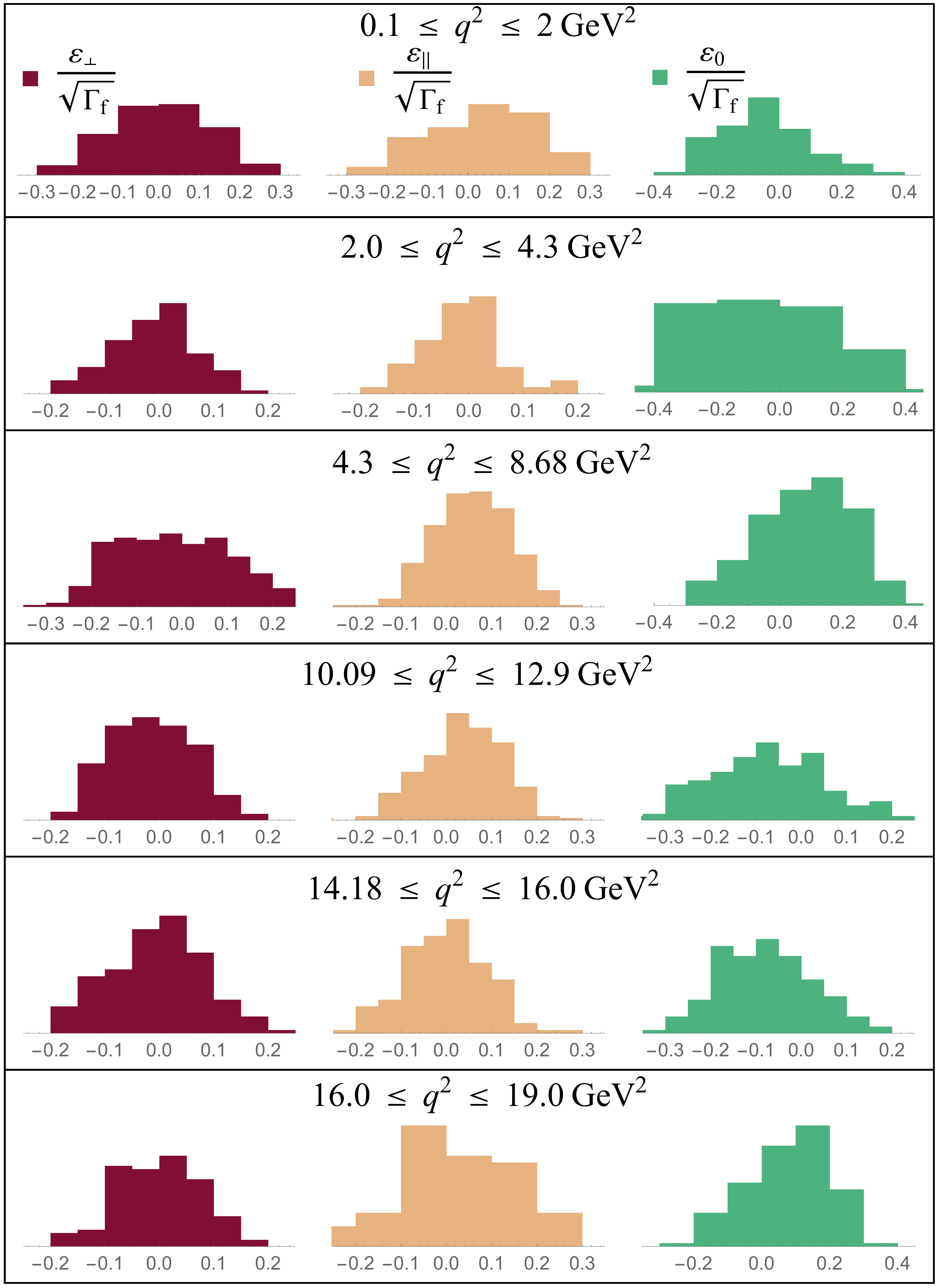}
	 \caption{The solutions for $\varepsilon_\perp/\sqrt{\Gf}$, 
	 $\varepsilon_\|/\sqrt{\Gf}$ and $\varepsilon_0/\sqrt{\Gf}$ using 
	 distributions 
	 with $140$, $78$, $275$, $175$, $113$ and $113$ 
	 events for first through sixth $q^2$ bins. The number of events 
	 are chosen to be statistically consistent with the number of events 
	 observed by LHC$b$~\cite{Aaij:2013iag} in each bin for this decay mode. 
	 All the $\varepsilon_\lambda$'s are consistent with zero and even at 
	 extreme cases 
	 $\varepsilon_\lambda^2/\Gf$ values are less than $0.2$.}
	\label{fig:eps}
  \end{center}
\end{figure}
\end{center}

\begin{center}
\begin{figure*}[hbtp!]
 \begin{center}
	\includegraphics*[width=0.3\textwidth]{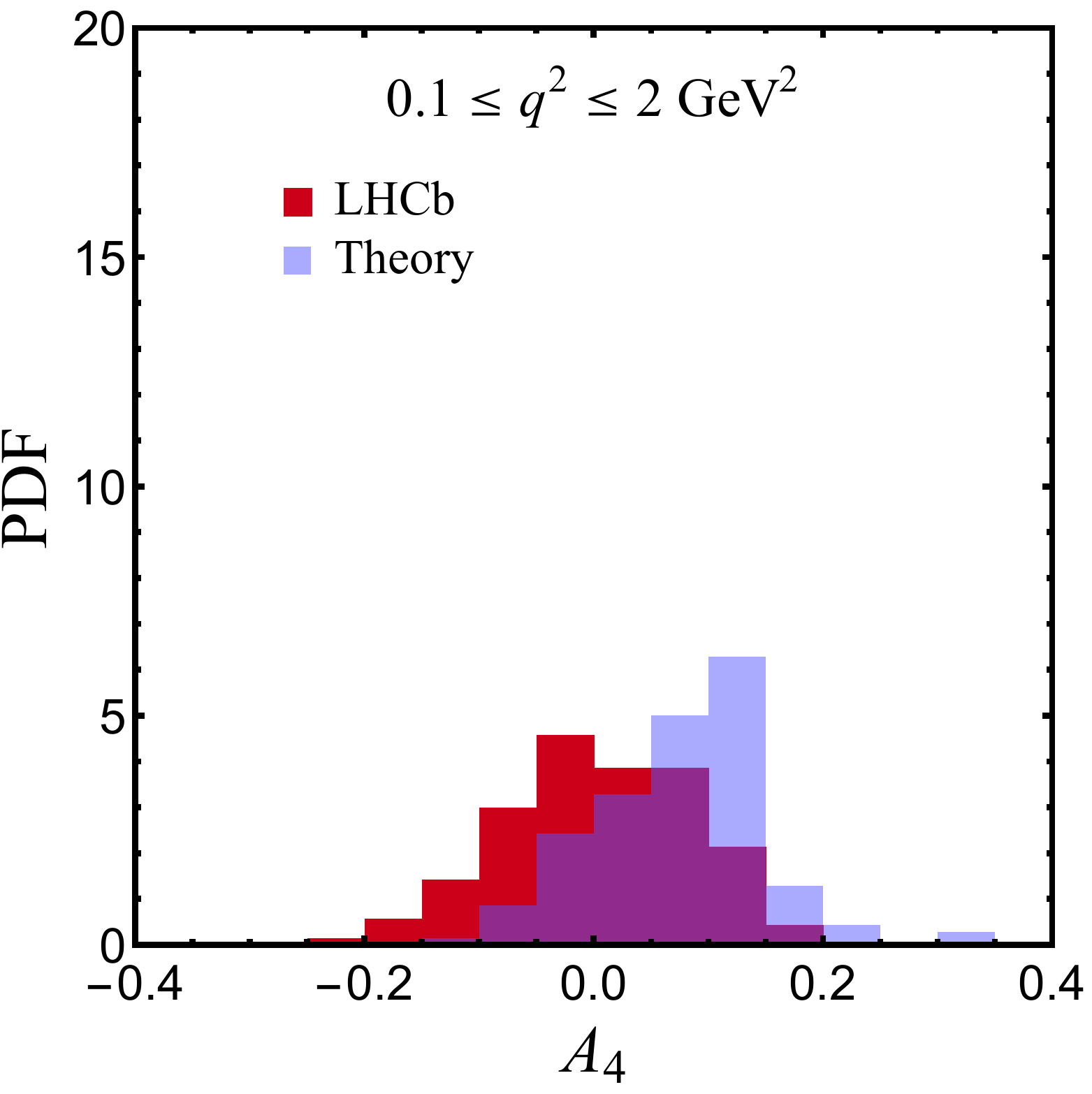}%
	 \includegraphics*[width=0.3\textwidth]{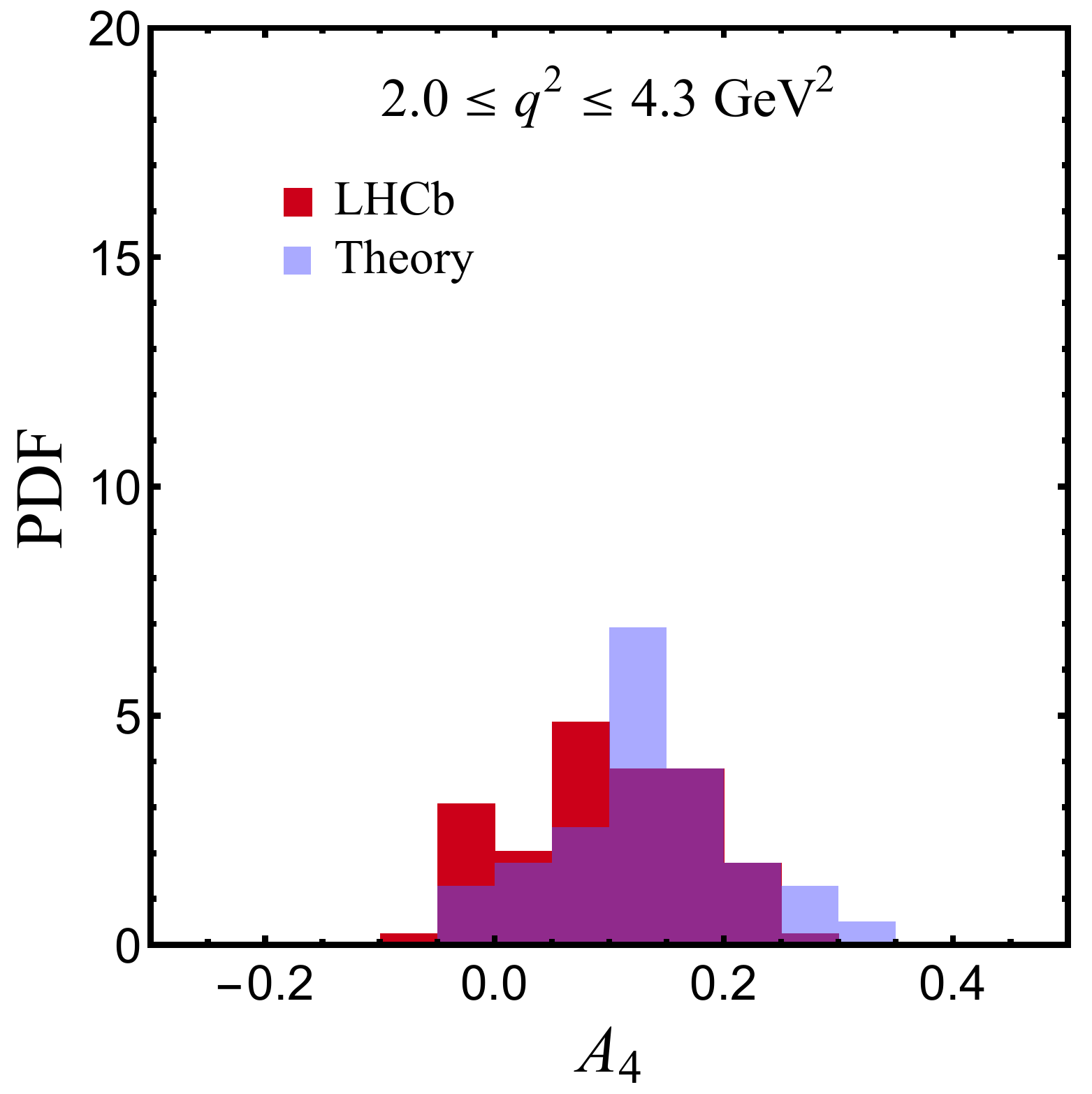}
	 \includegraphics*[width=0.3\textwidth]{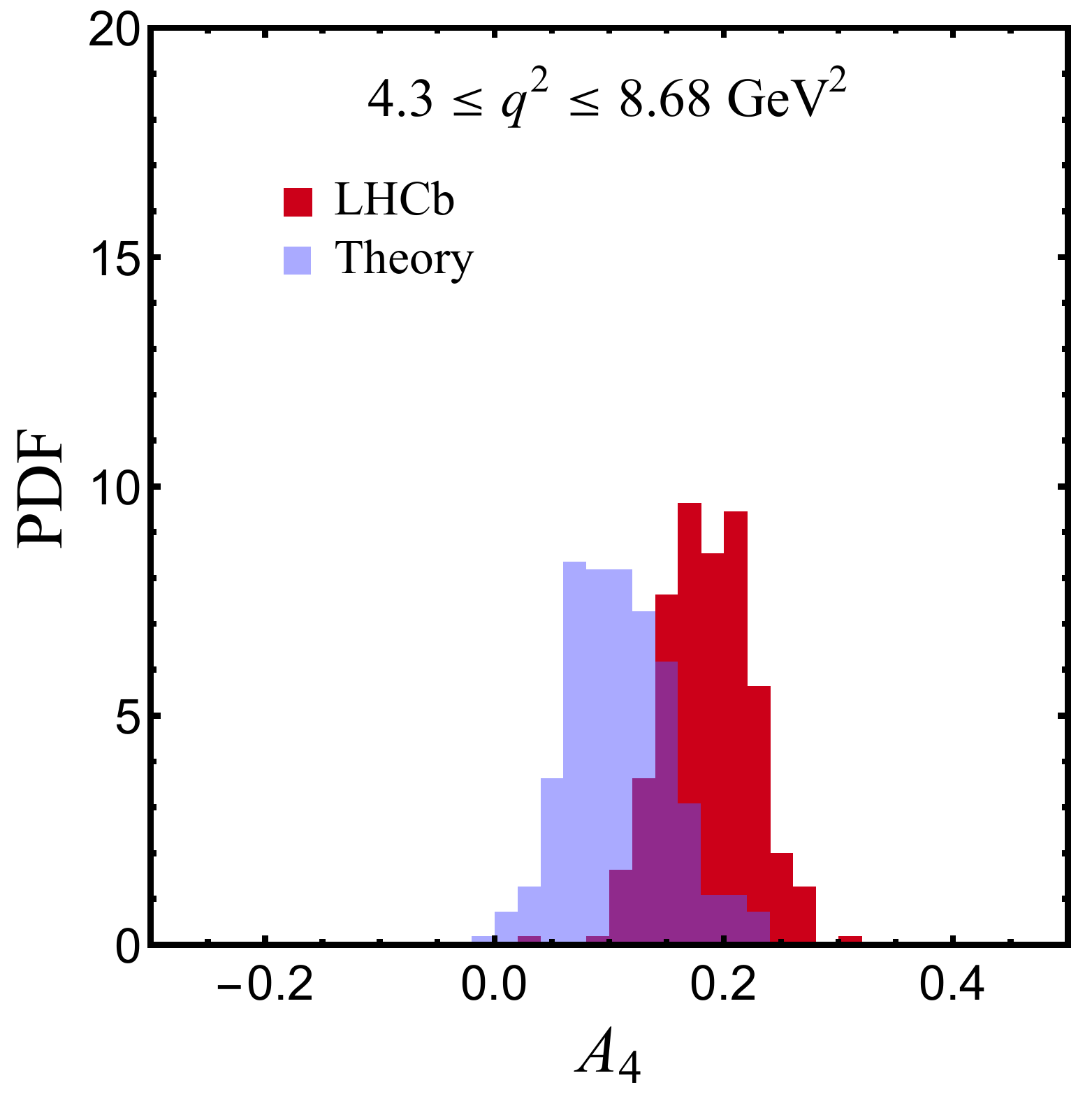}
	 \includegraphics*[width=0.3\textwidth]{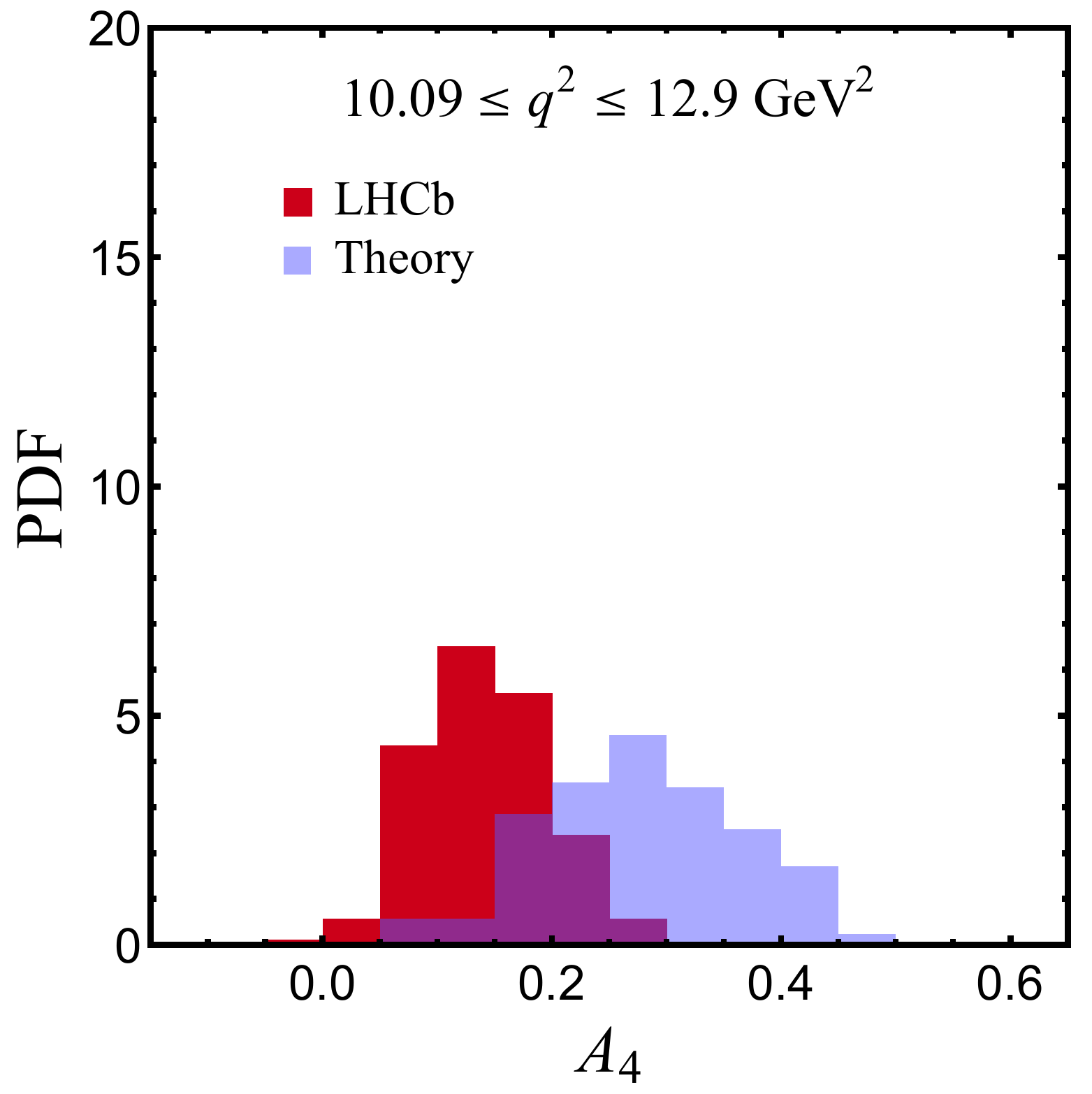}%
	 \includegraphics*[width=0.3\textwidth]{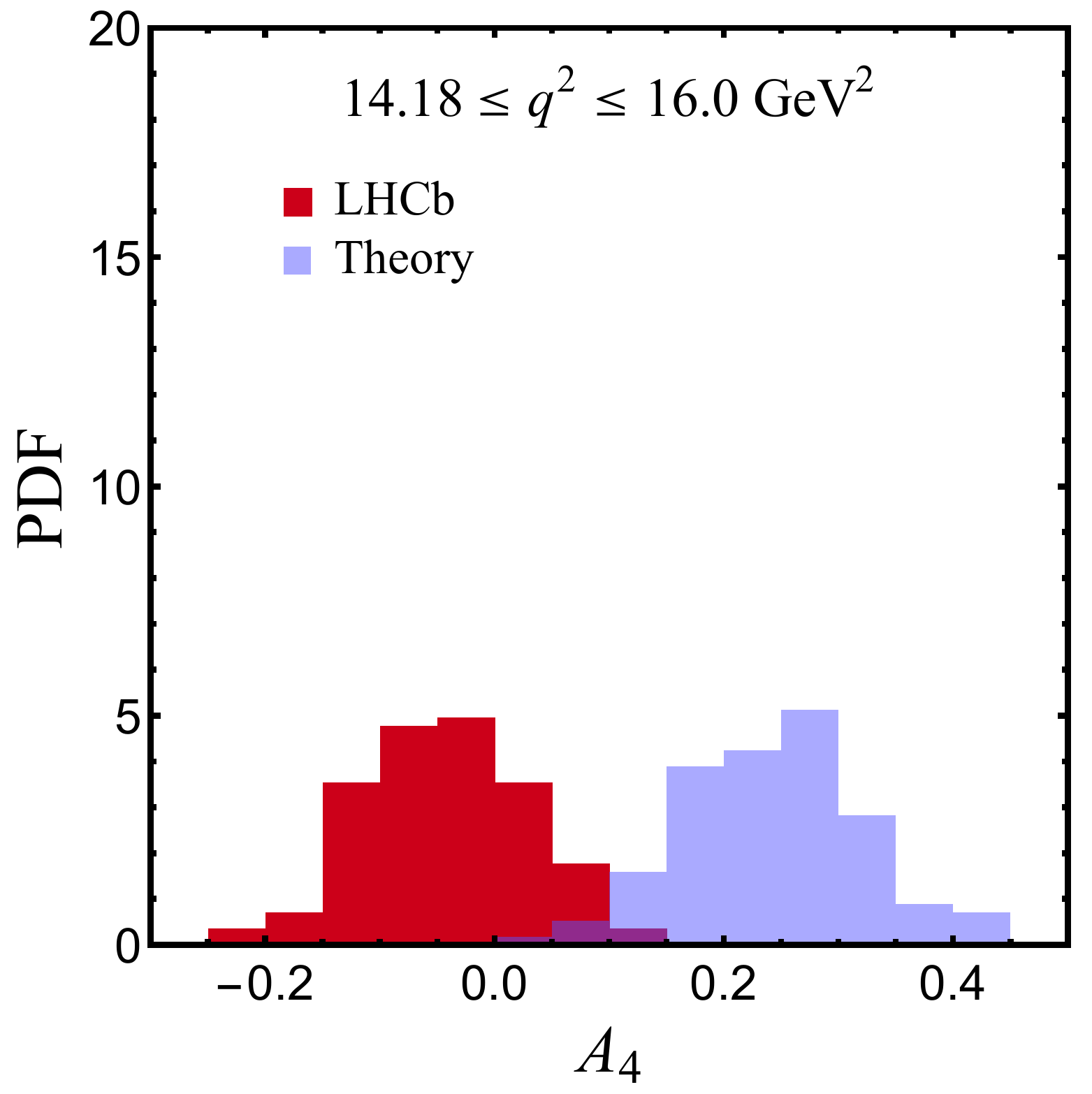}
	 \includegraphics*[width=0.3\textwidth]{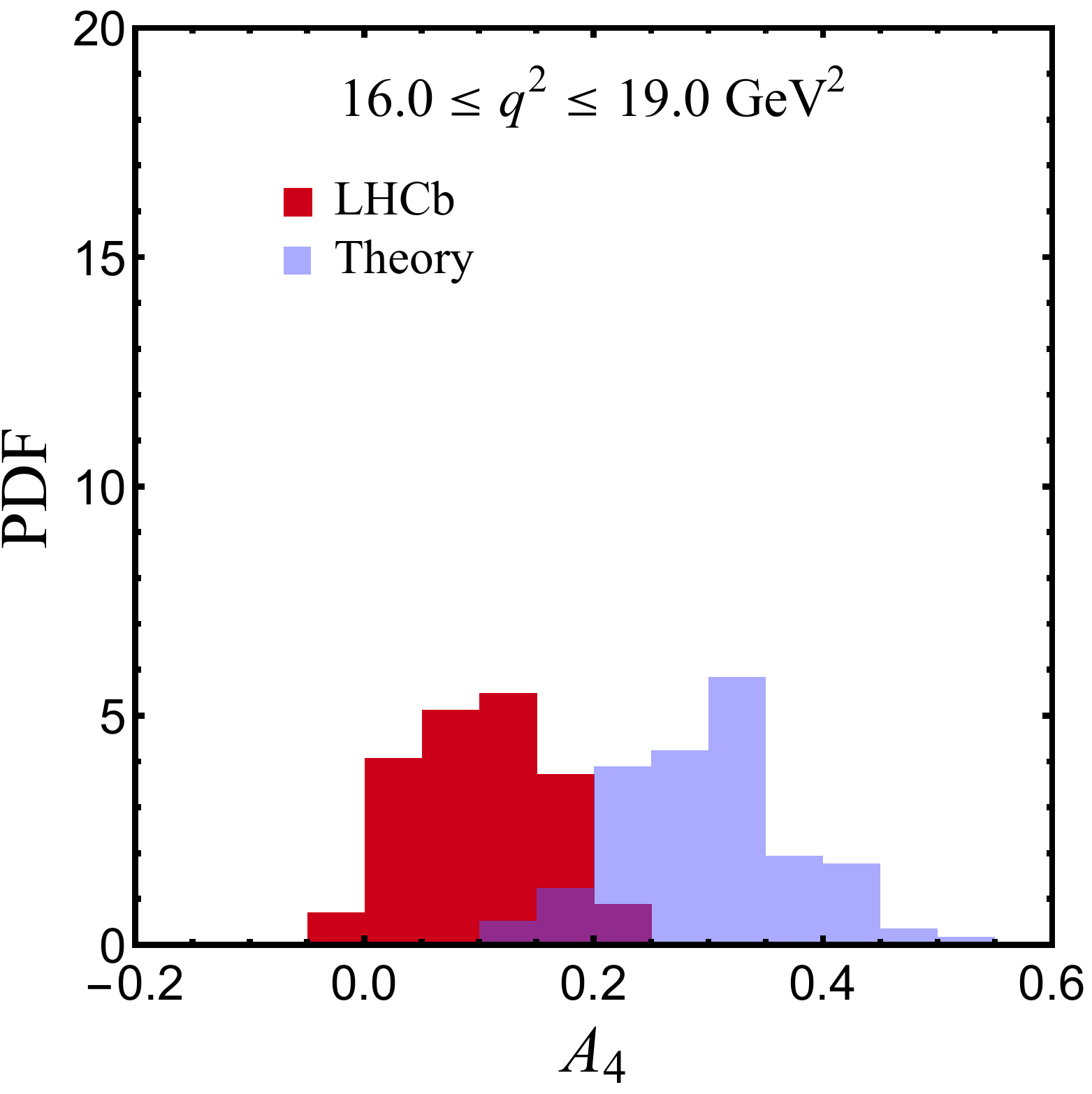}
	 \caption{A comparison of the measured and predicted $A_4$ values for 
	 the six $q^2$ bins considering all the measured observables. The simulated 
	 values of $A_4$ assuming Gaussian error in the LHC$b$ data are shown in 
	 red (dark), whereas the Blue (light) distributions referred to as 
	 ``Theory'' correspond to the values of $A_4^\text{pred}$ computed using 
	 Eq.~\eqref{eq:Obs-relationnew}. The plots correspond to a simulated theory
	 (LHC$b$ $1\invfb$ data~\cite{Aaij:2013iag}) sample of  $140$ ($140 $), 
	 $78$ ($73$), $275$ ($271$), $175$ ($168$), $113$ ($115$) and $113$ ($116$) 
	 events corresponding to first through sixth $q^2$ bins as depicted in the 
	 figure. The number of events are chosen to be consistent statistically 
	 with the number of events observed by LHC$b$ in each bin for this decay 
	 mode. The values of all other observables used in the two equations are 
	 randomly generated using LHC$b$ data assuming Gaussian measurements. We 
	 find that the $P$-values obtained using the Mathematica routine 
	 ``DistributionFitTest''~\cite{fittest} comparing the two distributions are 
	 always less than $10^{-9}$ for all bins except the second bin 
	 where the $P$-value is $6.78\times 10^{-3}$.}
	\label{fig:A4full}
  \end{center}
\end{figure*}
\end{center}

\begin{center}
\begin{figure}[htb]
 \begin{center}
	\includegraphics*[width=02.4in]{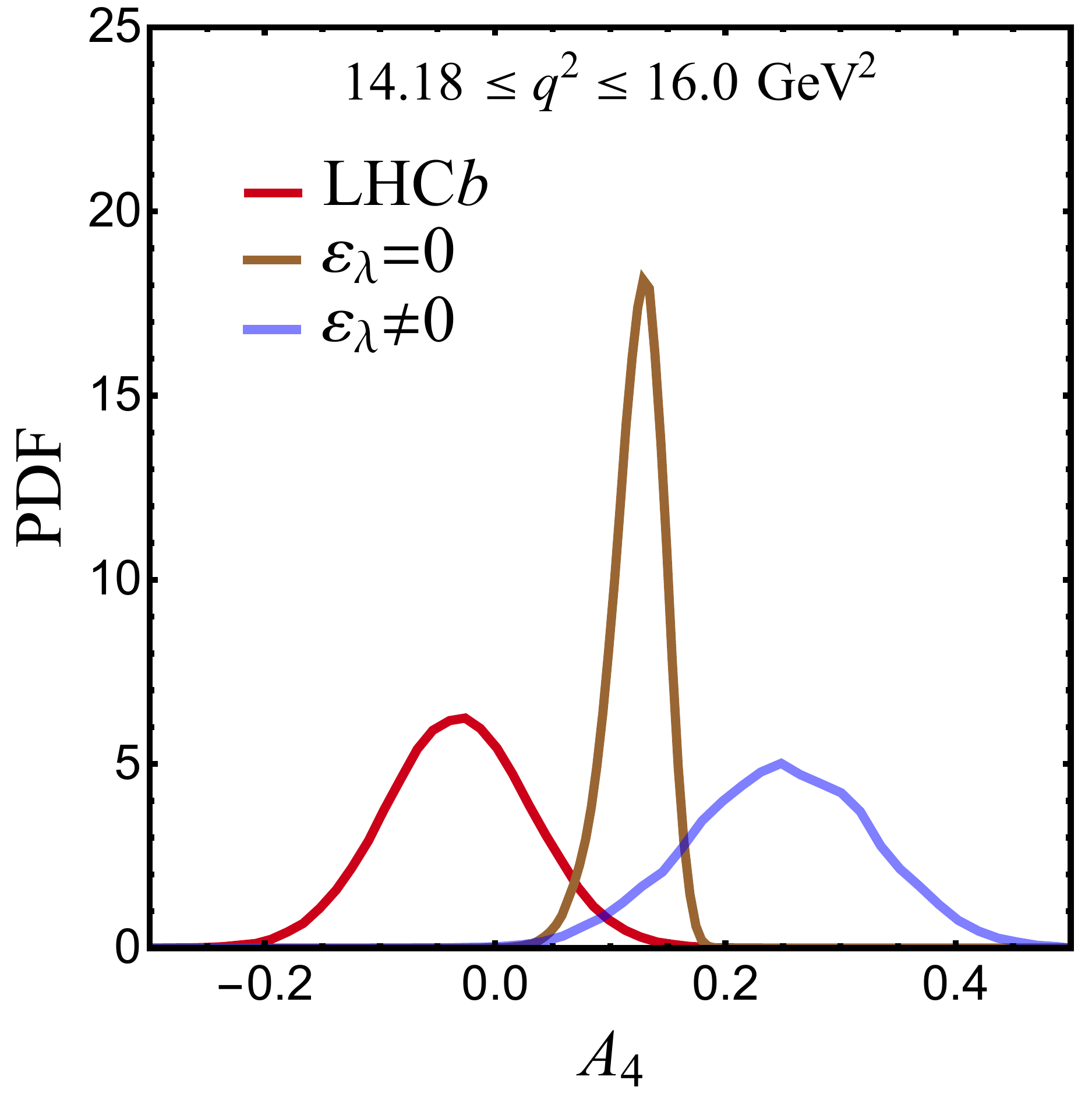}
	 \caption{A PDF plot comparing the measured fifth bin ($14.0\le q^2\le 
	 16.0~\gev^2$) value of $A_4$ with the two theoretically predicted values.
	 One assuming $\varepsilon_\lambda=0$ or completely real transversity 
	 amplitudes and the other with $\varepsilon_\lambda\ne0$ or complex 
	 transversity  amplitudes. The mean and errors for all the observables are 
	 assumed to be those measured by LHC$b$ using $1\invfb$ data set. All 
	 errors are assumed to be Gaussian. The PDF's depicted in the figure are 
	 generated using  $4\times 10^{5}$ random events resulting in the simulated 
	 values of $A_4$ for each curve. If $\varepsilon_\lambda\neq 0$ only $6708$ 
	 of the points survived the constraints. The plot corresponding to LHC$b$ 
	 $A_4$ measurement is shown in left most red (dark) plot, whereas the 
	 central brown (lighter) distribution corresponds to the theoretically 
	 calculated $A_4$ using Eq.~\eqref{eq:Obs-relation2} and the right most 
	 blue (light) distribution is for $A_4$ predicted using 
	 Eq.~\eqref{eq:Obs-relationnew}.}
	\label{fig:Three}
  \end{center}
\end{figure}
\end{center}

In Fig.~\ref{fig:A4large} the  measured Gaussian $A_4$ distribution is compared
with the distribution of $A_4^\text{pred}$ computed using
Eq.~\eqref{eq:Obs-relation}. In evaluating the right hand side of
Eq.~\eqref{eq:Obs-relation} we have used a Gaussian distribution of the
observables $F_L$, $F_\perp$, $A_5$ and $\AFB$ with experimental central value
as the mean and errors as the standard deviation from Ref.~\cite{Aaij:2013iag}.
The plots correspond to a simulated theory sample of  $144$, $76$, $281$, $169$,
$114$ and $124$ events corresponding to first through sixth $q^2$ bins. These
may be compared with $140$, $73$, $271$, $168$, $115$ and $116$ events obtained
for the respective bins by LHC$b$ using $1\invfb$ data~\cite{Aaij:2013iag}. We
have randomly chosen the number of events to be statistically consistent with
the LHC$b$ observation in each bin for this decay mode.  As should be expected
fewer events survive the constraint of Eq.~\eqref{eq:Obs-relation} when the best
fit points are at the edge of the permissible contour regions in
Figs.~\ref{fig:FLFP}, \ref{fig:FLAFB} and \ref{fig:A5AFB}. The simulated $A_4$
values corresponding to the LHC$b$ measurement for all six bins are shown in red
(dark) historgam and the yellow (light) histogram corresponds to the values of
$A_4^\text{pred}$ computed using Eq.~\eqref{eq:Obs-relation}. For a comparison,
the probability distribution function (PDF) curves  corresponding to $1000$
times more events are also shown for theory using brown (light) curve and data
using red (dark) curve.

The mean and 1$\sigma$ regions for the theoretically calculated
$A_4^\text{pred}$ distributions are shown in Fig.~\ref{fig:comp}. We compare
the two cases where lepton masses is ignored (Eq.~\eqref{eq:Obs-relation}) with
the case where lepton mass is finite (Eq.~\eqref{eq:Obs-relationnew}). The
purple (light) bands correspond to the massless case and the gray (dark) band
correspond to the massive case.  The error bars in red correspond to the
experimentally measured \cite{Aaij:2013iag} central values and errors in $A_4$
for the respective $q^2$ bins. The values of $A_4^\text{pred}$ obtained from the
Eq.~\eqref{eq:Obs-relation} seem to visually agree reasonably with the
experimental measurements  within the error bands in all the bins except for the
first and the fifth bin. A large discrepancy in fifth bin can also be seen here.
There is also a slight tension in first bin, which could be partly due to the
lepton mass may affect the first bin. The corrections due to mass terms can be
incorporated if the asymmetries $A_{10}$ and $A_{11}$ are measured in the
future. In the absence of such measurements we have used the theoretical
estimate of form-factors \cite{exclusiveB} to evaluate the effect of the finite
mass contribution. Details are depicted in Fig.~\ref{fig:comp}. The mass
contributions only effects the first bin, the other bins are unaffected. As
expected the agreement improves for the first bin if the mass contributions are
added. While Fig.~\ref{fig:comp} indicates only a mild disagreement between the
measured and predicted values of $A_4$, the distributions in
Fig.~\ref{fig:A4large} carry much more information than the mean and averages.
We have compared the two simulated distributions shown in the Histograms using
the Mathematica routine ``DistributionFitTest''~\cite{fittest}. The $P$-values
obtained by comparing the two are found to be less than $10^{-9}$ for each of
the bins, except the second and fourth bins, where the $P$-values obtained are
$2.54\times 10^{-5}$ and $6.47\times 10^{-6}$ respectively. A small $P$-value
indicates that one should reject the hypothesis that all observables are
consistent with the SM relation of Eq.~\eqref{eq:Obs-relation}.

In order to ascertain that the discrepancy in the $A_4$ enunciated using the
$P$-values is not due to the imaginary contributions being ignored we have also
preformed a simulation of all observables, including $A_7$ $A_8$ and $A_9$. We
solved for $\varepsilon_\perp$, $\varepsilon_{\|}$ and $\varepsilon_0$ using
Eqs.~\eqref{eq:eps_perpo} -- \eqref{eq:eps_0o}. These values of
$\varepsilon_\perp$, $\varepsilon_{\|}$ and $\varepsilon_0$ depend only on
observables and $\mathsf{P_1}$. We assume $\mathsf{P_1}$ values (see
Ref.~\cite{Das:2012kz}) to be $\mathsf{P_1} =-0.9395$, $-0.9286$, $-0.9034$,
$-0.8337$, $-0.7156$ and $-0.4719$ for the first through the sixth bin
respectively. We only remark that if $A_7$, $A_8$ and $A_9$ are measured to be
small the results are even more insensitive to the choice of the $\mathsf{P_1}$
value. Nevertheless, we also studied the effect of varying $\mathsf{P_1}$ within
the range  $\mathsf{P_1}\pm 0.5$, to ascertain our claim. Details will be
presented else where. The $\varepsilon_\lambda$ were solved iteratively and it
was found that they always converged in just a few iterations. If iteration led
to a value of $\varepsilon_\lambda$ larger than the derived constraints
permitted, a smaller allowed value was assigned and the iteration continued.  In
some cases an oscillatory or randomly varying pattern was observed but in these
cases the starting values of the observables could not be reproduced, indicating
that further constraints imposed by the chosen values of $A_7$, $A_8$ and $A_9$
could not be satisfied. The solutions obtained for
$\varepsilon_\lambda/\sqrt{\Gamma_f}$ are shown for each of the six bins in
Fig.~\ref{fig:eps}. It can be seen that all the $\varepsilon_\lambda$'s are
consistent with zero and even the tails of $\varepsilon_\lambda^2/\Gf$ do not
cross $0.2$. Having obtained the values of $\varepsilon_\lambda/\sqrt{\Gamma_f}$
we can now use the exact relation in Eq.~\eqref{eq:Obs-relationnew} to estimate
$A_4^{\text{pred}}$. A comparison between the measured $A_4$ and the predicted
value $A_4^{\text{pred}}$ including contributions from $A_7$, $A_8$ and $A_9$ is
done in Fig.~\ref{fig:A4full}. It must be emphasized that $A_4^{\text{pred}}$
obtained using Eq.~\eqref{eq:Obs-relationnew} is exact and takes into account
all the contributions in  SM. The asymmetries $A_{10}$ and $A_{11}$  (see
Eqs.~\eqref{eq:A10} and \eqref{eq:A11}) have not yet been measured and
Fig.~\ref{fig:comp} indicates that the lepton mass effects are negligible for
all but the first bin. We hence set $\mathbb{T}_1=0$ in evaluating
$A_4^{\text{pred}}$. This ensures that our results depend on only one
theoretical parameter, the ratio of form-factors $\mathsf{P_1}$ and that
parameter resulting in unmeasurable tiny effects do not complicate the
calculations. As predicted above, an even smaller number of events are now
consistent with the constraints derived in the paper. Interestingly,
$A_4^{\text{pred}}$ now fits better to a Gaussian distribution as indicated by a
Kolmogorov-Smirnov test, compared to the previous case where transversity
amplitudes were assumed to be real. This is indicative of the fact that the
transversity amplitudes are complex. However, the values of
$\varepsilon_\lambda/\sqrt{\Gamma_f}$ are not large as indicted in
Fig.~\ref{fig:eps}. We have simulated numbers of events consistent statistically
with the number of events observed by LHC$b$ in each bin. The plots as depicted
in Fig.~\ref{fig:A4full} correspond to a simulated theory (LHC$b$ $1\invfb$
data~\cite{Aaij:2013iag}) sample of  $140$ ($140 $), $78$ ($73$), $275$ ($271$),
$175$ ($168$), $113$ ($115$) and $113$ ($116$) events for the first through
sixth $q^2$ bins. The values of $A_4^{\text{pred}}$ predicted using
Eq.~\eqref{eq:Obs-relationnew} have a larger mean and variance as compared to
values obtained using Eq.~\eqref{eq:Obs-relation}. The $P$-values still continue
to be smaller than $10^{-9}$ for all the bins, except the second bin where the
$P$-value is $6.78\times 10^{-3}$, indicating that we reject the hypothesis that
all observables are consistent with the exact SM relation of
Eq.~\eqref{eq:Obs-relationnew}.

The PDF curves comparing the measured 
value of $A_4$ with both the theoretically predicted values assuming completely
real transversity amplitudes ($\varepsilon_\lambda=0$) and most general complex
transversity amplitudes ($\varepsilon_\lambda\ne0$) are shown in 
Fig.~\ref{fig:Three} for fifth bin ($14.0\le q^2\le 16.0~\gev^2$).
We have chosen the fifth bin for this detailed study since the tension between the
predicted value and experimentally observed value appears to be the largest as
can be seen from Figs.~\ref{fig:A4large}, \ref{fig:comp} and \ref{fig:A4full}.
The PDF's depicted in the figure are generated using $4\times 10^{5}$ random
events resulting in the simulated values of $A_4$ for each curve. If
$\varepsilon_\lambda\neq 0$ only $6708$ of the points survived the constraints
of Eq.~\eqref{eq:Obs-relationnew}.  LHC$b$ data assuming Gaussian error is shown
in left most red (dark) plot, whereas the central brown (lighter) distribution
corresponds to the theoretically calculated $A_4$ using
Eq.~\eqref{eq:Obs-relation2} and the right most blue (light) distribution is for
$A_4$ predicted using Eq.~\eqref{eq:Obs-relationnew}. The values of all other
observables used in the two equations are randomly generated assuming Gaussian
measurements of the LHC$b$ $1\invfb$ data.

In this section we have discussed the constraints already imposed by the
$1\invfb$ LHC$b$ data~\cite{Aaij:2013iag} on the parameter space of observables.
We also compare the measured values of $A_4$ with those predicted using the new
relations derived in this paper. We made several observations that indicate
possibly sizable non-factorizable contributions and imaginary contribution and
also possible higher order corrections in HQET to the transversity amplitudes. In
addition, the $P$-values comparing the measured $A_4$ with the predicted value
indicates new physics. However, we refrain from drawing even the obvious
conclusions given that, results for $3\invfb$ data will soon be presented by the
LHC$b$ collaboration. However, we emphasize that the approach developed in this
paper could not only conclusively indicate presence of significant
non-factorizable contributions and need for higher order power corrections to
form-factors but also the presence of NP with larger statistics.

\section{Conclusions}
\label{sec:conclusions}

In this paper we have derived a new relation involving all the $CP$ conserving
observables that can be measured in the decay $B\to \kstar \ell^+\ell^-$ using
an angular study of the final state for the decay. The relation provides a very
clean and sensitive way to test SM and search for NP by probing consistency
between the measured observables. The relation reduces to the one derived in
Ref.~\cite{Das:2012kz}, when certain reasonable assumptions were made. Since,
the relation is intended to be used as probe in search for NP, it is imperative
that no avoidable assumptions be made. We have generalized previous results with
this objective in mind. The new derivation is parametrically {\em exact in the
SM limit} and incorporates finite lepton and quark masses, complex amplitudes
enabling resonance contributions to be included, electromagnetic correction to
hadronic operators at all orders and all factorizable and non-factorizable
contributions to the decay.

We write the most general form factors and amplitudes in
Sec.~\ref{sec:framework} based only on  Lorentz invariance and gauge invariance.
Our approach differs from what usually done in literature as we make no attempt
to evaluate hadronic parameters but eliminate them in favour of measured
observables to the extent possible. Hence, our conclusions are not limited in
general by the order of accuracy up to  which the calculations are done.

The decay is described by six transversity amplitudes which survive in the
massless lepton case. If the mass of the lepton is finite yet another amplitude
contributes to the decay. We have shown in Sec.~\ref{sec:massive_case} that the
corrections to the amplitude arising from finite lepton mass can be determined
completely from observables measured using angular analysis. These contributions
are suppressed by $m^2/q^2$ and may be difficult to measure. A theoretical
estimate also shows that they are insignificant in all but the first bin. We
therefore began by focusing attention on the massless case which is described by
the six transversity amplitudes alone. The massive lepton case was considered
later to derive an exact relation valid in the SM limit. Even if the mass
effects are too tiny to distinguish an attempt to measure them would ensure that
the predictions are reliable and free from theoretical parameters.

We started by writing the most general parametric form of the transversity
amplitude in the SM given in Eq.~\eqref{eq:amp-def1} that takes into account
comprehensively all the contributions within SM. Unlike the derivations in
Ref.~\cite{Das:2012kz} the general transversity  amplitude is now allowed to be
complex, by introducing three additional parameters $\varepsilon_\lambda$. This,
however, poses no problem since there are three extra observables $A_7$, $A_8$
and $A_9$ given in Sec.~\ref{sec:angular}, which are non-vanishing in the
complex transversity amplitudes limit. Hence, dealing with complex amplitude
introduces only a technical difficulty of solving for additional variables
iteratively.

Using this general amplitude a new relation  (see Eq.~\eqref{eq:Obs-relation})
involving all the nine $CP$ conserving observables is derived in
Sec.~\ref{sec:massless_case}, that is exact in the SM limit assuming massless
leptons. The new derivation incorporates the effect of electromagnetic correction of
hadronic operators to all orders and all factorizable and non-factorizable
contributions including resonance effects to the decay. In addition to the nine
observables, this new relation depends only on one form-factors ratio:
$\mathsf{P_1}$. The new relation becomes independent of $\mathsf{P_1}$ and
reduces to the one derived in Ref.~\cite{Das:2012kz} in the limit that the
asymmetries $A_7$, $A_8$ and $A_9$ are all zero.

As mentioned repeatedly the inclusion of lepton mass contribution is trivial in
our approach; the effect on all the observables is directly obtained in terms of
asymmetries given in Eqs.~\eqref{eq:A10} and \eqref{eq:A11} that can be measured
as shown in Sec.~\ref{sec:massive_case}.  The new relation obtained is
generalized to include the lepton mass effects in
Eq.~\eqref{eq:Obs-relationnew}. It is important to note that it involves only
observables and the form-factor ratio $\mathsf{P_1}$ and is free from any
assumption within the SM framework. This relation also implies three inequalities
given in Eqs.~\eqref{eq:constraint-Z1o}--\eqref{eq:constraint-Z3o} which impose
constraints on the parameter space of observables. We also presented  three new 
relations between the observables that are exact at the zero crossings of
angular asymmetries $\AFBo$, $A_5^\o$ and $\AFBo+\sqrt{2}A_5^\o$. These are
particularly interesting if the mass effect and the imaginary contributions to
the Wilson coefficients $\widehat{C}_7$ and $\widehat{C}_9$ are ignored, as they
reduce to simple form presented in Eq.~\eqref{eq:zeroassym0}. Another
interesting aspect is that the form-factor ratios $\mathsf{P_1}$, $\mathsf{P_2}$
and $\mathsf{P_3}$ can each be written in terms of observables and
$\mathsf{P_1}$. In the limit of vanishing $A_7$, $A_8$ and $A_9$ (i.e negligible
imaginary contributions), the form-factor ratios can be measured purely in terms
of helicity fractions.

The limiting values of the observables at the minimum and maximum values of
$q^2$ are discussed in Sec.~\ref{sec:observables} based on very general
arguments. It is interesting to note that at $q^2=4m^2$ all angular asymmetries
vanish and each of the helicity fraction approaches
$1/3$. At the maximum value of $\qmax$ similar results can be obtained.

In Sec.~\ref{sec:NPAnalysis}, we have highlighted the possible ways to check the
consistency of the measured observables using the SM relation derived. It was
noted that the inclusion of non-zero $\varepsilon_\lambda$ indicating complex
contributions to the amplitudes invariably reduces the allowed parameter space
of the observables. Hence, in order to check the consistency of measured
observables we take a conservative approach and set all the
$\varepsilon_\lambda$'s to be equal to zero for the analysis. This was
necessary since $A_7$, $A_8$ and $A_9$ are all consistent with zero. The
relation among the observables, hence, reduces to Eq.~\eqref{eq:Obs-relation}
which is in terms of six observables $F_L$, $F_\|$, $F_\perp$, $\AFB$, $A_4$,
$A_5$ and is completely free from any form factor dependence. The $\chi^2$
function in Eq.~\eqref{eq:chisq} was minimized in the $4$-dimensional parameter space of
the observables $F_L$, $F_\perp$, $\AFB$ and $A_5$ to check the consistency
between the experimentally measured values  by varying each of them
simultaneously within the permissible domain and $A_4^\text{pred}$ was evaluated
using the relation in Eq.~\eqref{eq:Obs-relation}. The projections of the
minimized $\chi^2$ function are studied for the various pairs of observables as
shown in the contour plots of Figs.~\ref{fig:FLFP}--\ref{fig:A5AFB}. In most of
the contour plots the best fit (green) points lie at the edge of the boundaries
except for the third bin. The experimental measured central values (black
squares) generally lie within the contours except for the fourth and sixth bin.
It is interesting to note that the best fits are always in the $1\sigma$ region
perhaps validating the LHC$b$ data set.

We compared the two distributions generated by experimental measurement  and
theoretical prediction of the observable $A_4$, assuming that $A_7$, $A_8$ and
$A_9$ are all zero in Fig.~\ref{fig:A4large}. The number of events for the
``Theory'' histogram are chosen to be consistent statistically with the number
of events observed by LHC$b$ in the $1\invfb$~\cite{Aaij:2013iag} data set for
each of the bins. The mean values together with $1\sigma$ error bands are shown
in Fig.~\ref{fig:comp} with a comparison between the massless and massive lepton
case. It is found that lepton mass can be ignored except for the first $q^2$
bin. The fifth bin shows a large discrepancy whereas the other bins are in
reasonable agreement. Since the $A_4$ distributions in Fig.~\ref{fig:A4large}
carry much more information than the mean and averages, we compare the two
simulated distributions shown in the Histograms using the Mathematica routine
``DistributionFitTest''~\cite{fittest}. The $P$-values obtained by comparing the
two are found to be less than $10^{-9}$ for each of the bins, except the second
and fourth bins, where the $P$-values obtained are $2.54\times 10^{-5}$ and
$6.47\times 10^{-6}$ respectively.

In order to understand better the role of the imaginary contributions that were
earlier ignored, we have also preformed a simulation of all observables
including $A_7$ $A_8$ and $A_9$. The solutions for $\varepsilon_\perp$,
$\varepsilon_{\|}$ and $\varepsilon_0$ shown in Fig.~\ref{fig:eps} indicate that
all the $\varepsilon_\lambda$'s are consistent with zero and even the tails of
$\varepsilon_\lambda^2/\Gf$ do not cross $0.2$. A comparison of the measured and
predicted $A_4$ values for the six $q^2$ bins considering all the measured
observables (including $A_7$, $A_8$ and $A_9$) are shown in
Fig.~\ref{fig:A4full}. Interestingly, $A_4^{\text{pred}}$ now fits better  to a
Gaussian distribution than the $\varepsilon_\lambda=0$ case as indicated by a
Kolmogorov-Smirnov test, implying possible imaginary contributions to the
transversity amplitudes.  The $P$-values still continue to be smaller than
$10^{-9}$ for all the bins, except the second bin where the $P$-value is
$6.78\times 10^{-3}$, indicating that we reject the hypothesis that all
observables are consistent with the exact SM relation of
Eq.~\eqref{eq:Obs-relationnew}. Since the discrepency seems to be the largest
for the fifth bin ($14.0\le q^2\le 16.0~\gev^2$), we have performed a detailed
comparison of the PDF curves for both the theoretically predicted values using
$\varepsilon_\lambda=0$ and $\varepsilon_\lambda\ne0$ with the measured value of
$A_4$ as shown in Fig.~\ref{fig:Three}.

In this paper we have derived a relation among the observables by taking into
account all possible effects within Standard Model by restricting ourselves to
rely only on one hadronic input. The violation of this relation will provide a
smoking gun signal of New Physics. We have explicitly shown how the relation can
be used to test SM, and confirm our understanding of the hadronic effects. We
used the $1\invfb$ LHC$b$ measured values of the observables to highlight the
possible ways for the search of new physics that might contribute to this decay
with the derived relations.

\acknowledgments The work of Diganta Das is supported by the DFG Research Unit 
FOR 1873 “Quark Flavour Physics and Effective Field Theories”.
We thank Sheldon Stone for discussion on LHC$b$ observation of 
$B^+\to K^+\mu^+\mu^-$, which motivated our more detailed study.
\appendix

\section{Derivation of $\bf{r_\lambda}$ Solutions} 
\label{sec:appendix-1}

Here we present the derivation of $r_\|$, $r_\perp$ and $r_0$ solutions
defined in Eq.~\eqref{eq:rlambda}. Starting with the first set of equations 
(Set-I) 
involving $r_\|$ and $r_\perp$ in terms of the observables given in 
Eqs.~\eqref{eqI:Fparallel},
\eqref{eqI:Fperp} and \eqref{eqI:AFB} we have
\begin{eqnarray}
  \label{eq:def1}
r_\|^2+\widehat{C}_{10}^2 &=&\frac{F_\|^\prime\Gf 
  \mathsf{P_1^2}}{2\mathcal{F}_\perp^2} ,\\
\label{eq:def2}
r_\perp^2+\widehat{C}_{10}^2 &=&\frac{F_\perp^\prime\Gf}{2\mathcal{F}_\perp^2} 
,\\
\label{eq:def3}
\widehat{C}_{10}(r_\|+r_\perp) &=&\frac{\AFB\Gf 
\mathsf{P_1}}{3\mathcal{F}_\perp^2}.
  \end{eqnarray}
Multiplying Eq.~\eqref{eq:def1} and \eqref{eq:def2}  we can write 
\begin{eqnarray}
  \label{eq:16}
\frac{F_\|^\prime F_\perp^\prime \Gf^2\mathsf{P_1^2}}{4\mathcal{F}_\perp^4}&=&
(r_\|r_\perp-\widehat{C}_{10}^2)^2+\widehat{C}_{10}^2(r_\|+r_\perp)^2\nn\\   
   &=&(r_\|r_\perp-\widehat{C}_{10}^2)^2+\frac{\AFB^2\Gf^2\mathsf{P_1^2}} 
   {9\mathcal{F}_\perp^4}\nn 
\end{eqnarray}
hence,
\begin{eqnarray}
\label{eq1:uv}
r_\|r_\perp-\widehat{C}_{10}^2 &=&\pm\frac{\Gf \mathsf{P_1}}{2
          \mathcal{F}_\perp^2} \sqrt{F_\|^\prime 
          F_\perp^\prime-\frac{4\AFB^2}{9}}.
\end{eqnarray}
Now expressing $\widehat{C}_{10}^2$ in terms of $r_\|^2$  using
Eq.~(\ref{eq:def1}) and in terms of $r_\perp^2$ using 
Eq.~(\ref{eq:def2}) we can write
\begin{eqnarray}
  \label{eq2:uv}
2r_\|r_\perp -2\widehat{C}_{10}^2&=& 2 r_\|r_\perp
 -\Big(\frac{F_\|^\prime\Gf\mathsf{P_1^2}}{2\mathcal{F}_\perp^2}-r_\|^2\Big)-
 \Big(\frac{F_\perp^\prime\Gf}{2\mathcal{F}_\perp^2}-r_\perp^2\Big)\nn\\ 
  &=&\Big[(r_\|+r_\perp)^2-\frac{F_\|^\prime\Gf\mathsf{P_1^2}}
  {2\mathcal{F}_\perp^2}-\frac{F_\perp^\prime\Gf}{2\mathcal{F}_\perp^2}\Big]  
\end{eqnarray}
Equating Eqs.~(\ref{eq1:uv}) and (\ref{eq2:uv}) we get 
\begin{align}
  \label{eq:19}
  r_\|+r_\perp&=\pm\Bigg[\frac{F_\|^\prime\Gf 
  \mathsf{P_1^2}}{2\mathcal{F}_\perp^2}
  +\frac{F_\perp^\prime\Gf}{2\mathcal{F}_\perp^2}\pm  
  \frac{\Gf \mathsf{P_1}}{2 \mathcal{F}_\perp^2} 
  Z_1^\prime\Bigg]^{\nicefrac{1}{2}}\nn
  \\[2.ex]
&=\frac{\pm\sqrt{\Gf}}{\sqrt{2}\mathcal{F}_\perp}\Big[\mathsf{P_1^2}
F_\|^\prime+F_\perp^\prime \pm \mathsf{P_1}Z_1^\prime \Big]^{\nicefrac{1}{2}}
\end{align}
where $Z_1^\prime=\sqrt{4F_\|^\prime F_\perp^\prime -\tfrac{16}{9}\AFB^2}$.
Now, Eqs.~(\ref{eq:def1}) and ~(\ref{eq:def2}) imply: 
\begin{equation}
  \label{eq:20}
r_\|^2-r_\perp^2=
\frac{F_\|^\prime\Gf \mathsf{P_1^2}}{2\mathcal{F}_\perp^2}-
\frac{F_\perp^\prime \Gf}{2\mathcal{F}_\perp^2},
\end{equation}
which gives $r_\|-r_\perp$ to be,
\begin{equation}
  \label{eq:21}
r_\|-r_\perp= \dsp\frac{\pm\sqrt{\Gf}}{\sqrt{2}\mathcal{F}_\perp}\frac{\dsp
  \mathsf{P_1^2} F_\|^\prime -F_\perp^\prime}{\dsp\Big[\mathsf{P_1^2}
F_\|^\prime +F_\perp^\prime \pm \mathsf{P_1} Z_1^\prime 
\Big]^{\nicefrac{1}{2}}}.
\end{equation}

To fix the sign ambiguity of the radical let us consider the zero crossing point
of the observable $\AFB$ where,
\begin{align}
  \label{eq1:u+v}
r_\|+r_\perp\big|_{
\AFB=0}&=\pm\frac{\sqrt{\Gf}}{\sqrt{2}\mathcal{F}_\perp}
\Big(\sqrt{F_\perp^\prime}\pm \mathsf{P_1}\sqrt{F_\|^\prime}~\Big)=0  
\end{align}
It can be easily seen from Appendix.~\ref{sec:form-factors} that $\mathsf{P}_1$
is always negative and thus the positive sign ambiguity has to be chosen within
the radical. Solving Eqs.~\eqref{eq:19} and \eqref{eq:21} we get the expressions
for $r_\|$ and $r_\perp$ given in Eqs.~\eqref{eq:rparallel} and
\eqref{eq:rperp-1}. Similarly, following all the steps stated above for the
other two sets of equations (Set-II and Set-III) we get the solutions for $r_0$
(in Eq.~\eqref{eq:r0}) and two more expressions for the variable $r_\perp$ 
(Eqs.~\eqref{eq:rperp-2} and \eqref{eq:rperp-3}).

Generalization of Eqs.~\eqref{eq:19} and \eqref{eq:21} for the massive case 
in Sec.\ref{sec:massive_case} is trivial from here. Below we present the 
explicit 
expressions for both massless and massive cases. 
\begin{widetext}

\begin{equation}
\label{eq:22}
r_\|+r_\perp=\begin{cases} 
  \dsp \frac{\pm\sqrt{\Gf}}{\sqrt{2}\mathcal{F}_\perp}\Big[\mathsf{P_1^2}
  \Big(F_\|-\dsp\frac{2\varepsilon_\|^2}{\Gf}\Big)+ 
  \Big(F_\perp-\dsp\frac{2\varepsilon_\perp^2}{\Gf}\Big)
  + \mathsf{P_1}Z_1^\prime \Big]^{\nicefrac{1}{2}} ~~~& \mbox{massless case} \\[2.5ex]
  \dsp \frac{\pm\sqrt{\Gf^\o}}{\sqrt{2}\mathcal{F}_\perp 
  \beta}\Big[\mathsf{P_1^2}
  \Big(F_\|^\o-\dsp\frac{\mathcal{T}_\|}{\Gf^\o}\Big)+\Big(F_\perp^\o
  -\dsp\frac{\mathcal{T}_\perp}{\Gf^\o}\Big)
  + \mathsf{P_1}Z_1^\o \Big]^{\nicefrac{1}{2}} ~~~& \mbox{massive case}
\end{cases}
\end{equation}

\begin{equation}
r_\|-r_\perp=\begin{cases} 
\dsp\frac{\pm\sqrt{\Gf}}{\sqrt{2}\mathcal{F}_\perp}\frac{\dsp
\mathsf{P_1^2} \Big(F_\|-\dsp\frac{2\varepsilon_\|^2}{\Gf}\Big)
-\Big(F_\perp-\dsp\frac{2\varepsilon_\perp^2}{\Gf}\Big)}{\dsp\Big[\mathsf{P_1^2}
  \Big(F_\|-\dsp\frac{2\varepsilon_\|^2}{\Gf}\Big)+\Big(F_\perp
  -\dsp\frac{2\varepsilon_\perp^2}{\Gf}\Big)
  + \mathsf{P_1}Z_1^\prime \Big]^{\nicefrac{1}{2}}} ~~~& \mbox{massless case} 
  \\[2ex] 
  \dsp \frac{\pm\sqrt{\Gf^\o}}{\sqrt{2}\mathcal{F}_\perp \beta} \dsp 
  \frac{\mathsf{P_1^2}
  \Big(F_\|^\o-\dsp\frac{\mathcal{T}_\|}{\Gf^\o}\Big)-\Big(F_\perp^\o
  -\dsp\frac{\mathcal{T}_\perp}{\Gf^\o}\Big)}{\dsp\Big[\mathsf{P_1^2}
  \Big(F_\|^\o-\dsp\frac{\mathcal{T}_\|}{\Gf^\o}\Big)+\Big(F_\perp^\o
  -\dsp\frac{\mathcal{T}_\perp}{\Gf^\o}\Big)
  + \mathsf{P_1}Z_1^\o \Big]^{\nicefrac{1}{2}}} ~~~& \mbox{massive case}
\end{cases}
\end{equation} Using Eqs.~\eqref{eq:def3} and \eqref{eq:22} we can write
\begin{equation}
\label{eq:C10hat}
\widehat{C}_{10} =\begin{cases}
\dsp\frac{\pm\AFB\sqrt{2\Gf} \mathsf{P_1}}{\dsp 
3\mathcal{F}_\perp\Big[\mathsf{P_1^2}
  \Big(F_\|-\dsp\frac{2\varepsilon_\|^2}{\Gf}\Big)+ 
  \Big(F_\perp-\dsp\frac{2\varepsilon_\perp^2}{\Gf}\Big)
  + \mathsf{P_1}Z_1^\prime \Big]^{\nicefrac{1}{2}}} ~~~& \mbox{massless 
  case}\\
\dsp\frac{\pm\AFBo\sqrt{2\Gf^\o} \mathsf{P_1}}{\dsp 
3\mathcal{F}_\perp\Big[\mathsf{P_1^2}
  \Big(F_\|^\o-\dsp\frac{\mathcal{T}_\|}{\Gf^\o}\Big)+\Big(F_\perp^\o
  -\dsp\frac{\mathcal{T}_\perp}{\Gf^\o}\Big)
  + \mathsf{P_1}Z_1^\o \Big]^{\nicefrac{1}{2}}} ~~~& \mbox{massive case}  
\end{cases}
\end{equation}

\end{widetext}

\section{Form-factors}
\label{sec:form-factors}

The form-factors $\mathcal{F}_\lambda$ and
$\widetilde{\mathcal{G}}_\lambda$ can be related to the form-factors
$\mathcal{X}_i$ and $\mathcal{Y}_i$ introduced in Eqs.~(\ref{eq:formfactor1})
and (\ref{eq:formfactor2}) by comparing the
expressions for ${\cal A}_\lambda^{L,R}$ in Eqs.~\eqref{eq:trans-amp1}
-- \eqref{eq:trans-amp3} with Eq.~\eqref{eq:amp-def1} as follows:
\begin{subequations}
\begin{align}
\mathcal{F}_\perp=& N\sqrt{2}\sqrt{\lambda(m_B^2,m_\kstar^2,q^2)}
\mathcal{X}_3, \\
\widetilde{\mathcal{G}}_\perp=& N\sqrt{2}\sqrt{\lambda(m_B^2,m_\kstar^2,q^2)}~
  \frac{2(m_b+m_s)}{q^2}\widehat{C}_7 \mathcal{Y}_3\nn \\
  &\qquad +\cdots, \\
\mathcal{F}_\|= & 2\sqrt{2}N \mathcal{X}_1, \\
\widetilde{\mathcal{G}}_\|= & 2\sqrt{2}N~\frac{2(m_b-m_s)}{q^2}\widehat{C}_7 
\mathcal{Y}_1+\cdots, \\
\mathcal{F}_0=& \frac{N}{2 m_\kstar \sqrt{q^2}} 
    \big[ 4k.q \mathcal{X}_1 +\lambda(m_B^2,m_\kstar^2,q^2)\mathcal{X}_2\big],  
    \\ 
\widetilde{\mathcal{G}}_0=& \frac{N}{2 m_\kstar \sqrt{q^2}}
  \frac{2(m_b-m_s)}{q^2}\widehat{C}_7\big[4k.q \mathcal{Y}_1 \nn \\
    & \qquad+\lambda(m_B^2,m_\kstar^2,q^2) \mathcal{Y}_2\big] +\cdots,
\end{align}
\end{subequations}
where these $\mathcal{X}_i$'s and $\mathcal{Y}_i$'s can be related to the well
known form-factors $V$, $A_{0,1,2}$ and $T_{1,2,3}$  by comparing with
ref.~\cite{Beneke:2001at} which are known up to order NNLO in HQET.
However, it should be noted that the $\mathcal{F}_\lambda$ and
$\widetilde{\mathcal{G}}_\lambda$ values are not directly used anywhere
throughout our paper. Only the value of $\mathsf{P_1}$ is used to solve for
$\varepsilon_\lambda$ using Eqs.~\eqref{eq:eps_perpo}--\eqref{eq:eps_0o}.

\begin{subequations}
\begin{align}
\mathcal{X}_0=&-\frac{2 m_\kstar}{q^2} A_0(q^2), \\[2.5ex]
\mathcal{X}_1=& -\frac{1}{2}(m_B + m_{K^*}) A_1(q^2), \\
\mathcal{X}_2=& \frac{A_2(q^2)}{m_B + m_{K^*}},  \\
\mathcal{X}_3=& \frac{V(q^2)}{m_B + m_{K^*}}, \\
\mathcal{Y}_1=& \frac{1}{2}(m_B^2-m^2_{K^*}) T_2(q^2), \\
\mathcal{Y}_2=& -T_2(q^2)- \frac{q^2}{m_B^2-m^2_{K^*}} T_3(q^2), \\ 
\mathcal{Y}_3=& -T_1(q^2).
\end{align}
\end{subequations}
Here a point to be noted that as the form-factors $A_1$ and $A_2$ are always
positive the ratio 
\begin{align}
\frac{2k.q (m_B + m_{K^*})^2}{\lambda(m_B^2,m_\kstar^2,q^2)}\frac{A_1}{A_2} \ge 
0
\end{align}
giving rise to the fact that $\mathcal{F}_\|$ and $\mathcal{F}_0$ always have 
the 
same sign which is negative.

\end{document}